\documentclass{article}

\usepackage{chicago}

\newcommand{\iec}{\mbox{i.\,e.\,}}

\newcommand{\eg}{\mbox{e.\,g.\,\ }}
\newcommand{\egc}{\mbox{e.\,g.\,}}
\newcommand{\etc}{etc.\,\ }

\newcommand{\be}{\begin{equation}}
\newcommand{\ee}{\end{equation}}

\newcommand{\pb}[2]{\ensuremath{\left\{ #1 , #2 \right\} }}
\newcommand{\vctr}[1]{\ensuremath{\mathbf{ #1 }}}
\newcommand{\dr}[1]{\ensuremath{\mathrm{d} #1\,}}
\newcommand{\mc}[1]{\ensuremath{\mathcal{#1}}}
\newcommand{\dbd}[2]{\ensuremath{\frac{\dr{#1}}{\dr{#2}}}}
\newcommand{\pbp}[2]{\ensuremath{\frac{\partial #1}{\partial #2}}}
\newcommand{\ddt}{\ensuremath{\frac{\dr{}}{\dr{t}}}}

\newcommand{\ket}[1]{\ensuremath{\left|  #1 \right\rangle}}
\newcommand{\bra}[1]{\ensuremath{\left\langle #1 \right|}}
\newcommand{\bk}[2]{\ensuremath{\left\langle #1 | #2 \right\rangle}}
\newcommand{\proj}[2]{\ensuremath{\ket{#1} \bra{#2}}}
\newcommand{\tpk}[2]{\ensuremath{\ket{#1}\!\otimes\!\ket{#2}}}
\newcommand{\matel}[3]{\ensuremath{\bra{#1} #2 \ket{#3}}}
\newcommand{\op}[1]{\ensuremath{\widehat{\textsf{\ensuremath{#1}}}}}
\newcommand{\opad}[1]{\ensuremath{\op{#1}^{\dagger}}}
\newcommand{\id}{\op{\mathsf{1}}}
\newcommand{\comm}[2]{\ensuremath{\left[ #1 , #2 \right]}} 
\newcommand{\tr}{\textsf{Tr}}
\newcommand{\denop}{\op{\rho}}

\begin{document}

\title{The Quantum Measurement Problem: State of Play}
\author{David Wallace\thanks{Balliol College, Oxford OX1 3BJ; \texttt{david.wallace@balliol.ox.ac.uk}}}
\date{July 2007}
\maketitle

\begin{abstract}
This is a preliminary version of an article to appear in the forthcoming \emph{Ashgate Companion to the New Philosophy of Physics}. In it, I aim to review, in a way accessible to foundationally interested physicists as well as physics-informed philosophers, just where we have got to in the quest for a solution to the measurement problem. I don't advocate any particular approach to the measurement problem (not here, at any rate!) but I do focus on the importance of decoherence theory to modern attempts to solve the measurement problem, and I am fairly sharply critical of some aspects of the ``traditional'' formulation of the problem.
\end{abstract} 

\section*{Introduction}

By some measures, quantum mechanics (QM) is the great success story of modern physics: no other physical theory has come close to the range and accuracy of its predictions and explanations. By other measures, it is instead the great scandal of physics: despite these amazing successes, we have no satisfactory physical theory at all --- only an ill-defined heuristic which makes unacceptable reference to primitives such as ``measurement'', ``observer'' and even ``consciousness''.

This is the measurement problem, and it dominates philosophy of quantum mechanics. The great bulk of philosophical  work on quantum theory over the last half-century has been concerned either with the strengths and weaknesses of particular interpretations of QM --- that is, of particular proposed solutions to the measurement problem --- or with general constraints  on interpretations. Even questions which are notionally not connected to the measurement problem are hard to disentangle from it: one cannot long discuss the ontology of the wavefunction\footnote{Here and afterwards I follow the physicists' standard usage by using ``wavefunction'' to refer, as appropriate, either to the putatively \emph{physical} entity which evolves according to the Schr\"{o}dinger equation or to complex-valued function which represents it mathematically. I adopt a similar convention for ``state vector''.}, or the nature of locality in relativistic quantum physics, without having to make commitments which rule out one interpretation or another.

So I make no apologies that this review of ``the philosophy of quantum mechanics'' is focussed sharply on the measurement problem. Section \ref{DMWWproblem} sets up the problem from a modern perspective; section   \ref{DMWWdecoherence} is a self-contained discussion of the phenomenon of decoherence, which has played a major part in physicists' recent writings on the measurement problem. In sections \ref{DMWW3candidates}--\ref{DMWWhidden} I discuss the main approaches to solving the measurement problem currently in vogue: modern versions of the Copenhagen interpretation; the Everett interpretation; dynamical collapse; and hidden variables. Finally, in section \ref{DMWWRQM} I generalise the discussion beyond non-relativistic physics. I give a self-contained, non-mathematical introduction to quantum field theory, discuss some of \emph{its} conceptual problems, and draw some conclusions for the measurement problem.

\section{The Measurement Problem: a modern approach}\label{DMWWproblem}

The goal of this section is a clean statement of what the measurement problem actually is. Roughly speaking, my statement will be that QM provides a very effective \emph{algorithm} to predict macroscopic phenomena (including the results of measurements which purportedly record microscopic phenomena) but that it does not provide a satisfactorily formulated physical \emph{theory} which explains the success of this algorithm. We begin by formulating this algorithm.

\subsection{QM: formalism and interpretation}\label{DMWWquantumalgorithm}

To specify a quantum system, we have to give three things:
\begin{enumerate}
\item A Hilbert space \mc{H}, whose normalised vectors represent the possible states of that system.\footnote{More accurately: whose \emph{rays} --- that is, equivalence classes of normalised vectors under phase transformations --- represent the possible states of the system. }
\item Some additional structure on \mc{H} (all Hilbert spaces of the same dimension are isomorphic, so we need additional structure in order to describe specific systems). The additional structure is given by one or both of
\begin{itemize}
\item Certain \emph{preferred operators} on Hilbert space (or, certain \emph{preferred sets of basis vectors}).
\item A \emph{preferred decomposition} of the system into subsystems.
\end{itemize}
\item A \emph{dynamics} on \mc{H}: a set of unitary transformations which take a state at one time to the state it evolves into at other times. (Normally the dynamics is specified by the \emph{Hamiltonian}, the self-adjoint operator which generates the unitary transformations).
\end{enumerate}
For instance:
\begin{enumerate}
\item Non-relativistic one-particle QM is usually specified by picking a particular triple of operators and designating them as the position operators, or equivalently by picking a particular representation of states as functions on $\mathbf{\mathrm{R}}^3$ and designating it as the configuration-space representation. More abstractly, it is sometimes specified by designating pairs of operators $\langle \op{Q}_i,\op{P}_i\rangle$ (required to satisfy the usual commutation relations) as being the position and momentum observables.
\item In quantum computation a certain decomposition of the global Hilbert space into 2-dimensional component spaces is designated as giving the Hilbert spaces of individual qubits (normally taken to have spatially definite locations); sometimes a particular basis for each qubit is also designated as the basis in which measurements are made.
\item In quantum field theory (described algebraically) a map is specified from spatial regions to subalgebras of the operator algebras of the space, so that the operators associated with region $R$ are designated as representing the observables localised in $R$. At least formally, this can be regarded as defining a component space comprising the degrees of freedom at $R$.
\end{enumerate}
(Note that, although the specification of a quantum system is a bit rough-and-ready, the quantum systems \emph{themselves} have perfectly precise mathematical formalisms).

I shall use the term \emph{bare quantum formalism} to refer to a quantum theory picked out in these terms, prior to any notion of probability, measurement etc. Most actual calculations done with quantum theory --- in particle physics, condensed matter physics, quantum chemistry, \etc  --- can be characterised as calculations of certain mathematical properties of the bare quantum formalism (the expectation values of certain functions of the dynamical variables, in the majority of cases.)

Traditionally, we extract \emph{empirical content} from the bare formalism via some notion of \emph{measurement}. The standard ``textbook'' way to do this is to associate measurements with self-adjoint operators: if \ket{\psi} is the state of the system and \op{M} is the operator associated with some measurement, and if 
\be
\op{M}=\sum_i m_i \op{P}_i
\ee
is \op{M}'s spectral decomposition into projectors, then the probability of getting result $m_i$ from the measurement is $\matel{\psi}{\op{P}_i}{\psi}$. 

Now the really important thing here is the set of projectors $\op{P}_i$ and not $\op{M}$ itself: if we associate the measurement with $f(\op{M})$, which has spectral decomposition
\be
f(\op{M})=\sum_i f(m_i) \op{P}_i
\ee
then the only difference is that the different possible measurement outcomes are labeled differently. Hence it has become normal to call this sort of measurement a \emph{projection-valued measurement}, or PVM.

It has become widely accepted that PVMs are not adequate to represent all realistic sorts of measurement. In the more general  \emph{Positive-operator-valued measurement} formalism (see, \egc, \citeN[282--9]{peres}, \citeN[pp.\,90--92]{nielsenchuang}) a measurement process is associated with a family of positive operators $\{\op{E}_1, \ldots \op{E}_n\}$.
Each possible outcome of the measurement is associated with one of the $\op{E}_i$, and the probability of getting result $i$ when a system in state \ket{\psi} is measured is \matel{\psi}{\op{E}_i}{\psi}. PVMs are a special case, obtained only when each of the $\op{E}_i$ is a projector. This framework has proved extremely powerful in analyzing actual measurements (see, for instance, the POVM account of the Stern-Gerlach experiment given by \citeN{buschmeasurement}).

How do we establish \emph{which} PVM or POVM should be associated with a particular measurement? There are a variety of more-or-less \emph{ad hoc} methods used in practice, e.g.
\begin{enumerate}
\item In non-relativistic particle mechanics we assume that the probability of finding the system in a given spatial region \mc{R} is given by the usual formula 
\be
\Pr(x \in \mc{R})=\int_R |\psi|^2.
\ee
\item In high-energy particle physics,  if the system is in a state of definite particle number and has momentum-space expansion
\be
\ket{\psi}=\int \dr{^3 k} \alpha(\vctr{k}) \ket{\vctr{k}}
\ee
then we assume that the probability of finding its momentum in the vicinity of some $\vctr{k}$ is proportional to $|\alpha(\vctr{k})|^2$.
\item Again in non-relativistic particle mechanics, if we are making a joint measurement of position and momentum then we take the probability of finding the system in the vicinity of some phase-space point $(\vctr{q},\vctr{p})$ is given by one of the various ``phase-space POVMs'' \cite{buschmeasurement}.
\end{enumerate}

But ``measurement'' is a physical process, not an unanalyzable primitive, and  physicists routinely apply the formalism of quantum physics to the analysis of measurements themselves. Here we encounter a regress, though: if we have to construct a quantum-mechanical description of measurement, how do we extract empirical content from \emph{that description}? In actual physics, the answer is:  the regress ends when the measurement process has been magnified up to have \emph{macroscopically large} consequences. That is: if we have some microscopic system in some superposed state then the empirical content of that state is in principle determined by careful analysis of the measurement process applied to it. If the superposition is between macroscopically different states, however, we may directly read empirical content from it: a system in state
\be
\alpha \ket{\mathrm{\mbox{Macroscopic state 1}}}+ \beta\ket{\mathrm{\mbox{Macroscopic state 2}}}
\ee
 is interpreted, directly, as having probability $|\alpha|^2$ of being found in macroscopic state 1.
 
Let us get a little more precise about this. 
\begin{enumerate}
\item We identify some of the system's dynamical variables (that is, some of its self-adjoint operators) somehow as being the positions $\op{Q}_i$ and momenta $\op{P}_i$ of some macroscopic degrees of freedom of the system . For instance, for a simple system such as a macroscopic pointer, the centre-of-mass position and conjugate momentum of the system will suffice. For something more complicated (such as a fluid) we normally take the macroscopic degrees of freedom to be the density of the fluid averaged over some spatial regions large compared to atomic scales but small compared to macroscopic ones.
\item We decompose the Hilbert space of the system into a component space $\mc{H}_{macro}$ described by these macroscopic variables, and a space $\mc{H}_{micro}$ for the remaining degrees of freedom:
\be \mc{H}=\mc{H}_{macro}\otimes \mc{H}_{micro}.\ee
\item We construct \emph{wave-packet states} \ket{q^i,p_i} in $\mc{H}_{macro}$ --- Gaussian states, fairly localised around particular values ($q^i,p_i$)of $\op{Q}_i$ and $\op{P}_i$. These are the states which physicists in practice regard as ``macroscopically definite'': that is, located at the phase-space point ($q^i,p_i$). (We leave aside the \emph{conceptual } problems with regarding them thus: for now, we are interested in explicating only the pragmatic method used to extract empirical content from QM.)
\item Next, we expand the state in terms of them:
\be \ket{\psi}=\int \dr{p_i} \dr{q^i} \alpha(q^i,p_i)\tpk{q^i,p_i}{\psi(q^i,p_i)}.\ee
\item We regard \ket{\psi}, expanded thus, as a probabilistic mixture. That is, we take the probability density of finding the system's macroscopic variables to be in the vicinity of ($q^i,p_i$) to be $|\alpha(q^i,p_i)|^2$. Or to be (slightly) more exact, we take the probability of finding the system's macroscopic variables to be  in some reasonably large set $V$ to be
\be \  \int_V \dr{p_i} \dr{q^i} |\alpha(q^i,p_i)|^2.\ee 
\end{enumerate}
We might call this the \emph{Quantum Algorithm}. Empirical results are extracted from the Bare Quantum Formalism by applying the Quantum Algorithm to it.

 \subsection{The Measurement Problem}\label{DMWWinterpretation}

 The Bare Quantum Formalism (for any given theory) is an elegant piece of mathematics; the Quantum Algorithm is an ill-defined and unattractive mess.  And this is the Measurement Problem.
\begin{quote}
\textbf{The Measurement Problem:} Applying the Quantum Algorithm to the Bare Quantum Formalism produces extremely accurate predictions about macroscopic phenomena: from the results of measurement processes to the boiling points of liquids. But we have no satisfactorily formulated scientific theory which reproduces those predictions.
\end{quote}

A solution of the measurement problem, then, is a satisfactorily formulated scientific theory (``satisfactorily formulated'', that is, relative to your preferred philosophy of science) from which we can explain why the Quantum Algorithm appears to be correct. Most such solutions do so by providing theories from which we can prove that the Algorithm \emph{is} correct, at least in the vast majority of experimental situations. There is no requirement here that different solutions are empirically indistinguishable; two solutions may differ from one another, and from the predictions of the Algorithm, in some exotic and so-far-unexplored experimental regime.

(Why call it the \emph{measurement} problem? Because traditionally it has been the measurement process which has been taken as the source of macroscopic superpositions, and because only when we have such superpositions do we have any need to apply the Quantum Algorithm. But processes other than formal measurements --- the amplification of classical chaos into quantum-mechanical indeterminateness, in particular --- can also give rise to macroscopic superpositions.)

Solutions of the measurement problem are often called ``interpretations of QM'', the idea being that all such ``interpretations'' agree on the formalism and thus on the experimental predictions. But in fact, different proposed solutions of the measurement problem are often different physical theories with different formalism. Where possible, then, I avoid using ``interpretation'' in this way (though often tradition makes it unavoidable). 

There is, however, a genuinely interesting distinction between those proposed solutions which do, and those which do not, modify the formalism. It will be helpful to make the following definition: a \emph{pure interpretation} is a (proposed) solution of the measurement problem which has \emph{no mathematical formalism} other than the Bare Quantum Formalism. Proposed solutions which are \emph{not} pure interpretations I call \emph{modificatory}: a modificatory solution either adds to the bare formalism, or modifies it (by changing the dynamics, for instance), or in principle eliminates it altogether.

 \subsection{Against the traditional account of quantum mechanics}\label{DMWWagainst}

 There is a more traditional way to formulate QM, which goes something like this:
\begin{enumerate}
\item A quantum system is represented by a vector \ket{\psi} in a Hilbert space \mc{H}.
\item Properties of the system are represented by projectors on \mc{H}.
\item If and only if \ket{\psi} is an eigenstate of some projector, the system possesses the property associated with that projector; otherwise the value is `indefinite' or `indeterminate' or somesuch. (The `eigenvalue-eigenvector link')
\item A measurement of some property associated with projector \op{P} will find that property to be possessed by the system with probability \matel{\psi}{\op{P}}{\psi}.
\end{enumerate} 
From this perspective, the ``measurement problem'' is the problem of understanding what `indefinite' or `indeterminate' property possession means (or modifying the theory so as to remove it) and of reconciling the indefiniteness with the definite, probabilistically determined results of quantum measurements.

However,  this ``traditional account'' is not an ``interpretation-neutral'' way of stating the basic assumptions of QM; it is a false friend. Primarily, this is because it fails  to give a good account of how physicists in practice apply QM:  it assumes that measurements can be treated as PVMs, whereas as we have seen, it is now generally accepted that many practical measurement processes are best understood via the more general POVM formalism.

This is particularly clear where continuous variables are concerned --- that is, where almost all the quantities measured in practice are concerned. Here, physicists will normally regard a system as ``localised'' at some particular value of some continuous variable --- position, usually --- if its wavefunction is strongly peaked around that value. The fact that the wavefunction strictly speaking vanishes nowhere does not seem to bother them. In particular,  measurements frequently measure continuous variables, and frequently output the result using further continuous variables (such as a pointer position). The practical criterion for such measurements is that if the system being measured is localised in the vicinity of $x$, the pointer displaying the result of the measurement should end up localised near whatever pointer position is supposed to display ``$x$''. This is straightforwardly represented via a POVM, but there is no natural way to understand it in terms of projections and the properties which they are supposed to represent.

Independent of recent progress in physics, there are reasons internal to philosophy of QM to be skeptical about the traditional account. As we shall see, very few mainstream interpretations of QM fit this framework: mostly they either treat the wavefunction as a physical thing (whose ``properties'' are then any properties at all of that thing, not just the property of being an eigenstate of some particular operator); or they associate physical properties to some additional ``hidden variables''; or they deny that the system has observer-independent properties at all.

One of the recurring themes of this chapter will be that the traditional account, having been decisively rejected in the practice of physicists, should likewise be discarded by philosophers: it distorts the philosophy of QM, forcing interpretations into Procrustean beds and encouraging wild metaphysics.

\section{Decoherence theory}\label{DMWWdecoherence}

Quite apart from its conceptual weaknesses, it is \emph{prima facie} surprising that the Quantum Algorithm is well-defined enough to give any determinate predictions at all. For the division between `macroscopic' and `microscopic' degrees of freedom, essential to its statement, was defined with enormous vagueness. Over \emph{how large} a spatial region must we average to get macroscopic density? --- $10^{-5} \mathrm{m}$? $10^{-4} \mathrm{m}$? Fortunately, it is now fairly well understood how to think about this question, thanks to one of the most important quantum-foundational developments of recent years: decoherence theory. 

\subsection{The concept of decoherence}\label{DMWWdecoherencegeneral}

Suppose we have some unitarily-evolving quantum system, with Hilbert space \mc{H}, and consider some decomposition of the system into component subsystems:
\be \mc{H}=\mc{H}_{sys}\otimes \mc{H}_{env},\ee
which we will refer to as the \emph{system} and the \emph{environment}.
Now suppose that $\{\ket{\alpha}\}$ is some (not-necessarily-orthogonal) basis of $\mc{H}_{sys}$ and that the dynamics of the joint system is such that, if we prepare it in a product state
\be
\tpk{\alpha}{\psi}
\ee
then it evolves rapidly into another pure state
\be
\tpk{\alpha}{\psi;\alpha}
\ee
with $\bk{\psi;\alpha}{\psi;\beta}\simeq \delta(\alpha-\beta)$. (Here, ``rapidly'' means rapidly relative to other relevant dynamical timescales). In other words, we suppose that the environment measures the system in the $\{\ket{\alpha}\}$ basis and records the result.

Suppose further that this ``recording'' is reasonably robust, so that subsequent system-environment interactions do not tend to erase it: that is, we don't get evolutions like
\be
\lambda_1 \tpk{\alpha_1}{\psi;\alpha_1}+\lambda_2 \tpk{\alpha_2}{\psi;\alpha_2}
\longrightarrow 
\tpk{\phi}{\chi}.
\ee
In this (loosely-defined) situation, we say that the environment \emph{decoheres} the system, and that the basis $\{\ket{\alpha}\}$ is a \emph{preferred} basis or \emph{pointer} basis. The timescale on which the recording of the system's state occurs is called the \emph{decoherence timescale}.

Much follows from decoherence. The most obvious effects are \emph{synchronic} (or at least, have a consequence which may be expressed synchronically):  the system cannot stably be prepared in superpositions of pointer-basis states. Such superpositions very rapidly become entangled with the environment. Conversely, if the system is prepared in a pointer-basis state, it will remain stably in that pointer-basis state (at least for times long compared to the decoherence timescale). Equivalently, the density operator of the system, when expressed in the pointer basis, will be diagonal or nearly so.

However, the more important consequence is diachronic. If the environment is keeping the density operator almost diagonal, then interference terms between elements of the pointer basis must be being very rapidly suppressed, and the evolution is effectively \emph{quasi-classical}. 

To see this more clearly, suppose that the dynamics of the system is such that after time $t$, we have the evolution
\be \label{DMWWdeco1}\tpk{\alpha_1}{\psi_1} \longrightarrow \ket{\Lambda_1}=\lambda_{11}\tpk{\alpha_1}{\psi_{11}}+\lambda_{12}\tpk{\alpha_2}{\psi_{12}};
\ee
\be \label{DMWWdeco2}\tpk{\alpha_2}{\psi_2} \longrightarrow \ket{\Lambda_2}=\lambda_{21}\tpk{\alpha_1}{\psi_{21}}+\lambda_{22}\tpk{\alpha_2}{\psi_{22}}.
\ee
By linearity, the superposition
\be \label{DMWWdecosup}
\ket{\Psi}=\mu_1 \tpk{\alpha_1}{\psi_1}
+\mu_2 \tpk{\alpha_2}{\psi_2}
\ee
evolves in the same time to
\[
\mu_1 \left( \lambda_{11}\tpk{\alpha_1}{\psi_{11}}+\lambda_{12}\tpk{\alpha_2}{\psi_{12}}\right)
+
\mu_2 \left( \lambda_{21}\tpk{\alpha_1}{\psi_{21}}+\lambda_{22}\tpk{\alpha_2}{\psi_{22}}\right)
\]
\be \label{DMWWdecoresult}
= \ket{\alpha_1}\left(   \mu_1 \lambda_{11}\ket{\psi_{11}}  +\mu_2 \lambda_{21}\ket{\psi_{21}}\right)
+\ket{\alpha_2}\left(   \mu_1 \lambda_{12}\ket{\psi_{12}}  +\mu_2 \lambda_{22}\ket{\psi_{22}}\right).
\ee

Now, suppose that we want to interpret states \ket{\Psi}, \ket{\Lambda_1} and \ket{\Lambda_2} probabilistically with respect to the $\{\ket{\alpha}\}$ --- for example, in \ket{\Psi} we want to interpret $|\mu_1|^2$ as the probability of finding the system in state \ket{\alpha_1}. Generally speaking, interference makes this impossible: (\ref{DMWWdeco1}) and (\ref{DMWWdeco2}) would entail that if the joint system is initially  in state \tpk{\alpha_i}{\psi_i}, after time $t$ there is probability $|\lambda_{i1}|^2$ of finding the system in state $\ket{\alpha_1}$. Applying the probabilistic interpretation to \ket{\Psi} tells us that the joint system initially has probability $|\mu_i|^2$ of indeed being initially in state \tpk{\alpha_i}{\psi_i}, and hence the system has probability
\be P= |\mu_1|^2 |\lambda_{11}|^2 + |\mu_2|^2 |\lambda_{21}|^2
\ee
of being found in $\ket{\alpha_1}$ after time $t$. But if we apply the probabilistic interpretation directly to (\ref{DMWWdecoresult}), we get a contradictory result:
\be
P' = |\mu_1|^2 |\lambda_{11}|^2 + |\mu_2|^2 |\lambda_{21}|^2 + 2\mathrm{\mbox{Re}}(\mu_1^*\lambda_{11}^*\mu_2\lambda_{21}\bk{\psi_{11}}{\psi_{21}}).
\ee
Crucially, though, the contradiction is eliminated and we get the same result in both cases (irrespective of the precise values of the coefficients) \emph{provided} that $\bk{\psi_{11}}{\psi_{21}}=0$. And this is exactly what decoherence, approximately speaking, guarantees: the states \ket{\psi_{11}} and \ket{\psi_{21}} are approximately-orthogonal records of the distinct states of the system in the original superposition.

So: we conclude that in the presence of decoherence, and provided that we are interested only in the state of the system and not of the environment, it is impossible to distinguish between a \emph{superposition} of states like $\tpk{\alpha}{\psi_\alpha}$ and a mere probabilistic \emph{mixture} of such states. 

\subsection{Domains and rates of decoherence}\label{DMWWdecoherencerates}

When does decoherence actually occur? Some clear results have been established:
\begin{enumerate}
\item The macroscopic degrees of freedom of a system are decohered by the microscopic degrees of freedom.
\item The pointer basis picked out in $\mc{H}_{macro}$ is a basis of quasi-classical, Gaussian states.
\end{enumerate}
This should not be surprising. Decoherence occurs because the state of the system is recorded by the environment; and, because the dynamics of our universe are spatially local, the environment of a macroscopically large system in a given spatial position will inevitably record that position. (A single photon bouncing off the system will do it, for instance.)  So superpositions of systems in macroscopically distinct positions will rapidly become entangled with the environment. And superpositions of states with macroscopically distinct \emph{momentums} will very rapidly evolve into states of macroscopically distinct positions. The only states which will be reasonably stable against the decoherence process will be wave-packets whose macroscopic degrees of freedom have reasonably definite position \emph{and} momentum.

Modelling of this decoherence process (both computationally and mathematically) shows that\footnote{These figures are derived from data presented in \citeN{joosetal}.} 
\begin{enumerate}
\item The process is extremely rapid. For example:
\begin{enumerate}
\item A dust particle of size $\sim 10^{-3} \mathrm{cm}$ in a superposition of states $\sim 10^{-8} \mathrm{m}$ apart will become decohered by sunlight after $\sim 10^{-5}$ seconds, and by the Earth's atmosphere after $\sim 10^{-18} \mathrm{s}$; the same particle in a superposition of states $\sim 10^{-5} \mathrm{m}$ apart will become decohered by sunlight in $\sim 10^{-13} \mathrm{s}$ (and by the atmosphere in $10^{-18}\mathrm{s}$ again: once the separation is large compared to the wavelength of particles in the environment then the separation distance becomes irrelevant.)  
\item A kitten in a superposition of states $10^{-10}\mathrm{m}$ apart is decohered by the atmosphere in $\sim 10^{-25}\mathrm{s}$ and by sunlight in $\sim 10^{-8}\mathrm{s}$; the same kitten in a superposition of states $10^{-5} \mathrm{m}$ apart is decohered by the atmosphere in $\sim 10^{-26}\mathrm{s}$, by sunlight in $\sim 10^{-21}\mathrm{s}$, and by the microwave background radiation in $\sim 10^{-15}\mathrm{s}$.
\end{enumerate}
\item In general there is no need for the ``environment'' to be in some sense \emph{external} to the system. In general, the macroscopic degrees of freedom of a system can be decohered by the residual degrees of freedom of that same system: in fluids, for instance, the `hydrodynamic' variables determined by averaging particle density and velocity over regions large compared to particle size are decohered by the remaining degrees of freedom of the fluid. 
\item The dynamics of the macroscopic degrees of freedom seem, in general, to be `quasi-classical' not just in the abstract sense that they permit a probabilistic interpretation, but in the more concrete sense that they approximate classical equations of motion. To be more precise: 
\begin{enumerate}
\item If the classical limit of the system's dynamics  is classically regular (\iec, non-chaotic), as would be the case for a heavy particle moving inertially, then the pointer-basis states evolve, to a good approximation, like the classical states they are supposed to represent. That is, if the classical-limit dynamics would take the phase-space point $(\vctr{q},\vctr{p})$ to (\vctr{q}(t),\vctr{p}(t)), then the quantum dynamics are approximately
\be
\tpk{\vctr{q},\vctr{p}}{\psi}\longrightarrow \tpk{\vctr{q}(t),\vctr{p}(t)}{\psi(t)}.
\ee
\item If the classical limit of the system's dynamics is chaotic, then classically speaking a localised region in phase space will become highly fibrillated, spreading out over the energetically available part of phase-space (while retaining its volume). The quantum system is unable to follow this fibrillation: on timescales comparable to those on which the system becomes classically unpredictable, it spreads itself across the entire available phase space region:
\be \tpk{\vctr{q},\vctr{p}}{\psi}\longrightarrow \int_\Omega \dr{q}\dr{p}\tpk{\vctr{q},\vctr{p}}{\psi_{q,p}(t)}
\ee
(where $\Omega$ is the available region of phase space). In doing so, it still tracks the coarse-grained behaviour of the classical system, but fails to track the fine details: thus, classical unpredictability is transformed into quantum indeterminacy.
\end{enumerate}
\end{enumerate}
For our purposes, though, the most important point is this: decoherence gives a criterion for applicability of the Quantum Algorithm. For the `quasi-classical' dynamics that it entails for macroscopic degrees of freedom is a guarantee of the \emph{consistency} of that algorithm: provided `macroscopic' is interpreted as `decohered by the residual degrees of freedom beyond our ability to detect coherence', then the algorithm will give the same results regardless of exactly when, and at what scales, the algorithm is deployed to make a probabilistic interpretation of the quantum state.

\subsection{Sharpening decoherence: consistent histories}\label{DMWWhistoryframework}
The presentation of decoherence given in the previous section was somewhat loosely defined, and it will be useful to consider the most well-developed attempt at a cleaner definition: the \emph{consistent histories} formalism. To motivate this formalism, consider a decoherent system with pointer basis $\{\ket{\alpha}\}$, as above, and suppose (as is not in fact normally the case) that the pointer basis is discrete and orthonormal: $\bk{\alpha}{\beta}$=$\delta_{\alpha,\beta}$.
Suppose also that we consider the system only at discrete times $t_0,t_1,\ldots t_n$.
Now, decoherence as we defined it above is driven by the establishment of records of the state of the system (in the pointer basis) made by the environment. Since we are discretising time it will suffice to consider this record as made only at the discrete times (so the separation $(t_{n+1}-t_n)$ must be large compared with the decoherence timescale).  Then if the system's state at time $t_0$ is 
\be \ket{\Psi_0}=\sum_{i_0} \mu_{i_0} \tpk{\alpha_{i_0}}{\psi(i_0)}
\ee
it should evolve by time $t_1$ into some state like
\be \ket{\Psi_1}=\sum_{i_0,i_1}\mu_{i_0}\Lambda_1(i_0,i_1)\tpk{\alpha_{i_1}}{\psi(i_0,i_1)}
\ee
(for some transition coefficients $\Lambda_1(i_0,i_1)$), with the states \ket{\psi(i_0,i_1)} \emph{recording} the fact that the system (relative to that state) was in state $\ket{\alpha_{i_0}}$ and is now into $\ket{\alpha_{i_1}}$ (and thus being orthogonal to one another). Similarly, by time $t_2$ the system will be in state
\be \ket{\Psi_1}=\sum_{i_0,i_1,i_2}\mu_{i_0}\Lambda_1(i_0,i_1)\Lambda_2(i_1,i_2)\tpk{\alpha_{i_2}}{\psi(i_0,i_1,i_2)}
\ee
and (iterating)
by time $t_n$ will finish up in state
\be \ket{\Psi_n}=\sum_{i_0,i_1,\cdots i_n}\mu_{i_0}\Lambda_1(i_0,i_1)\cdots\Lambda_n(i_{n-1},i_n)\tpk{\alpha_{i_n}}{\psi(i_0,i_1,\ldots,i_n)}.
\ee
Since we require (by definition) that record states are orthogonal or nearly so, we have
\be
\bk{\psi(i_0,\ldots i_n)}{\psi(j_0,\ldots j_n)}\simeq \delta_{i_0,j_0}\cdots \delta_{i_n,j_n}.
\ee
There is an elegant way to express this, originally due to \citeN{griffiths} and developed by \citeN{gellmannhartle} and others. For each
$\ket{\alpha_i}$, and each of our discrete times $t_0, \ldots t_n$, let $\op{P}_i$ be the projector
\be \op{P}_i(t_j)=\opad{U}(t_j,t_0)\left( \proj{\alpha_i}{\alpha_i}\otimes \id \right) \op{U}(t_j,t_0),
\ee
where $\op{U}(t_j,t_0)$ is the unitary evolution operator taking states at time $t_0$ to states at time $t_j$ (unless the Hamiltonian is time-dependent, $\op{U}(t_j,t_0)=\exp(-i(t_j-t_0)\op{H}/\hbar)$).
Then for any sequence $\vctr{i}=(i_0,i_1,\ldots i_n)$ of indices we may define the \emph{history operator} $\op{C}_\vctr{i}$
by
\be \label{DMWWhistdef}
\op{C}_\vctr{i}= \op{P}_{i_n}(t_n)\cdots \op{P}_{i_0}(t_0).
\ee
Now,
\[ \op{P}_{i_0}(t_0)\ket{\Psi_0}=\mu_{i_0}\tpk{\alpha_{i_0}}{\psi(i_0)};\]
\[\op{P}_{i_1}(t_1)\op{P}_{i_0}(t_0)\ket{\Psi_0}=\mu_{i_0}\op{P}_{i_1}(t_1)\tpk{\alpha_{i_0}}{\psi(i_0)}
=\mu_{i_0}\Lambda_{t_1}(i_0,i_1)\tpk{\alpha_{i_1}}{\psi(i_0,i_1)}
\] 
\[\cdots\]
\be
\op{C}_\vctr{i}\ket{\Psi_0}=\mu_{i_0}\Lambda_1(i_0,i_1)\cdots\Lambda_n(i_{n-1},i_n)\tpk{\alpha_{i_n}}{\psi(i_0,i_1,\ldots,i_n)}.
\ee

This has an immediate corollary:
\be
\matel{\Psi_0}{\opad{C}_{\vctr{j}}\op{C}_{\vctr{i}}}{\Psi_0}\propto \bk{\alpha_{j_n}}{\alpha_{i_n}}\bk{\psi(j_0,j_1,\ldots,j_n)}{\psi(i_0,i_1,\ldots,i_n)}
\ee
and hence 
\be \label{DMWWmeddec}
\matel{\Psi_0}{\opad{C}_{\vctr{j}}\op{C}_{\vctr{i}}}{\Psi_0}\simeq 0 \mathrm{\mbox{  unless  }}\vctr{i}=\vctr{j}.
\ee
Furthermore, if we apply the Quantum Algorithm, it tells us that the probability of the system being found successively in states (corresponding to)
$\ket{\alpha_{i_0}}, \ldots \ket{\alpha_{i_n}}$ is given by $\matel{\Psi_0}{\opad{C}_{\vctr{i}}\op{C}_{\vctr{i}}}{\Psi_0}$ . The condition (\ref{DMWWmeddec}) then has a natural interpretation: it tells us that there is no interference between distinct histories, so that the Quantum Algorithm can be applied at successive times without fear of contradiction.

Now let us generalise. Given an \emph{arbitrary} complete set of projectors $\op{P}_i(t_j)$ for each time $t_j$ in our finite set
we can define histories $\op{C}_{\vctr{i}}$ via (\ref{DMWWhistdef}).
 We say that these histories satisfy the \emph{medium decoherence condition} (\cite{gellmannhartle93}) with respect to some state \ket{\Psi} if  $\matel{\Psi}{\opad{C}_{\vctr{j}}\op{C}_{\vctr{i}}}{\Psi}\simeq 0$ whenever $\vctr{i}\neq\vctr{j}$.

A set of histories satisfying medium decoherence  has the following attractive properties:
\begin{enumerate}
\item If (as above) the quantities $\matel{\Psi}{\opad{C}_{\vctr{i}}\op{C}_{\vctr{i}}}{\Psi}$ are interpreted as probabilities of a given history being realised  then medium decoherence guarantees that this can be done consistently, at least within the limits of what we can experimentally determine. In particular, it guarantees that if we define coarse-grained histories (by, \egc, leaving out some intermediate time $t_i$ or amalgamating some projectors into a single joint projector), the coarse-graining obeys the probability calculus:
\be\label{DMWWconsistent}\Pr(\sum_{\vctr{i}\in I}\op{C}_{\vctr{i}})\simeq \sum_{\vctr{i}\in I}\Pr(\op{C}_{\vctr{i}}).\ee
For 
\[ \Pr(\sum_{\vctr{i}\in I}\op{C}_{\vctr{i}})\simeq \matel{\Psi}{(\sum_{\vctr{j}\in I}\opad{C}_{\vctr{j}})(\sum_{\vctr{i}\in I}\op{C}_{\vctr{i}})}{\Psi}
\]
\be
\simeq\sum_{\vctr{j}\in I}\sum_{\vctr{i}\in I}\matel{\Psi}{\opad{C}_{\vctr{j}}\op{C}_{\vctr{i}}}{\Psi}
\ee
which in the presence of medium decoherence is just equal to $\sum_{\vctr{i}\in I}\matel{\Psi}{\opad{C}_{\vctr{i}}\op{C}_{\vctr{i}}}{\Psi}$.

(Actually a weaker condition --- that the \emph{real part} of $\matel{\Psi}{\opad{C}_{\vctr{j}}\op{C}_{\vctr{i}}}{\Psi}=0$ --- is sufficient to deliver (\ref{DMWWconsistent}). This condition is called consistency;  it does not seem to occur in natural situations other than those which also deliver medium decoherence.)
\item Medium decoherence guarantees the existence of records (in an abstract sense). The probabilistic interpretation tells us that at time $t_n$ the system should be thought of as having one of the states
\be \ket{\Psi(\vctr{i})}=\mc{N}\op{C}_\vctr{i}\ket{\Psi}\ee
(where \mc{N} is a normalising factor). These states are mutually orthogonal: as such, a single measurement (in the traditional sense) suffices, in principle, to determine the entire history and not just the current state. 
\end{enumerate}
In light of its elegance, it is tempting to adopt the criterion of medium decoherence of a set of histories as the \emph{definition} of decoherence, with the decoherence of the previous section only a special case (and an ill-defined one at that). And in fact the resultant formalism (call it the `decoherent histories' formalism) has more than just elegance to recommend it. For one thing, it makes explicit the state-dependence of decoherence. This was in any case implicit in the previous section's analysis: for the `environment' to decohere the system, it must be in an appropriate state. (If the state of the `environment' is a quintillion-degree plasma, for instance, the system will certainly not undergo quasi-classical evolution!) For another, it allows for a system/environment division which is not imposed once and for all, but can vary from history to history. 

It would be a mistake, however, to regard the decoherent histories formalism as \emph{conceptually} generalising the environment-induced decoherence discussed in section \ref{DMWWdecoherencegeneral}. In both cases, the mechanism of decoherence is the same: some subset of the degrees of freedom are recorded by the other degrees of freedom, with the local nature of interactions picking out a phase-space-local basis as the one which is measured; this recording process breaks the coherence of the macroscopic degrees of freedom, suppressing interference and leading to dynamics which are quasi-classical  and admit of a probabilistic interpretation (at least approximately). And although the decoherent-histories formalism in theory has the power to incorporate history-dependent system/environment divisions, in practice even simple models where this actually occurs have proven elusive, and actual applications of the decoherent-histories formalism have in the main been restricted to the same sort of system/environment split considered in section \ref{DMWWdecoherencegeneral} (although the `environment' is often taken to be microscopic degrees of freedom of the same system).

Furthermore, there are some infelicities of the decoherent-histories formalism as applied to realistic cases of decoherence. In particular, the natural pointer basis for realistic systems seems to be non-orthonormal wave-packet states and the rate of decoherence of superpositions of such states depends smoothly on the spatial distance between them. This does not sit altogether easily with the decoherent-histories formalism's  use of discrete times and an orthonormal pointer basis. 

Perhaps most importantly, though, the consistency condition alone is insufficient to restore quasi-classical dynamics in the `concrete' sense of section \ref{DMWWdecoherencerates} --- that is, it is insufficient to provide approximately classical equations of motion. I return to this point in section \ref{DMWWsolutionthatisnt}.

In any case, for our purposes what is important is that (no matter how `decoherence' is actually defined)  the basis of quasi-classical states of a macroscopic system is very rapidly decohered by its environment. This guarantees the consistency, for all practical purposes, of the Quantum Algorithm; whether it goes further and actually solves the measurement problem is a matter to which I will return in sections \ref{DMWW3candidates} and \ref{DMWWeverett}. 

\subsection{Further Reading}\label{DMWWdecoherencefurtherreading}

Joos \emph{et al}~\citeyear{joosetal} and \citeN{zurek01review} provide detailed reviews of decoherence theory; \citeN{zurek91} is an  accessible short introduction. \citeN{bacciagaluppiencyclopedia} reviews the philosophical implications of decoherence.

\section{Three candidates for orthodoxy}\label{DMWW3candidates}

In philosophy of QM, terminology is not the least source of confusion. Authors frequently discuss the ``orthodox'' interpretation of QM as if everyone knew what they meant, even though different authors ascribe different and indeed contradictory properties to their respective versions of orthodoxy. It does not help that physicists use the term ``Copenhagen interpretation'' almost interchangeably with ``orthodox interpretation'' or ``textbook interpretation'', while philosophers tend to reserve the term for Bohr's actual, historical position, and use a term like ``Dirac-von Neumann interpretation'' for the textbook version.

In this section --- which aims to present the ``orthodox interpretation'' --- I follow the sage advice of Humpty Dumpty, who reminded Alice that words mean what we want them to mean. There are at least three mainstream positions on the measurement problem which are often described as ``orthodoxy''. Two of them --- operationalism and the consistent-histories formalism --- are highly controversial pure interpretations of QM, which their proponents nonetheless often describe as the ``orthodox'' or indeed the only possible interpretation. (In their different ways, both are also claimed to capture the true spirit of Copenhagen). The third (which I call the ``new pragmatism'') is not actually regarded by anyone as a \emph{successful} solution to the measurement problem but, arguably, best captures the pragmatic quantum theory actually used by working physicists. It is best understood by considering, first, an attractive but failed solution.

\subsection{The solution that isn't: non-uniqueness of decoherent histories}\label{DMWWsolutionthatisnt}

Suppose that there was \emph{exactly one} finest-grained set of decoherent histories --- defined, say, by projectors $\op{P}_i(t_i)$ which satisfied the medium decoherence condition exactly; suppose also that this set of histories picked out a preferred basis reasonably close to the ``quasi-classical'' states used in the Quantum Algorithm, so that each $\op{P}_i(t_i)$ projected onto those states interpreted by the Quantum Algorithm as saying: the macroscopic degrees of freedom of the system will certainly be found to have some particular values $(q^i,p_i)$.

In this case, a solution of sorts to the measurement problem would be at hand. It would simply be a stochastic theory of the macroscopic degrees of freedom, specified as follows:
\begin{quote}
Given that:
\begin{enumerate}
\item the \emph{universal state} is \ket{\Psi};
\item the unique finest-grained decoherent-history space consistent with \ket{\Psi} is generated by projectors $\op{P}_i(t_j)$, associated with values $(q^i,p_i)$ for the macroscopic degrees of freedom at time $t_j$;
\item the macroscopic degrees of freedom at time $t_j$ have values $(q^i,p_i)$, corresponding to projector $\op{P}_i(t_j)$
\end{enumerate} 
then the probability of the macroscopic degrees of freedom at time $t_{j'}$ having values $(q^{i'},p_{i'})$
is given by 
\be
\Pr (q^{i'},p_{i'};t_{j'}| q^i,p_i;t_i)=\frac{\matel{\Psi}{\op{P}_{i'}(t_{j'})\op{P}_i(t_i)\op{P}_i(t_i)\op{P}_{i'}(t_{j'})}{\Psi}}{\matel{\Psi}{\op{P}_i(t_i)\op{P}_i(t_i)}{\Psi}}.
\ee
(It follows from this and the decoherence condition, of course, that the probability of a given \emph{history} $\op{C}_\vctr{i}$ is just \matel{\Psi}{\opad{C}_\vctr{i}\op{C}_\vctr{i}}{\Psi}.)
\end{quote}
How satisfactory is this as an interpretation of QM? It is not a \emph{pure} interpretation; on the other hand, since it is (ex hypothesi) a successful interpretation, it is unclear that this matters. It is not obviously compatible with relativity, since it makes explicit use of a preferred time; perhaps this could be avoided via a spacetime-local version of the preferred projectors, but it seems unlikely that genuinely pointlike degrees of freedom would decohere. The role of the `universal state' is pretty unclear --- in fact, the ontology as a whole is pretty unclear, and the state-dependent nature of the preferred set of histories is at least odd. 

These questions are moot, though. For the basic assumption which grounds the interpretation --- that there exists a unique (finest-grained) exactly-decoherent history space --- is badly mistaken, as has been shown by \citeN{dowkerkent} and \citeN{kenthistory}. The problem does not appear to be \emph{existence}: as section \ref{DMWWdecoherence} showed, there are good reasons to expect the histories defined by macroscopic degrees of freedom of large systems to approximately decohere, and Dowker and Kent have provided plausibility arguments to show that in the close vicinity of any almost-decoherent family of histories we can find an exactly-decoherent one. It is \emph{uniqueness}, rather, that causes the difficulties: there are excellent reasons to believe that the set of exactly decoherent history spaces is huge, and contains (continuously) many history spaces which are not remotely classical. Indeed, given a family of decoherent histories defined up to some time $t$, there are continuously many distinct ways to continue that family. As such, the simple decoherence-based interpretation above becomes untenable.

The temptation, for those seeking to solve the measurement problem via decoherence, is to introduce some additional criterion stronger than medium decoherence --- some $X$ such that there is a unique finest-grained history space satisfying medium-decoherence-plus-$X$.  And in fact there is a popular candidate in the literature: \emph{quasi-classicality} \cite{gellmannhartle93}. That is: the preferred history space not only decoheres: the decohering degrees of freedom obey approximately classical equations of motion.

It is plausible (though to my knowledge unproven) that this condition is essentially unique; it is \emph{highly} plausible that there are not continuously many essentially different ways to vary a quasi-classical decoherent history space. But as a candidate for $X$, quasi-classicality is pretty unsatisfactory. For one thing, it is essentially vague: while we have good theoretical reasons to expect exactly-decoherent histories in the vicinity of approximately decoherent ones, we have no reason at all to expect exactly classical histories in the vicinity of quasi-classical ones. For another, it is a high-level notion all-but-impossible to define in microphysical terms. It is as if we were to write a theory of atomic decay which included ``density of multicellular organisms'' as a term in its equations.

As such, it seems that no satisfactory decoherent-history-based interpretation can be developed along the lines suggested here.

\subsection{The new pragmatism}

However, an interpretation need not be \emph{satisfactory} to be \emph{coherent} (so to speak). No-one who took the measurement problem seriously regarded the Dirac-von Neumann formulation of QM, with its objective collapse of the wavefunction at the moment of measurement, as a \emph{satisfactory} physical theory; yet it was widely discussed and used in physics when one wanted a reasonably clear statement of the theory being applied (and never mind its foundational problems). The quasi-classical condition discussed in the previous section lets us improve on the Dirac-von Neumann interpretation by making (somewhat) more precise and objective its essential appeal to `measurement' and `observation'. The resultant theory has been called the `unknown set' interpretation by \citeN{kentbohmhistory}; I prefer to call it the \emph{New Pragmatism}, to emphasise that no-one really regards it as acceptable. It is, nonetheless, one of our three ``candidates for orthodoxy''; though it has not been explicitly stated in quite the form which I shall use, it seems to conform quite closely to the theory that is in practice appealed to by working physicists.

\begin{quote}
\textbf{The New Pragmatism (decoherent-histories version):}
The state of the Universe at time $t$ is given by specifying some state vector \ket{\Psi(t)}, which evolves unitarily, and some 
\emph{particular} quasi-classical, approximately decoherent consistent-history space, generated by
the projectors $\op{P}_i(t_j)$
The state \ket{\Psi(t)} is to be interpreted as a probabilistic mixture of eigenstates of the quasi-classical projectors: that is, expanding it as
\be
\ket{\Psi(t)}=\sum_{i}\op{P}_i(t)\ket{\Psi(t)}\matel{\Psi(t)}{\op{P}_i(t)}{\Psi(t)},
\ee
the probability that the state of the Universe is (up to normalisation) $\op{P}_i(t)\ket{\Psi(t)}$ is 
$|\matel{\Psi(t)}{\op{P}_i(t)}{\Psi(t)}|^2$. Because the history space is approximately decoherent, any interference-generated inconsistencies caused by this probabilistic reading of the state will be undetectable; if that is felt to be unsatisfactory, just require that the history space is exactly decoherent (some such will be in the vicinity of any given approximately-decoherent history space).
\end{quote}

According to the New Pragmatism, then, the quantum state vector is physical --- is, indeed, the complete physical description of the system. It evolves in some mixture of unitary steps and stochastic jumps, and at any given time it assigns approximately-definite values of position and momentum to the macroscopic degrees of freedom of the system. We do not know the actual decoherent-history space used (hence `unknown set), but we know it well enough to predict all probabilities to any reasonably-experimentally-accessible accuracy.

The New Pragmatism, it will be apparent, is a pretty minimal step beyond the Quantum Algorithm itself: if we were to ask for the most simple-minded way to embed the Algorithm into a theory, without any concern for precision or elegance, we would get something rather like the New Pragmatism. This is even more obvious if we reformulate it from the language of decoherent histories to the environment-induced decoherence of section \ref{DMWWdecoherence}:
\begin{quote}
\textbf{The New Pragmatism (wave-packet version):}
Fix some \emph{particular} division of Hilbert space into macroscopic and microscopic degrees of freedom: $\mc{H}=\mc{H}_{macro}\otimes \mc{H}_{micro}$; and fix some particular basis $\ket{\vctr{q},\vctr{p}}$ of wave-packet states for $\mc{H}_{macro}$. Then the state vector \ket{\Psi(t)} of the Universe always evolves unitarily, but is to be understood as a probabilistic mixture of approximately-macroscopically-definite states: if the universal state is the superposition
\be
\ket{\Psi(t)}=\int\dr{\vctr{p}}\dr{\vctr{q}}\alpha(\vctr{q},\vctr{p})\tpk{\vctr{q},\vctr{p}}{\psi(\vctr{q},\vctr{p};t)}
\ee
then the actual state is one of the components of this superposition, and has probability
$|\alpha(\vctr{q},\vctr{p})|^2$ of being $\tpk{\vctr{q},\vctr{p}}{\psi(\vctr{q},\vctr{p};t)}$.
(And of course this state in turn is somehow to be understood as having macroscopic phase-space location $(\vctr{q},\vctr{p})$.)
\end{quote} 
It is an interesting philosophy-of-science question to pin down exactly what is unacceptable about the New Pragmatism. And it is not obvious at all that it \emph{is} unacceptable from some anti-realist standpoints (from the standpoint of \citeN{vanfraassenscientificimage}, for instance). Nonetheless, it is accepted as unsatisfactory. Unlike our other two candidates for orthodoxy, and despite the frequency with which it is in fact used, no-one really takes it seriously. 

\subsection{The consistent-histories interpretation}\label{DMWWhistoryinterpretation}

A more `serious' interpretation of QM, still based on the decoherent-histories formalism, has been advanced by Griffiths \citeyear{griffiths,griffithsbook} and  Omnes \citeyear{omnes,omnesbook}: it might be called the `consistent histories' interpretation,\footnote{Terminology is very confused here. Some of those who advocate `consistent-histories' interpretations --- notably Gell-Mann and Hartle  --- appear to mean something very different from Griffiths and Omnes, and much closer in spirit to the Everett interpretation.} and its adherents claim that it incorporates the essential insights of Bohr's complementarity, and should be viewed as the natural modern successor to the Copenhagen interpretation. 

The positions of Griffiths and Omnes are subtle, and differ in the details. However, I think that it is possible to give a general framework which fits reasonably well to both of them. We begin, as with the impossible single-decoherent-history-space theory, with some universal state \ket{\Psi}. Now, however, we consider \emph{all} of the maximally-fine-grained consistent history spaces associated with \ket{\Psi}. (Recall that a history space is \emph{consistent} iff the \emph{real} part of \matel{\Psi}{\opad{C}_\vctr{i}\op{C}_\vctr{j}}{\Psi} vanishes for $\vctr{i}\neq \vctr{j}$; it is a mildly weaker condition than decoherence, necessary if the probabilities of histories are to obey the probability calculus.) 

Now in fact, these ``maximally fine-grained'' history spaces are actually constructed from \emph{one-dimensional} projectors. For any exactly-consistent history which does not so consist can always be fine grained, as follows: let it be constructed as usual from projectors $\op{P}_{i_j}(t_j)$, and define the state \ket{i_k, \ldots i_0} by
\be
\ket{i_k, \ldots i_0}= \mc{N}\op{P}_{i_k}(t_k)\cdots \op{P}_{i_0}(t_0)\ket{\Psi}
\ee
(where \mc{N} is just a normalising factor.) Then define a fine-graining $\op{P}^m_{i_k}(t_k)$ as follows: 
\be \op{P}^0_{i_k}(t_k)=\proj{i_k, \ldots i_0}{i_k, \ldots i_0};\ee
the other $\op{P}^m_{i_k}(t_k)$ are arbitrary one-dimensional projectors chosen to satisfy
\be
\sum_m \op{P}^m_{i_k}(t_k)=\op{P}_{i_k}(t_k).
\ee
It is easy to see that
\[
\op{P}^{m_n}_{i_n}(t_n)\cdots \op{P}^{m_0}_{i_0}(t_0)\ket{\Psi}=0\mathrm {\mbox{      whenever any }}m_k \neq 0
\]
\be
\op{P}^{0}_{i_n}(t_n)\cdots \op{P}^{0}_{i_0}(t_0)\ket{\Psi}=\op{P}_{i_n}(t_n)\cdots \op{P}_{i_0}(t_0)\ket{\Psi},
\ee
and hence the fine-graining also satisfies the consistency condition. Notice (this is why I give the proof explicitly, in fact) how sensitive this fine-graining process is to the universal state \ket{\Psi} (by contrast, when we are dealing with the coarse-grained approximately-decoherent histories given by dynamical decoherence, the history space is fairly insensitive to all but broad details of \ket{\Psi}).

Griffiths and Omnes now regard each consistent history space as providing some valid \emph{description} of the quantum system under study. And under a given description $\{\op{C}_\vctr{i}\}$, they take the probability of the system's actual history being $\op{C}_\vctr{i}$ to be given by the usual formula $\matel{\Psi}{\opad{C}_\vctr{i}\op{C}_\vctr{i}}{\Psi}$. 

Were there in fact only one consistent history space, this would reduce to the `impossible' interpretation which I discussed in section \ref{DMWWsolutionthatisnt} . But of course this is not the case, so that a great deal of conceptual work must be done by the phrase `under a given description'.

It is very unclear how this work is in fact to be done. There are of course multiple descriptions of even classical systems, but these descriptions can in all cases be understood as coarse-grainings of a single exhaustive description (\citeN{griffithsbook} dubs this the \emph{principle of unicity}). By contrast, in the consistent-histories form of QM this is not the case:
\begin{quote}
The principle of unicity does not hold: there is not a unique exhaustive description of a physical system or a physical process. Instead, reality is such that it can be described in various alternative, incompatible ways,using descriptions which cannot be combined or compared. \cite[p.\,365]{griffithsbook}
\end{quote}

There is a close analogy between this `failure of unicity' and Bohrian complementarity, as proponents of the consistent-histories interpretation recognise. The analogy becomes sharper in the concrete context of measurement: which histories are `consistent' in a given measurement process depends sensitively on what the measurement device is constructed to measure.  If, for instance, we choose to measure a spin-half particle's spin in the $x$ direction, then schematically the process looks something like
\[
(\alpha\ket{+_x}+\beta \ket{-_x})\otimes\ket{\mathrm{\mbox{untriggered device}}}
\]
\be
\longrightarrow
\alpha \tpk{+_x}{\mathrm{\mbox{device reads `up'}}}
+
\beta
\tpk{-_x}{\mathrm{\mbox{device reads `down'}}}.
\ee
A consistent-history space for this process might include histories containing the projectors
\[
\proj{\pm_x}{\pm_x}\otimes\proj{\mathrm{\mbox{untriggered device}}}{\mathrm{\mbox{untriggered device}}},
\]
\[
\proj{\pm_x}{\pm_x}\otimes\proj{\mathrm{\mbox{device reads `up'}}}{\mathrm{\mbox{device reads `up'}}},
\]
and
\[
\proj{\pm_x}{\pm_x}\otimes\proj{\mathrm{\mbox{device reads `down'}}}{\mathrm{\mbox{device reads `down'}}},
\]
But if the experimenter instead chooses to measure the $z$ component of spin, then this set will no longer be consistent
and we will instead need to use a set containing projectors like
\[
\proj{\pm_z}{\pm_z}\otimes\proj{\mathrm{\mbox{device reads `down'}}}{\mathrm{\mbox{device reads `down'}}},
\]
So while for Bohr the classical context of \emph{measurement} was crucial, for the consistent-histories interpretation this just emerges as a special case of the consistency requirement, applied to the measurement process.  (Note that it is, in particular, perfectly possible to construct situations where the consistent histories at time $t$ are fixed by the experimenter's choices at times far later than $t$ --- cf \citeN[p.\,255]{griffithsbook}, \citeN[p.\,54--5]{dicksonbook} --- in keeping with Bohr's response to the EPR paradox.)

But serious conceptual problems remain for the consistent-histories interpretation:
\begin{enumerate}
\item What is the ontological status of the universal state vector \ket{\Psi}? It plays an absolutely crucial role in the theory in determining which histories are consistent: as we have seen, when we try to fine-grain histories down below the coarse-grained level set by dynamical decoherence then the details of which histories are consistent becomes extremely sensitively dependent on \ket{\Psi}. Perhaps it can be interpreted as somehow `lawlike'; perhaps not. It is certainly difficult to see how it can be treated as physical without letting the consistent-histories interpretation collapse into something more like the Everett interpretation.
\item Does the theory actually have predictive power? The criticisms of Kent and Dowker continue to apply, and indeed can be placed into a sharp form here: they prove that a given consistent history can be embedded into two different history spaces  identical up to a given time and divergent afterwards, such that the probabilities assigned to the history vary sharply from one space to the other. In practice, accounts of the consistent-history interpretation seem to get around this objection by foreswearing cosmology and falling back on some externally-imposed context to fix the correct history; shades of Bohr, again.
\item Most severely, is Griffith's `failure of unicity' really coherent? It is hard to make sense of it; no wonder that many commentators on the consistent-history formalism (\egc, \citeN[p.788]{penroseroadtoreality}) find that they can make sense of it only by regarding every history in every history space as actualised: an ontology that puts Everett to shame. 
\end{enumerate}

\subsection{Operationalism}\label{DMWWoperationalist}

The consistent-histories interpretation can make a reasonable case for being the natural home for Bohr's complementarity. But there is another reading of the Copenhagen interpretation which arguably has had more influence on physicists' attitude to the measurement problem: the operationalist doctrine that physics is concerned not with an objective `reality' but only with the result of experiments. This position has been sharpened in recent years into a relatively well-defined interpretation (stated in particularly clear form by \citeN{peres}; see also \citeN{fuchsperes}): the \emph{operationalist interpretation} that is our third candidate for orthodoxy.

Following Peres, we can state operationalism as follows:
\begin{quote}
\textbf{The operationalist interpretation:}
Possible measurements performable on a quantum system are represented by the POVMs of that system's Hilbert space. 
All physics tells us is the probability, for each measurement, of a given outcome: specifically, it tells us that the probability of the outcome corresponding
to positive operator \op{A} obtaining is $\tr(\denop \op{A})$ (or $\matel{\psi}{\op{A}}{\psi}$ in the special case where a pure state may be used). As such, the state of the system is not a physical thing at all, but simply a shorthand way of recording the probabilities of various outcomes of measurements; and the evolution rule
\be \ket{\psi(t)}=\exp(-it \op{H}/\hbar)\ket{\psi}\ee
is just a shorthand way of recording how the various probabilities change over time for an isolated system.
\end{quote}
In fact, we do not even need to \emph{postulate} the rule $\Pr(\op{A})=\tr(\denop \op{A})$. It is enough to require \emph{non-contextuality}: that is, to require that the probability of obtaining the result associated with \op{A} is independent of which POVM \op{A} is embedded into. Suppose $\Pr$ is any non-contextual probability measure on the positive operators: that is, suppose it is a function from the positive operators to $[0,1]$ satisfying
\be
\sum_i\op{A}_i=\id \longrightarrow \sum_i\\Pr(\op{A}_i)=1.
\ee
Then it is fairly simple (Caves \emph{et al}~\citeyearNP{cavesetalgleason}) to prove that $\Pr$ must be represented by a density operator: $\Pr(\op{A})=\tr(\denop \op{A})$ for some \denop. 

Modifications of the operationalist interpretation are available. The probabilities may be taken to be subjective \cite{cavesetalprobability}, as referring to an ensemble of systems (\citeNP{ballentine}, \citeNP{taylorghost}), or as irreducible single-case chances \cite{fuchsperes}. The `possible measurements' may be taken to be given by the PVMs alone rather than the POVMs (in which case Gleason's theorem must be invoked in place of the simple proof above to justify the use of density operators). But the essential content of the interpretation remains: the `quantum state' is just a way of expressing the probabilities of various measurement outcomes, and --- more generally --- quantum theory itself is not in the business of supplying us with an objective picture of the world. Fuchs and Peres put this with admirable clarity: 
\begin{quote}
We have learned something new when we can distill from the accumulated data a compact description of all that was seen and an indication of which further experiments will corroborate that description. This is what science is about. If, from such a description, we can \emph{further} distill a model of a free-standing ``reality'' independent of our interventions, then so much the better. Classical physics is the ultimate example of such a model. However, there is no logical necessity for a realistic worldview to always be obtainable. If the world is such that we can never identify a reality independent of our experimental activity, then we must be prepared for that, too. \ldots [Q]uantum theory does \emph{not} describe physical reality. What it does is provide an algorithm for computing \emph{probabilities} for the macroscopic events (``detector clicks'') that are the consequences of our experimental interventions. This strict definition of the scope of quantum theory is the only interpretation ever needed, whether by experimenters or theorists. \cite{fuchsperes}

\ldots

Todd Brun and Robert Griffiths point out [in \citeN{fuchsperescomments}] that ``physical theories have always had as much to do with providing a coherent picture of reality as they have with predicting the results of experiment.'' Indeed, have always had. This statement was true in the past, but it is untenable in the present (and likely to be untenable in the future). Some people may deplore this situation, but we were not led to reject a freestanding reality in the quantum world out of a predilection for positivism. We were led there because this is the overwhelming message quantum theory is trying to tell us. 
\cite{fuchsperesreply}
\end{quote}
Whether or not Fuchs and Peres were led to their position `out of a predilection for positivism', the operationalist interpretation is nonetheless positivist in spirit, and is subject to many of the same criticisms. However, in one place it differs sharply. Where the positivists were committed to a once-and-for-all division between observable and unobservable, a quantum operationalist sees no difficulty in principle with applying QM to the measurement process itself. In a measurement of spin, for instance, the state
\be
\alpha\ket{+_x}+\beta \ket{-_x}
\ee
may just be a way of expressing that (among other regularities)  the probability of getting result `up' on measuring spin in the $x$ direction is $|\alpha|^2$. But the measurement process may itself be modeled in QM in the usual way ---
\[
(\alpha\ket{+_x}+\beta \ket{-_x})\otimes\ket{\mathrm{\mbox{untriggered device}}}
\]
\be
\longrightarrow
\alpha \tpk{+_x}{\mathrm{\mbox{device reads `up'}}}
+
\beta
\tpk{-_x}{\mathrm{\mbox{device reads `down'}}}.
\ee
--- \emph{provided} that it is understood that this state is itself just a shorthand way of saying (among other regularities)  that the probability of finding the measurement device to be in state ``reads up' '' is $|\alpha|^2$. It is not intended to describe a physical superposition any more than $\alpha\ket{+_x}+\beta \ket{-_x}$ is. 

In principle, this can be carried right up to the observer:
\be
\alpha\ket{\mathrm{\mbox{Observer sees `up' result}}}
+
\beta\ket{\mathrm{\mbox{Observer sees `down' result}}}
\ee
is just a shorthand expression of the claim that if the `observer' is themselves observed, they will be found to have seen `up' with probability $|\alpha|^2$.

Of course, if analysis of any given measurement process only gives dispositions for certain results in subsequent measurement processes, then
there is a threat of infinite regress. The operationalist interpretation responds to this problem by adopting an aspect of the Copenhagen interpretation essentially lost in the consistent-histories interpretation: the need for a separate classical language to describe measurement devices, and the resultant ambiguity (\citeN[p.\,373]{peres} calls it \emph{ambivalence}) as to which language is appropriate when.

To spell this out (here I follow \citeN[pp.\,376--7]{peres}) a measurement device, or any other macroscopic system, may be described either via a density operator \denop\ on Hilbert space
 (a quantum state, which gives only probabilities of certain results on measurement) or a probability distribution $W(\vctr{q},\vctr{p})$ over phase-space points (each of which gives an actual classical state of the system). These two descriptions then give different formulae for the probability of finding the system to have given position and momentum:
\begin{itemize}
\item The quantum description just \emph{is} a shorthand for the probabilities of getting various different results on measurement. In particular, there will exist some POVM $\op{A}_{q,p}$ such that the probability density of getting results $(q,p)$ on a joint measurement of position and momentum is $\tr(\denop \op{A}_{q,p})$.
\item According to the classical description, the system actually has some particular values of $q$ and $p$, and the probability density for any given values is just $W(q,p)$ 
\end{itemize}
If the two descriptions are not to give contradictory predictions for the result of experiment, then we require that $W(q,p)\simeq\tr(\denop \op{A}_{q,p})$; or, to be more precise, we require that the integrals of $W(q,p)$ and $\tr(\denop \op{A}_{q,p})$ over sufficiently large regions of phase space are equal to within the limits of experimental error. This gives us a recipe to construct the classical description from the quantum: just set $W(q,p)$ equal to $\tr(\denop \op{A}_{q,p})$. If this is done at a given time $t_0$, then at subsequent times $t>t_0$ the classical dynamics applied to $W$ and the quantum dynamics applied to \denop\ will break the equality:
\be
\tr(\denop(t) \op{A}_{q,p})\neq W(q,p;t)
\ee
(where $W(q,p;t)$ is the distribution obtained by time-evolving $W(q,p)$ using Hamilton's equations.) But if the system is sufficiently large, decoherence guarantees that the equality continues to hold approximately when $W(q,p;t)$ and $\tr(\denop(t) \op{A}_{q,p})$ are averaged over sufficiently large phase-space volumes.

The `operationalism' of this interpretation is apparent here. There is no \emph{exact} translation between classical and quantum descriptions, only one whose imprecisions are too small to be detected empirically.\footnote{A further ambiguity in the translation formula is the POVM $A(q,p)$: in fact, no unique POVM for phase-space measurement exists. Rather, there are many equivalently-good candidates which essentially agree with one another provided that their predictions are averaged over phase-space volumes large compared to $\hbar^n$, where $n$  is the number of degrees of freedom of the system.}
But if QM --- if science generally --- is merely a tool to predict results of experiments, it is unclear at best that we should be concerned about ambiguities which are empirically undetectable in practice. Whether this indeed a valid conception of science --- and whether the operationalist interpretation really succeeds in overcoming the old objections to logical positivism --- I leave to the reader's judgment.

\subsection{Further Reading}

Standard references for consistent histories are \citeN{griffithsbook} and \citeN{omnesbook}; for critical discussion, see \citeN[pp.\,52--57]{dicksonbook}, \citeN[212--236]{bubbook} and \citeN{ghirardiconsistent}. The best detailed presentation of operationalism is \citeN{peres}; for a briefer account see \citeN{fuchsperes}. For two rather different reappraisals of the original Copenhagen interpretation, see \citeN{cushingcopenhagen} and \citeN{saunderscopenhagen}.

Recently, operationalist approaches have taken on an ``information-theoretic'' flavour, inspired by quantum information. See Chris Timpson's contribution to this volume for more details. 

Though they cannot really be called ``orthodox'', the family of interpretations that go under the name of ``quantum logic'' are also pure interpretations which attempt to solve the measurement problem by revising part of our pre-quantum philosophical picture of the world. In this case, though, the part to be revised is classical logic. Quantum logic is not especially popular at present, and so for reasons of space I have omitted it, but for a recent review see \citeN{dicksonlogic}.

\section{The Everett interpretation}\label{DMWWeverett}

Of our three `candidates for orthodoxy', only two are pure interpretations in the sense of section \ref{DMWWinterpretation}, and neither of these are `realist' in the conventional sense of the world. The consistent-histories interpretation purports to describe an objective reality, but that reality is unavoidably perspectival, making sense only when described from one of indefinitely many contradictory perspectives; whether or not this is coherent, it is not how scientific realism is conventionally understood! The operationalist interpretation, more straightforwardly, simply denies explicitly that it describes an independent reality. And although the new pragmatism does describe such a reality, it does it in a way universally agreed to be ad hoc and unacceptable.

There is, however, one pure interpretation which purports to be realist in a completely conventional sense: the Everett interpretation. Unlike the three interpretations we have considered so far, its adherents make no claim that it is any sort of orthodoxy; yet among physicists if not philosophers it seems to tie with operationalism and consistent histories for popularity. Its correct formulation, and its weaknesses, are the subject of this section.

\subsection{Multiplicity from indefiniteness?}

At first sight, applying straightforward realism to QM without modifying the formalism seems absurd. Undeniably, unitary QM produces superpositions of macroscopically distinct quasi-classical states; whether or not such macroscopic superpositions even make sense, their existence seems in flat contradiction with the fact that we actually seem to observe macroscopic objects only in definite states.

The central insight in the Everett interpretation is this: \emph{superpositions} of macroscopically distinct states are somehow to be understood in terms of \emph{multiplicity}. For instance (to take the time-worn example)
\be \label{DMWWschrodingercat}\alpha\ket{\mathrm{\mbox{Live cat}}}+\beta\ket{\mathrm{\mbox{Dead cat}}}\ee
is to be understood (somehow) as representing not a single cat in an indefinite state, but rather  a multiplicity of cats, one (or more) of which is alive, one (or more) of which is dead. Given the propensity of macroscopic superpositions to become entangled with their environment, this `many cats' interpretation becomes in practice a `many worlds' interpretation: quantum measurement continually causes the macroscopic world to branch into countless copies.

The problems in cashing out this insight are traditionally broken in two:
\begin{enumerate}
\item The `preferred basis problem': \emph{how} can the superposition justifiably be understood as some kind of multiplicity?
\item The `probability problem': how is probability to be incorporated into a theory which treats wavefunction collapse as some kind of branching process?
\end{enumerate}

\subsection{The preferred-basis problem: solution by modification}\label{DMWWmodification}

If the preferred basis problem is a question (``how can quantum superpositions be understood as multiplicities?'') then there is a traditional answer, more or less explicit in much criticism of the Everett interpretation \cite{barrettbook,kent,butterfieldeverett}: they cannot. That is: it is no good just \emph{stating} that a state like (\ref{DMWWschrodingercat}) describes multiple worlds: the formalism must be explicitly \emph{modified} to incorporate them. This position dominated discussion of the Everett interpretation in the 1980s and early 1990s: even advocates like \citeN{deutsch85} accepted the criticism and rose to the challenge of providing such a modification.

Modificatory strategies can  be divided into two categories.
\emph{Many-exact-worlds theories} augment the quantum formalism by adding an ensemble of `worlds' to the state vector. The `worlds' are each represented by an element in some particular choice of `world basis' $\ket{\psi_i(t)}$ at each time $t$: the proportion of worlds in state $\ket{\psi_i(t)}$ at time $t$ is $|\bk{\Psi(t)}{\psi_i(t)}$, where \ket{\Psi(t)} is the (unitarily-evolving) universal state. Our own world is just one element of this ensemble. 
Examples of many-exact-worlds theories are the early Deutsch (\citeyearNP{deutsch85,deutschghost}), who tried to use the tensor-product structure of Hilbert space to define the world basis\footnote{A move criticised on technical grounds by \citeN{brownondeutsch}.}, and Barbour(\citeyearNP{barbour2,barbour99}), who chooses the position basis.

In \emph{Many-minds theories}, by contrast, the multiplicity is to be understood as illusory. A state like (\ref{DMWWschrodingercat}) really is indefinite, and when an observer looks at the cat and thus enters an entangled state like
\be\alpha\tpk{\mathrm{\mbox{Live cat}}}{\mathrm{\mbox{Observer sees live cat}}}+\beta\tpk{\mathrm{\mbox{Dead cat}}}{\mathrm{\mbox{Observer sees dead cat}}}\ee
then the observer too has an indefinite state. However: to each physical observer is associated not one mental state, but an ensemble of them: each mental state has a definite experience, and the proportion of mental states where the observer sees the cat alive is $|\alpha|^2$. Effectively, this means that in place of a global `world-defining basis' (as in the many-exact-worlds theories) we have a `consciousness basis' for each observer.\footnote{Given that an `observer' is represented in the quantum theory by some Hilbert space many of whose states are not conscious at all, and that conversely almost any sufficiently-large agglomeration of matter can be formed into a human being, it would be more accurate to say that we have a consciousness basis for all \emph{systems}, but one with many elements which correspond to no conscious experience at all.} When an observer's state is an element of the consciousness basis, all the minds associated with that observer have the same experience and so we might as well say that the observer is having that experience. But in all realistic situations the observer will be in some superposition of consciousness-basis states, and the ensemble of minds associated with that observer will be having a wide variety of distinct experiences. Examples of many-minds theories are \citeN{albertloewermm}, Lockwood (\citeyearNP{lockwoodbook,lockwoodbjps1}), \citeN{pagesensible} and Donald(\citeyearNP{donald90,donald92,donald02}).

It has increasingly become recognised, by supporters and detractors alike that there are severe problems with either of these approaches to developing the Everett interpretation. Firstly, and most philosophically, both the many-exact-worlds and the many-minds theories are committed to a very strong (and arguably very anti-scientific) position in philosophy of mind: the rejection of \emph{functionalism}, the view that mental properties should be ascribed to a system in accordance with the functional role of that system (see \egc, \citeN{armstrongmind}, \citeN{lewisradical}, \citeN{hofstadterdennett}, \citeN{levinencyclopedia} for various explications of functionalism). This is particularly obvious in the case of the Many-Minds theories, where some rule associating conscious states to physical systems is simply postulated in the same way that the other laws of physics are postulated. If it is just a \emph{fundamental law} that consciousness is associated with some given basis, clearly there is no hope of a functional \emph{explanation} of how consciousness emerges from basic physics (and hence much, perhaps all, of modern AI, cognitive science and neuroscience is a waste of time). And in fact many adherents of Many-Minds theories (\egc, Lockwood and Donald) embrace this conclusion, having been led to reject functionalism on independent grounds.

It is perhaps less obvious that the many-exact-worlds theories are equally committed to the rejection of functionalism. But
if the `many worlds' of these theories are supposed to include \emph{our world}, it follows that conscious observers are found within each world. This is only possible compatible with functionalism if the worlds are capable of containing independent complex structures which can instantiate the `functions' that subserve consciousness. This in turn requires that the world basis is decoherent (else the structure would be washed away by interference effects) and --- as we have seen --- the decoherence basis is not effectively specifiable in any precise microphysical way. (See \citeN{wallaceworlds} for further discussion of the difficulty of localising conscious agents within  `worlds' defined in this sense.)

There is a more straightforwardly physical problem with these approaches to the Everett interpretation. Suppose that a wholly satisfactory Many-Exact-Worlds or Many-Minds theory were to be developed, specifying an exact `preferred basis' and an exact transition rule defining identity for worlds or minds. Nothing would then stop us from taking that theory, discarding all but one of the worlds/minds\footnote{It would actually be a case of discarding all but one \emph{set} of minds --- one for each observer.} and obtaining an equally empirically effective theory without any of the ontological excess which makes Everett-type interpretations so unappealing. Put another way: an Everett-type theory developed along the lines that I have sketched would really just be a hidden-variables theory with the additional assumption that continuum many non-interacting sets of hidden variables exist, each defining a different classical world. (This point is made with some clarity by \citeN{bellqmforcosmologists} in his classic attack on the Everett interpretation.) 

In the light of these sorts of criticisms, these modify-the-formalism approaches to the Everett interpretation have largely fallen from favour. Almost no advocate of ``the Many-Worlds Interpretation'' actually advocates anything like the Many-Exact-Worlds approach\footnote{\citeN{barbour99} may be an exception.} (Deutsch, for instance, clearly abandoned it some years ago) and Many-Minds strategies which elevate consciousness to a preferred role continue to find favour mostly in the small group of philosophers of physics strongly committed for independent reasons to a non-functionalist philosophy of mind. Advocates of the Everett interpretation among physicists (almost exclusively) and philosophers (for the most part) have returned to Everett's original conception of the Everett interpretation as a pure interpretation: something which emerges simply from a realist attitude to the unitarily-evolving quantum state. 

\subsection{The Bare Theory: how not to think about the wave function}\label{DMWWbaretheory}

One way of understanding the Everett interpretation as pure interpretation --- the so-called `Bare Theory' --- was suggested by \citeN{albertqmbook}. It has been surprisingly influential among philosophers of physics --- not as a \emph{plausible interpretation of QM}, but as the correct reading of the Everett interpretation. 

\citeN[p.\,94]{barrettbook} describes the Bare Theory as follows:
\begin{quote}
The bare theory is simply the standard von Neumann-Dirac formulation of QM with the standard interpretation of states (the eigenvalue-eigenstate link) but stripped of the collapse postulate --- hence, \emph{bare}.
\end{quote}
From this perspective, a state like (\ref{DMWWschrodingercat}) is not an eigenstate of the `cat-is-alive' operator (that is, the projector which projects onto all states where the cat is alive); hence, given the eigenstate-eigenvalue link the cat is in an indefinite state of aliveness. Nor is it an eigenstate of the `agent-sees-cat-as-alive' operator, so the agent's mental state is indefinite between seeing the cat alive and seeing it dead. But it \emph{is} an eigenstate of the `agent-sees-cat-as-alive-\textbf{or}-agent-sees-cat-as-dead' operator: the states
\be
\tpk{\mathrm{\mbox{Live cat}}}{\mathrm{\mbox{Observer sees live cat}}}
\ee
and
\be
\tpk{\mathrm{\mbox{Dead cat}}}{\mathrm{\mbox{Observer sees dead cat}}}
\ee
are both eigenstates of that operator with eigenvalue one, so their superposition is also an eigenstate of that operator. Hence if we \emph{ask} the agent, `did you see the cat as either alive or dead' they will answer `yes'. 

That is: the bare theory --- without any flaky claims of `multiplicity' or `branching' --- undermines the claim that macroscopic superpositions contradict our experience. It predicts that we will \emph{think}, and \emph{claim}, that we do not observe superpositions at all, even when our own states are highly indefinite, and that we are simply mistaken in the belief that we see a particular outcome or other. That is, it preserves unitary QM --- at the expense of a scepticism that ``makes Descartes's demon and other brain-in-the-vat stories look like wildly optimistic appraisals of our epistemic situation'' \cite[p.\,94]{barrettbook}. As Albert puts it:
\begin{quote}
[M]aybe \ldots the linear dynamical laws are nonetheless the complete laws of the evolution of the \emph{entire world}, and maybe all the appearances to the contrary (like the appearance that experiments have outcomes, and the localised that the world doesn't evolve deterministically) turn out to be just the sorts of \emph{delusions} which \emph{those laws themselves} can be shown to \emph{bring on!} 
\end{quote}
A quite extensive literature has developed trying to point out exactly what is wrong with the Bare Theory (see, \egc, \citeN[pp.\,117--125]{albertqmbook}, \citeN{barrettbare}, \citeN[pp.\,92--120]{barrettbook}, \citeN{bubcliftonmonton}, \citeN[pp.\,45--47]{dicksonbook}). The consensus seems to be that:
\begin{enumerate}
\item If we take a `minimalist', pure-interpretation reading of Everett, we are led to the Bare Theory; and
\item The bare theory has some extremely suggestive features; but
\item It is not ultimately satisfactory as an interpretation of QM because it fails to account for probability/is empirically self-undermining/smuggles in a preferred basis (delete as applicable); and so
\item Any attempt to solve the measurement problem along Everettian lines cannot be `bare' but must add additional assumptions.
\end{enumerate}

From the perspective of this review, however, this line of argument is badly mistaken. It relies essentially on the assumption that the eigenstate-eigenvalue link is part of the basic formalism of QM, whereas --- as I argued in section \ref{DMWWagainst} --- it plays no part in modern practice and is flatly in contradiction with most interpretative strategies. It will be instructive, however, to revisit this point in the context of the state-vector realism that is essential to the Everett interpretation.

If the state vector is to be taken as physically real, the eigenstate-eigenvalue link becomes a claim about the properties of that state vector. Specifically: 
\begin{quote}All properties of the state vector are represented by projectors, and the state vector \ket{\psi} has the property represented by \op{P} if it is inside the subspace onto which \op{P} projects: that is, if $\op{P}\ket{\psi}=\ket{\psi}$. If $\op{P}\ket{\psi}=0$ then \ket{\psi} certainly lacks the property represented by \op{P}; if neither is true then either \ket{\psi} definitely lacks the property or it is indefinite whether \ket{\psi} has the property.
\end{quote}
As is well known, it follows that the logic of state-vector properties is highly non-classical: it is perfectly possible, indeed typical, for a system to have a definite value of the property $(p \vee q)$ without definitely having either $p$ or $q$. The quantum logic program developed this into a mathematically highly elegant formalism; see \citeN{bubbook} for an clear presentation.

What is less clear is why we should take this account of properties at all seriously. We have seen that it fails to do justice to modern analyses of quantum measurement; furthermore, from the perspective of state-vector realism it seems to leave out all manner of perfectly ordinary properties of the state. Why not ask:
\begin{itemize}
\item Is the system's state an eigenstate of energy?
\item Is its expectation value with respect to the Hamiltonian greater than $100$ joules?
\item Is its wavefunction on momentum space negative in any open set?
\item Does its wavefunction on configuration space tend towards any Gaussian?
\end{itemize}
If the state vector is physical, these all seem perfectly reasonable questions to ask about it. Certainly, each has a determinate true/false answer for any given state vector. Yet none correspond to any projector (it is, for instance, obvious that there is no projector that projects onto all and only eigenstates of some operator!)

Put more systematically: if the state vector is physical, then the set \mc{S} of normalised vectors in Hilbert space is (or at least represents) the set of possible states in which an (isolated, non-entangled) system can be found. If we assume standard logic, then a property is defined (at least for the purposes of physics) once we have specified the states in \mc{S} for which the property holds. That is: properties in quantum physics correspond to subsets of the state space, just as properties in classical physics correspond to subsets of the phase space. 

If we assume \emph{non}-standard logic, of course, we doubtless get some different account of properties; if we assume a particular non-standard logic, very probably we get the eigenstate-eigenvalue account of properties. The fact remains that if we wish to assume state-vector realism and standard logic (as did Everett) we do not get the Bare Theory. 

(There is, to be fair, an important question about what we \emph{do} get. That is, how can we think about the state vector (construed as real) other than through the eigenstate-eigenvalue link? This question has seen a certain amount of attention in recent years. The most common answer seems to be `wavefunction realism': if the state vector is a physical thing at all it should be thought of as a field on $3N$-dimensional space. Bell proposed this (in the context of the de Broglie-Bohm theory):
\begin{quote}
\emph{No one can understand this theory until he is willing to think of $\psi$ as a real physical field rather than a `probability amplitude'. Even though it propagates not in 3-space but in 3N-space.} (\citeNP{bellqmforcosmologists}; emphasis his)
\end{quote}
\citeN{albertmetaphysics} proposes it as the correct reading of the state vector in any state-vector-realist theory;\footnote{This would seem to imply that Albert would concur with my criticism of the Bare Theory, but I am not aware of any actual comment of his to this effect.} \citeN{lewisconfiguration} and \citeN{montonconfiguration} concur. (Monton argues that wavefunction realism is unacceptable, but he does so in order to argue against state-vector realism altogether rather than to advocate an alternative). 

There are alternatives, however. Chris Timpson and myself \citeyear{wallacetimpsonshort} suggest a more spatio-temporal ontology, in which each spacetime region has a (possibly impure) quantum state but in which, due to entanglement the state of region $A \cup B$ is not determined simply by the states of regions $A$ and $B$ separately (a form of nonlocality which we claim is closely analogous to what is found in the Aharonov-Bohm effect). \citeN{deutschhayden} argue for an ontology based on the Heisenberg interpretation which appears straightforwardly local (but see \citeN{wallacetimpsonshort} for an argument that this locality is more apparent than real). \citeN{saundersmetaphysics} argues for a thoroughly relational ontology reminiscent of Leibniz's monadology. To what extent these represent real metaphysical alternatives rather than just different ways of describing the quantum state's structure is a question for wider philosophy of science and metaphysics.)

\subsection{Decoherence and the preferred basis}\label{DMWWeverettdecoherence}

In any case, once we have understood the ontological fallacy on which the Bare Theory rests, it remains to consider whether multiplicity does indeed emerge from a realist reading of the quantum state, and if so how. The 1990s saw an emerging consensus on this issue, developed  by Zurek, Gell-Mann and Hartle, Zeh, and many others\footnote{See, \egc, Zurek~\citeyear{zurek91,zurekroughguide}, Gell-Mann and Hartle~ \citeyear{gellmannhartle,gellmannhartle93}), \citeN{zeh93}.}  and explored from a philosophical perspective by Saunders~\citeyear{saundersevolution,saundersmetaphysics,saundersdecoherence}: the multiplicity is a consequence of decoherence.  That is, the structure of ``branching worlds'' suggested by the Everett interpretation is to be identified with the branching structure induced by the decoherence process. And since the decoherence-defined branching structure is comprised of quasi-classical histories, it would follow that Everett branches too are quasi-classical. 

It is important to be clear on the nature of this ``identification''. It cannot be taken as an additional axiom (else we would be back to the Many-Exact-Worlds theory); rather, it must somehow be forced on us by a realist interpretation of the quantum state. \citeN{gellmannhartle93} made the first sustained attempt to defend why this is so, with their concept of an IGUS: an ``information-gathering-and-utilising-system'' (similar proposals were made by \citeN{saundersevolution} and \citeN{zurekroughguide}) An IGUS, argue Gell-Mann and Hartle, can only function if the information it gathers and utilises is information about particular decoherent histories. If it attempts to store information about superpositions of such histories, then that information will be washed out almost instantly by the decoherence process. As such, for an IGUS to function it must perceive the world in terms of decoherent histories: proto-IGUSes which do not will fail to function. Natural selection then ensures that if the world contains IGUSes at all --- and in particular if it contains intelligent life --- those IGUSes will perceive the decoherence-selected branches as separate realities.

The IGUS approach is committed, implicitly, to functionalism: it assumes that intelligent, conscious beings just are information-processing systems, and it furthermore assumes that these systems are instantiated in certain structures within the quantum state. (Recall that in the ontology I have defended, the quantum state is a highly structured object, with its structure being describable in terms of the expectation values of whatever the structurally preferred observables are in whichever bare quantum formalism we are considering.) In \citeN{wallacestructure} I argued that this should be made explicit, and extended to a general functionalism about higher-level ontology: quite generally (and independent of the Everett interpretation) we should regard macroscopic objects like tables, chairs, tigers, planets and the like as structures instantiated in a lower-level theory. A tiger, for instance, is a pattern instantiated in certain collections of molecules; an economy is a pattern instantiated in certain collections of agents. 

\citeN{dennettrealpatterns} proposed a particular formulation of this functionalist ontology: those formally-describable structures which deserve the name `real' are those which are predictively and explanatorily necessary to our account of a system (in endorsing this view in \citeN{wallacestructure}, I dubbed it `Dennett's criterion'). So for instance (and to borrow an example from \citeN{wallacestructure}) what makes a tiger-structure ``real'' is the phenomenal gain in our understanding of systems involving tigers, and the phenomenal predictive improvements that result, if we choose to describe the system using tiger-language rather than restricting ourselves to talking about the molecular constituents of the tigers. A variant of Dennett's approach has been developed by \citeN{rossinformation} and \citeN{ladymanbook}; ignoring the fine details of how it is to be cashed out, let us call this general approach to higher-level ontology simply \emph{functionalism}, eliding the distinction between the general position and the restricted version which considers only the philosophy of mind.

Functionalism in this sense is not an uncontroversial position. \citeN{kim98}, in particular, criticises it and develops a rival framework based on mereology (this framework is in turn criticised in \citeN{rossspurrett}; see also  \citeN{wallacebbs} and other commentaries following \citeN{rossspurrett}, and also the general comments on this sort of approach to metaphysics in chapter 1 of \citeN{ladymanbook}). Personally I find it difficult to see how any account of higher-level ontology that is not functionalist in nature can possibly do justice to science as we find it; as \citeN[p.\,17]{dennettsweetdreams} puts it, ``functionalism in this broadest sense is so ubiquitous in science that it is tantamount to a reigning presumption of all science''; but in any event, the validity or otherwise of functionalism is a very general debate, not to be settled in the narrow context of the measurement problem.

The reason for discussing functionalism here is that (as I argued in \citeN{wallacestructure}) it entails that the decohering branches really should be treated --- really \emph{are} approximately independent quasi-classical worlds. Consider: if a system happens to be in a quasi-classical state
$\tpk{\vctr{q}(t),\vctr{p}(t)}{\psi(t)}$ (as defined in section \ref{DMWWquantumalgorithm} and made more precise in section \ref{DMWWdecoherence}) then\footnote{I ignore the possibility of chaos; if this is included, then the quantum system would be better described as instantiating an ensemble of classical worlds.} its evolution will very accurately track the phase-space point $(\vctr{q}(t),\vctr{p}(t))$ in its classical evolution, and so instantiates the same structures. As such, insofar as that phase-space point actually represents a macroscopic system, and insofar is what it \emph{is} to represent a macroscopic system is to instantiate certain structures, it follows that $\tpk{\vctr{q}(t),\vctr{p}(t)}{\psi}$ represents that same macroscopic system. The fact that these `certain structures' are instantiated in the expectation values\footnote{Note that here `expectation value' of an operator $\op{P}$ simply denotes \matel{\psi}{\op{P}}{\psi}; no probabilistic interpretation is intended. } of some phase-space POVM rather than in the location of a collection of classical particles is, from a functionalist perspective, quite beside the point.

Now if we consider instead a superposition
\be \label{DMWWtwostructures}
\alpha\tpk{\vctr{q}_1(t),\vctr{p}_1(t)}{\psi_1(t)}+\beta\tpk{\vctr{q}_2(t),\vctr{p}(t)}{\psi_2(t)}
\ee
then not one but two structures are instantiated in the expectation values of that same phase-space POVM: one corresponding to the classical history $(\vctr{q}_1(t),\vctr{p}_1(t))$, one to $(\vctr{q}_2(t),\vctr{p}_2(t))$, with decoherence ensuring that the structures do not interfere and cancel each other out but continue to evolve independently, each in its own region of phase space. 

Generalising to arbitrary such superpositions, we deduce that functionalism applied to the unitarily-evolving, realistically-interpreted quantum state yields the result that decoherence-defined branches are classical worlds. Not worlds in the sense of universes, precisely defined and dynamically completely isolated, but worlds in the sense of planets --- very accurately defined but with a little inexactness, and not quite dynamically isolated, but with a self-interaction amongst constituents of a world which completely dwarfs interactions between worlds.

This functionalist account of multiplicity is not in conflict with the IGUS strategy, but rather contains it. For not only could IGUSes not process information not restricted to a single branch, they could not even \emph{exist} across branches. The structures in which they are instantiated will be robust against decoherence only if they lie within a single branch. 

\subsection{Probability: the Incoherence Problem}

The decoherence solution to the preferred-basis problem tells us that the quantum state is really a constantly-branching structure of quasi-classical worlds. It is much less clear how notions of probability fit into this account: if an agent knows for certain that he is about to branch into many copies of himself --- some of which see a live cat, some a dead cat --- then how can this be reconciled with the Quantum Algorithm's requirement that he should expect with a certain probability to see a live cat?

It is useful to split this problem in two:
\begin{description}
\item[The Incoherence Problem:] In a deterministic theory where we can have perfect knowledge of the details of the branching process, how can it even make sense to
assign probabilities to outcomes?
\item[The Quantitative Problem:]Even if it does make sense to assign
probabilities to outcomes, why should they be the probabilities given by the
Born rule?
\end{description}

The incoherence problem rests on problems with personal identity. In branching, one person is replaced by a multitude of (initially near-identical) copies of that person, and it might be thought that this one-to-many relation of past to future selves renders any talk of personal identity simply incoherent in the face of branching (see, \egc, \citeN{albertloewermm} for a defence of this point). 
However (as pointed out by \citeN{saundersprobability}) this charge of incoherence fails to take account of what grounds ordinary personal identity: namely (unless we believe in Cartesian egos) it is grounded by the causal and structural relations between past and future selves. These relations exist no less strongly between past and future selves when there exist \emph{additional} such future selves; as such, if it is rational to care about one's unique future self (as we must assume if personal identity in \emph{non}-branching universes is to be made sense of) then it seems no less rational to care about one's multiple future selves in the case of branching. This point was first made --- entirely independently of QM --- by Parfit; see his \citeyear{parfitbook}. 

This still leaves the question of how \emph{probability} fits in, and at this point there are two strategies available: the \emph{Fission Program} and the \emph{Subjective Uncertainty Program} \cite{wallaceepist}. The Fission Program works by considering situations where the interests of future selves are in \emph{conflict}. For instance, suppose the agent, about to observe Schr\"{o}dinger's Cat and thus to undergo branching, is offered an each-way bet on the cat being left alive. If he takes the bet, those future versions of himself who exist in live-cat branches will benefit and those who live in dead-cat branches will lose out. In deciding whether to take the bet, then, the agent will have to weigh the interests of some of his successors against those of others. Assigning a (formal) probability to each set of successors and choosing that action which benefits the highest-probability subset of successors is at least \emph{one way} of carrying out this weighing of interests. 

This strategy is implicit in \citeN{deutschprob} and has been explicitly  defended by Greaves (\citeyearNP{greaves}).
It has the advantage of conforming unproblematically to our intuition that ``I can feel uncertain over $P$ only if I think that there is a fact of the matter regarding $P$ of which I am ignorant'' \cite{greaves}; it has the disadvantage of doing violence to our intuitions that uncertainty about the future is generally justified; it  is open to question what epistemic weight these intuitions should bear.\footnote{See \citeN{wallacebranching}, especially section 6, for more discussion of this point.} There is, however, a more serious problem with the Fission Program: it is at best uncertain whether it solves the measurement problem. For recall: in the framework of this review, `to solve the measurement problem' is to construct a theory which entails the truth (exact or approximate) of the Quantum Algorithm, and that Algorithm dictates that we should regard macroscopic superpositions as probabilistic, and hence that an agent expecting branching should be in a state of uncertainty. The challenge for fission-program advocates is to find an alternative account of our epistemic situation according to which the Everett interpretation is nonetheless explanatory of our evidence. See \citeN{greavesepistemic} for Greaves' proposed account, which draws heavily on Bayesian epistemology.

The Subjective Uncertainty Program aims to establish that probability really, literally, makes sense in the Everett universe: that is, that an agent who knows for certain that he is about to undergo branching is nonetheless justified in being \emph{uncertain} about what to expect. (This form of uncertainty cannot depend on ignorance of some facts describable from a God's-eye perspective, since the relevant features of the universal state are \emph{ex hypothesi} perfectly knowable by the agent --- hence, \emph{subjective} uncertainty). 

Subjective uncertainty was first defended by \citeN{saundersprobability}, who asks: suppose that you are about to be split into multiple copies, then \emph{what} should you expect to happen? He argues that, given that each of your multiple successors has the same structural/causal connections to you as would have been the case in the absence of splitting, the only coherent possibility is \emph{uncertainty}: I should expect to be one of my future selves but I cannot know which. 

I presented an alternative strategy for justifying subjective uncertainty in \citeN{wallacebranching} (and more briefly in \citeN{wallaceepist}). My proposal is that we are led to subjective uncertainty by considerations in the philosophy of language: namely, if we ask how we would analyse the semantics of a community of language-users in a constantly branching universe, we conclude that claims like ``$X$ might happen'' come out true if $X$ happens in some but not all branches.

If the Subjective Uncertainty program can be made to work, it avoids the epistemological problem of the Fission Program,  for it aims to recover the quantum algorithm itself  (and not just to account for its empirical success.)  It remains controversial, however, whether subjective uncertainty really makes sense. For further discussion of subjective uncertainty and identity across branching, see \citeN{greaves}, \citeN{saunderswallace}, \citeN{wallaceepist}  and  \citeN{lewissu}.

\subsection{Probability: the Quantitative Problem}\label{DMWWprobabilityquantitative}

The Quantitative Problem of probability in the Everett interpretation is often posed as a paradox: the \emph{number} of branches has nothing to do with the \emph{weight} (\iec, modulus-squared of the amplitude) of each branch, and the only reasonable choice of probability is that each branch is equiprobable, so the probabilities in the Everett interpretation can have nothing to do with the Born rule.  

This sort of criticism has sometimes driven advocates of the Everett interpretation back to the strategy of modifying the formalism, adding a continuous infinity of worlds \cite{deutsch85} or minds \cite{albertloewermm,lockwoodbook}  in proportion to the weight of the corresponding branch. But this is unnecessary, for the criticism was mistaken in the first place: it relies on the idea that there is some sort of remotely well-defined branch number, whereas there is no such thing.

This can most easily be seen using the decoherent-histories formalism. Recall that the `branches' are decoherent histories in which quasi-classical dynamics apply, but recall too that the criteria of decoherence and quasi-classicality are approximate rather than exact. We can always fine-grain a given history space at the cost of slightly less complete decoherence, or coarse-grain it to ensure more complete decoherence; we can always replace the projectors in a history space by ever-so-slightly-different projectors and obtain an equally decoherent, equally quasi-classical space. These transformations do not  affect the \emph{structures} which can be identified in the decoherent histories (for those structures are themselves only approximately defined) but they wildly affect the \emph{number} of branches with a given macroscopic property.

The point is also apparent using the formalism of quasi-classical states discussed in section \ref{DMWWquantumalgorithm}. Recall that in that framework, a macroscopic superposition is written
\be
\int \dr{\vctr{q}}\dr{\vctr{p}}\alpha(\vctr{q},\vctr{p})\tpk{\vctr{q},\vctr{p}}{\psi_{q,p}}.
\ee
If the states $\tpk{\vctr{q},\vctr{p}}{\psi_{q,p}}$ are to be taken as each defining a branch, there are continuum many of them, but if they are too close to one another then they will not be effectively decohered. So we will have to define branches via some coarse-graining of phase space into cells $\mc{Q}_n$, in terms of which we can define states
\be \ket{n}=\int_{\mc{Q}_n}\dr{\vctr{q}}\dr{\vctr{p}}\alpha(\vctr{q},\vctr{p})\tpk{\vctr{q},\vctr{p}}{\psi_{q,p}}.
\ee
The coarse-graining must be chosen such that the states \ket{n} are effectively decohered, but there will be no precisely-determined `best choice' (and in any case no precisely-determined division of Hilbert space into macroscopic and microscopic degrees of freedom in the first place.)

As such, the `count-the-branches' method for assigning probabilities is ill-defined.\footnote{\citeN{wallace3branch} presents an argument that the count-the-branches rule is incoherent even if the branch number \emph{were} to be exactly definable.} But if this dispels the \emph{paradox} of objective probability, still a \emph{puzzle} remains: why use the Born rule rather than any other probability rule?

Broadly speaking, three strategies have been proposed to address this problem without modifying the formalism. The oldest strategy is to appeal to relative frequencies of experiments. It has long been known (\citeNP{everett}) that if many copies of a system are prepared and measured in some fixed basis, the total weight of those branches where the relative frequency of any result differs appreciably from the weight of that result tends to zero as the number of copies tends to infinity. But it has been recognised for almost as long that this account of probability courts circularity: the claim that a branch has \emph{very small weight} cannot be equated with the claim that it is \emph{improbable}, unless we assume that which we are trying to prove, namely that weight=probability.

It is perhaps worth noting, though, that precisely equivalent objections can be made against the frequentist definition of probability. Frequentists equate probability with long-run relative frequency, but again they run into a potential circularity. For we cannot prove that relative frequencies converge on probabilities, only that they \emph{probably} do: that is, that the probability of the relative frequencies differing appreciably from the probabilities tends to zero as the number of repetitions of an experiment tends to infinity (the maths is formally almost identical in the classical and Everettian cases). As such, it is at least arguable that anyone who is happy with frequentism \emph{in general} as an account of probability should have no additional worries in the case of the Everett interpretation.\footnote{\citeN{fgg} try to evade the circularity by direct consideration of infinitely many measurements, rather than just by taking limits; their work has recently criticised by \citeN{cavesschackfrequentism}. } 

The second strategy might be called \emph{primitivism}: simply postulate that weight=probability. This strategy is explicitly defended by \citeN{saundersprobability}; it is implicit in Vaidman's ``Behaviour Principle'' \cite{vaidmanencyclopedia}. It is open to the criticism of being unmotivated and even incoherent: effectively, to make the postulate is simply to stipulate that it is rationally compelling to care about one's successors in proportion to their weight (or to expect to be a given successor in proportion to his weight, in subjective-uncertainty terms), and it is unclear that we have any right to \emph{postulate} any such rationality principle, as if it were a law of nature. But again, it can be argued that classical probability theory is no better off here --- what is a ``propensity'', really, other than a primitively postulated rationality principle? (This is David Lewis's ``big bad bug'' objection to Humean supervenience; see \citeN[pp.\,xiv-xvii]{lewispapers2} and \citeN{lewis94} for further discussion of it). \citeN{papineaubjps} extends this objection to a general claim about probability in Everett: namely, although we do not understand it at all, we do not understand classical probability any better! --- so it is unfair to reject the Everett interpretation simply on the grounds that it has an inadequate account of probability.

The third, and most recent, strategy has no real classical analogue (though it has some connections with the `classical' program in philosophy of probability, which aims to derive probability from symmetry). This third strategy aims to derive the principle that weight=probability from considering the constraints upon rational action of agents living in an Everettian universe.\footnote{Given this, it is tempting to consider the Deutsch program as a form of subjectivism about probability, but --- as I argue more extensively in \citeN{wallaceepist} --- this is not the case. There was always a conceptual connection between objective probability and the actions of rational agents (as recognised in Lewis's Principal Principle \cite{lewischance}) --- what makes a probability `objective' is that all rational agents are constrained by it in the same way, and this is what Deutsch's proofs (purport to) establish for the quantum weight. In other words, there are objective probabilities --- and they have turned out to be the quantum weights.} It was initially proposed by \citeN{deutschprob}, who presented what he claimed to be a proof of the Born rule from decision-theoretic assumptions; this proof was criticised by Barnum \emph{et al}~\citeyear{barnumetal}, and defended by \citeN{decshort}. Subsequently, I have presented various expansions and developments on the proof (Wallace \citeyearNP{wallaceprobdec},\citeyearNP{wallace3branch}), and Zurek \citeyear{zurekenvariance03,zurekenvariance05} has presented another variant of it. It remains a subject of controversy whether or not these `proofs' indeed prove what they set out to prove.

\subsection{Further Reading}

\citeN{barrettbook} is an extended discussion of Everett-type interpretations (from a perspective markedly different from mine); \citeN{vaidmanencyclopedia} is a short (and fairly opinionated) survey.

\citeN{kent} is a classic criticism of ``old-style'' many-worlds theories; \citeN{baker}, \citeN{lewissu} and \citeN{hemmopitowsky07} criticise various aspects of the Everett interpretation as presented in this chapter.

\section{Dynamical-collapse theories}\label{DMWWdynamicalcollapse}

In this section and the next, we move away from pure \emph{interpretations} of the bare quantum formalism, and begin to consider substantive \emph{modifications} to it. There are essentially two ways to do this:
\begin{quote}
Either the wavefunction, as given by the Schr\"{o}dinger equation, is not everything, or it is not right (\citeNP{bellbook}, p.\,201)
\end{quote} 
That is, if unitary QM predicts that the quantum state is in a macroscopic superposition, then either 
\begin{enumerate}
\item the macroscopic world does not supervene on the quantum state alone but also (or instead) on so-called ``hidden variables'', which pick out one term in the superposition as corresponding to the macroscopic world; or 
\item the predictions of unitary QM are false: unitary evolution is an approximation, valid at the microscopic level but violated at the macroscopic, so that macroscopic superpositions do not in fact come into existence.
\end{enumerate}
The first possibility leads us towards hidden variable theories, the topic of section \ref{DMWWhidden}. This section is concerned with ``dynamical collapse'' theories, which modify the dynamics  to avoid macroscopic superpositions.

\subsection{The GRW theory as a paradigm of dynamical-collapse theories}\label{DMWWGRW}

How, exactly, should we modify the dynamics? Qualitatively it is fairly straightforward to see what is required. Firstly, given the enormous empirical success of QM at the microscopic level we would be well advised to leave the Schr\"{o}dinger equation alone at that level. At the other extreme, the Quantum Algorithm dictates that states like 
\be \label{DMWWcatagain}\alpha\ket{\mathrm{\mbox{dead cat}}}+\beta \ket{\mathrm{\mbox{live cat}}}\ee
must be interpretable probabilistically, which means that our modification must ``collapse'' the wavefunction rapidly into either \ket{\mathrm{\mbox{dead cat}}} or \ket{\mathrm{\mbox{live cat}}} --- and furthermore, they  must do it stochastically, so that the wavefunction collapses into \ket{\mathrm{\mbox{dead cat}}} with probability $|\alpha|^2$ and \ket{\mathrm{\mbox{live cat}}} with probability $|\beta|^2$.

Decoherence theory offers a way to make these qualitative remarks somewhat more precise. We know that even in unitary QM, probabilistic mixtures of pointer-basis states are effectively indistinguishable from coherent superpositions of those states. So we can be confident that our dynamical-collapse theory will not be in contradiction with the observed successes of quantum theory provided that coherent superpositions are decohered by the environment before they undergo dynamical collapse --- or, equivalently, provided that superpositions which are robust against decoherence generally \emph{do not} undergo dynamical collapse. Furthermore, dynamical collapse should leave the system in (or close to) a pointer-basis state --- this is in any case desirable, since the pointer-basis states are quasi-classical states, approximately localized in phase space.

The other constraint --- that macroscopic superpositions should collapse quickly --- is harder to quantify. \emph{How} quickly should they collapse? Proponents of dynamical-collapse theories --- such as \cite{bassighirardireview} --- generally require that the speed of collapse should be chosen so as to prevent ``the embarrassing occurrence of linear superpositions of appreciably different locations of a macroscopic object''. But it is unclear exactly when a given superposition counts as ``embarrassing''. One natural criterion is that the superpositions should collapse before humans have a chance to observe them. But the motivation for this is open to question. For suppose that a human observer looks at the state (\ref{DMWWcatagain}). If collapse is quick, the state rapidly collapses into
\be \ket{\mathrm{\mbox{dead cat}}} \,\,\,\,\,\,\,\mathrm{\mbox{or}}\,\,\,\,\,\, \ket{\mathrm{\mbox{live cat}}},
\ee
and observation puts the cat-observer system into the state
\be
\tpk{\mathrm{\mbox{dead cat}}}{\mathrm{\mbox{observer sees dead cat}}}\,\,\,\,\,\,\,\mathrm{\mbox{or}}\,\,\,\,\,\, 
\tpk{\mathrm{\mbox{live cat}}}{\mathrm{\mbox{observer sees live cat}}}.
\ee
Given the stochastic nature of the collapse, the probability of the observer being in a state where he remembers seeing a dead cat is $|\alpha|^2$.

Now suppose that the collapse is much slower, taking several seconds to occur. Then the cat-observer system enters the superposition
\be
\alpha \tpk{\mathrm{\mbox{dead cat}}}{\mathrm{\mbox{observer sees dead cat}}}+ \beta
\tpk{\mathrm{\mbox{live cat}}}{\mathrm{\mbox{observer sees live cat}}}.
\ee
Who knows what it is like to be in such a state?\footnote{According to the functionalist analysis of section \ref{DMWWeverettdecoherence} ``it is like'' there being two people, one alive and one dead; but we shall not assume this here.}  But no matter: in a few seconds the state collapses to 
\be
\tpk{\mathrm{\mbox{dead cat}}}{\mathrm{\mbox{observer sees dead cat}}}\,\,\,\,\,\,\,\mathrm{\mbox{or}}\,\,\,\,\,\, 
\tpk{\mathrm{\mbox{live cat}}}{\mathrm{\mbox{observer sees live cat}}}.
\ee
Once again, the agent is in a state where he remembers seeing either a live or dead cat, and the probability is $|\alpha|^2$ that he remembers seeing a dead cat --- since his memories are encoded in his physical state, he will have no memory of the superposition. So the fast and slow collapses appear indistinguishable empirically.

However, let us leave this point to one side. The basic constraints on a collapse theory remain: it must cause superpositions of pointer-basis states to collapse to pointer-basis states, and it must do so quickly enough to suppress ``embarrassing superpositions''; however, it must not have any appreciable affect on states which do not undergo decoherence.

Here we see again the difficulties caused by the approximate and ill-defined nature of decoherence. If decoherence were an exactly and uniquely defined process, we could just stipulate that  collapse automatically occurs when states enter superpositions of pointer-basis states. Such a theory, in fact, would be exactly our `solution that isn't' from section \ref{DMWWsolutionthatisnt}. But since decoherence is not at all like this, we cannot use it directly to define a dynamical-collapse theory.

The requirement on a dynamical collapse theory is then: find a modification to the Schr\"{o}dinger equation that is cleanly defined in microphysical terms, and yet which closely approximates collapse to the decoherence-preferred basis. And such theories can in fact be found. The classic example is the ``GRW theory'' of \citeN{grw}.  The GRW theory postulates that every particle in the Universe has some small spontaneous chance per unit time of collapsing into a localised Gaussian wave-packet:
\be\label{DMWWgrw}
\psi(x)\longrightarrow \mc{N}\exp(-(x-x_0)^2/2L^2) \psi(x)
\ee 
where $L$ is a new fundamental constant (and \mc{N} is just a normalisation factor). The probability of collapse defines another new constant: $\tau$, the mean time between collapses. Crucially, the `collapse centre' $x_0$ is determined stochastically: the probability that $\psi$ collapses to a Gaussian with collapse centre in the vicinity of $x_0 $ is proportional to $|\psi(x_0)|^2$. If the particle is highly localised (that is, localised within a region small compared with $L$) then the collapse will have negligible effect on it; if it is in a superposition of such states, it will  be left in just one of them, with the probability of collapse to a given state being equal to its mod-squared amplitude.

Now, $\tau$ is chosen to be extremely small, so that the chance of an isolated particle collapsing in a reasonable period of time is quite negligible. But things are otherwise if the particle is part of a macroscopic object. (The generalisation of (\ref{DMWWgrw}) to $N$-particle systems is just
\be\label{DMWWgrw2}
\psi(x_1,\ldots x_m, \ldots x_N)\longrightarrow \mc{N}\exp(-(x_m-x_0)^2/2L^2) \psi(x_1,\ldots x_m, \ldots x_N)
\ee 
where the collapse occurs on the $m$th particle.) For suppose that that macroscopic object is in a superposition: something like (schematically)
\be
\alpha \ket{\mathrm{\mbox{at }}X}\otimes \cdots \otimes \ket{\mathrm{\mbox{at }}X}
+ 
\beta \ket{\mathrm{\mbox{at }}Y}\otimes \cdots \otimes \ket{\mathrm{\mbox{at }}Y}.
\ee
If $N \gg 1/\tau$, then within a small fraction of a second one of these particles will undergo collapse. Then the collapse will kick that particle (roughly speaking) into either \ket{\mathrm{\mbox{at }}X} (with probability $|\alpha|^2$) or \ket{\mathrm{\mbox{at }}Y} (with probability $|\beta|^2$).  For convenience, suppose it in fact collapses to $X$. Then because of the entanglement, so do all of the other particles - the system as a whole collapses to a state very close to
\be
\ket{\mathrm{\mbox{at }}X}\otimes \cdots \otimes \ket{\mathrm{\mbox{at }}X}.
\ee
(Taking more mathematical care: if $\psi(x_1, \ldots x_N)$ is the wavefunction of a macroscopic $N$-particle body approximately localised at $x=0$, then 
\be
\alpha \psi(x_1-X, \ldots x_N-X)
+ \beta
\psi(x_1-Y, \ldots x_N-Y).
\ee
If the first particle undergoes collapse, then its collapse centre has a probability $\simeq |\alpha|^2$ to be in the vicinity of $X$. Assuming this is so, the post-collapse wavefunction is approximately proportional to
\be
\alpha 
\psi (x_1-X, \ldots x_N-X)
+
\beta \exp(-|X-Y|^2/L^2)\psi(x_1-Y, \ldots x_N-Y).
\ee
On the assumption that $|X-Y| \gg L$, the second term in the superposition is hugely suppressed compared with the first.)

So: the GRW theory causes superpositions of $N$ particles to collapse into localised states in a time $\sim \tau/N$, which will be very short if $\tau$ is chosen appropriately; but it has almost no detectable effect on small numbers of particles. From the perspective in which I have presented dynamical collapse, GRW incorporates two key observations:
\begin{enumerate}
\item Although the decoherence process is approximately defined and highly emergent, the actual pointer-basis states are fairly simple: they are Gaussians, approximately localised at a particular point in phase space. As such, it is sufficient to define collapse as suppressing superpositions of position states.
\item Similarly, although the \emph{definition} of `macroscopic system' given by decoherence is highly emergent, \emph{in practice} such systems can be picked out simply by the fact that they are compounds of a great many particles. So a collapse mechanism defined for single particles is sufficient to cause rapid collapse of macroscopic systems.
\end{enumerate}
The actual choice of GRW parameters is determined by the sorts of considerations discussed above. Typical choices are $L=10^{-5} \mathrm{cm}$, $\tau=10^{16} \mathrm{s}$, ensuring that an individual particle undergoes collapse only after $\sim 10^8$ years, but a grain of dust $\sim 10^{-2}$ cm across will undergo collapse within a hundredth of a second, and Schr\"{o}dinger's cat will undergo it after $\sim 10^{-11}$ seconds. (In fact, if the GRW theory holds then the cat never has the chance to get into the alive-dead superposition in the first place: dynamical collapse will occur in the cat-killing apparatus long before it begins its dread work.)

The GRW theory is not the ``last word'' on dynamical-collapse theories. Even in the non-relativistic domain it is not fully satisfactory: manifestly, the collapse mechanism does not preserve the symmetries of the wavefunction, and so it is not compatible with the existence of identical particles. These and other considerations led \citeN{pearle} to develop ``continuous state localisation'' (or CSL), a variant on GRW where the collapse mechanism preserves the symmetry of the wavefunction, and most advocates of dynamical collapse now support CSL rather than GRW. (See \citeN[section 8]{bassighirardireview} for a review of CSL.)

However, there seems to be a consensus  that foundational issues with CSL can be equally well understood in the mathematically simpler context of GRW. As such, conceptual and philosophical work on dynamical collapse is predominantly concerned with GRW, in the reasonable expectation that lessons learned there will generalise to CSL and perhaps beyond.

\subsection{The problem of tails and the Fuzzy Link}\label{DMWWtails}

The main locus of purely \emph{philosophical} work on the GRW theory in the past decade has been the so-called ``problem of tails''. 
As I shall argue (following \citeN{cordero} to some extent) there are actually two ``problems of tails'', only one of which is a particular problem of dynamical-collapse theories, but both are concerned with the stubborn resistance of the wavefunction to remain decently confined in a finite-volume region of space.

The original ``problem of tails'' introduced by \citeN{albertloewer1996}  works as follows. Suppose we have a particle in a superposition of two fairly localised states \ket{\mathrm{\mbox{here}}} and \ket{\mathrm{\mbox{there}}}:
\be \ket{\psi}=\alpha\ket{\mathrm{\mbox{here}}}+\beta\ket{\mathrm{\mbox{there}}}.\ee
Dynamical collapse will rapidly occur, propelling the system into something like
\be \ket{\psi'}=\sqrt{1-\epsilon^2}\ket{\mathrm{\mbox{here}}}+\epsilon\ket{\mathrm{\mbox{there}}}.\ee
But (no matter how small $\epsilon$ may be) this is not the same state as
\be\ket{\psi''}=\ket{\mathrm{\mbox{here}}}.\ee
Why should the continued presence of the `there' term in the superposition --- the continued indefiniteness of the system between `here' and `there' --- be ameliorated in any way at all just because the `there' term has low amplitude?

Call this the \emph{problem of structured tails} (the reason for the name will become apparent).  It is specific to dynamical collapse theories; it is a consequence of the GRW collapse mechanism, which represents collapse by multiplication by a Gaussian and so fails to annihilate terms in a superposition no matter how far they are from the collapse centre.

It is interesting, though, that most of the recent `tails' literature has dealt with a rather different problem which we might call the \emph{problem of bare tails}. Namely: even if we ignore the `there' state, the wave function of \ket{\mathrm{\mbox{here}}} is itself spatially highly delocalised. Its centre-of-mass wavefunction is no doubt a Gaussian, and Gaussians are completely delocalised in space, for all that they may be concentrated in one region or another. So how can a delocalised wave-packet possibly count as a localised particle?

\emph{This} problem has little or nothing to do with the GRW theory. Rather, it is an unavoidable consequence of using wave-packets to stand in for localised particles. For \emph{no} wave-packet evolving unitarily will remain in any finite spatial region for more than an instant (consider that infinite potentials would be required to prevent it tunneling to freedom.) 

Apparent force is added to this objection by applying the eigenvector-eigenvalue link. The latter gives a perfectly clear criterion for when a particle is localised in any spatial region $R$: it must be an eigenstate of the operator 
\be\op{P}_R=\int_R \dr{x}\proj{x}{x}.\ee
That is, it must have support within $R$; hence, no physically realisable state is every localised in a finite region.

One might be inclined to respond: so much the worse for the eigenvector-eigenvalue link, at least in the context of continuous observables. As we have seen in section \ref{DMWWagainst}, its motivation in modern QM is tenuous at best. But that simply transfers the problem: if the eigenvector-eigenvalue link is not to be the arbiter for which physical states count as localised, what is?

Albert and Loewer propose a solution: a natural extension of the eigenvector-eigenvalue link which they call the \emph{fuzzy link}. Recall that  the eigenvector-eigenvalue link associates (at least a subset of) properties 1:1 with projectors, and regards a state \ket{\psi} as possessing (the property associated with) projector \op{P} iff $\op{P}\ket{\psi}=\ket{\psi}$; that is, if
\be|\op{P}\ket{\psi}-\ket{\psi}|=0.\ee
The fuzzy link is a relaxation of this condition: the properties remain in one-to-one correspondence with the projectors, but now \ket{\psi} has property (associated with) \op{P} if, for some fixed small $p$,
\be\label{DMWWfuzzylink}|\op{P}\ket{\psi}-\ket{\psi}|<p.\ee
We shall return to the constant $p$ and the question of what determines it; for now note only that it must be chosen to be sufficiently large that wave-packets really count as localised, sufficiently small that intuitively `delocalised' states do not erroneously count as localised.

\subsection{The counting anomaly}\label{DMWWcounting}

But these constraints lead to a problem: the \emph{counting anomaly}, introduced by  \citeN{lewis94}. Suppose, with Lewis, that a (ridiculously\footnote{Somewhere in the vicinity of $10^{10^{20}}$ are required; recall that the number of particles in the visible universe is about $10^{10^2}.$}) large number $N$ of distinguishable non-interacting particles are confined within some box. The wavefunction of each will be strongly peaked inside the box, so that if $\op{P}_i$ is the `particle $i$ is in the box' operator (that is, if it projects onto states of the $i$th particle with support in the box) then $|\op{P}_i\ket{\psi}-\ket{\psi}|\sim\epsilon$ for extremely small $\epsilon$. (For instance, for a 1-metre box and an atom whose wavepacket has characteristic width $\sim 10^{-10}\mathrm{m}$, $\epsilon$ is of the order of $10^{-10^{10}}$.) 

But now consider the proposition `all $N$ particles are in the box'. By definition, this is represented by the operator $\op{P}=\Pi_{i=1}^N\op{P}_i$. Suppose that each particle has identical state \ket{\psi}; suppose that each \ket{\psi} is highly localised in the box, as above. Then the overall state of the $N$ particles is $\ket{\Psi}=\otimes_{i=1}^N\ket{\psi}$ . Then $|\op{P}\ket{\Psi}|=\Pi_{i=1}^N |\op{P}_i\ket{\psi}|=(1-\epsilon)^N$. 

And this is unfortunate for the Fuzzy Link. For no matter how small $\epsilon$ may be, there will be some value of $N$ for which $(1-\epsilon)^N<p$. And for that value of $N$, the Fuzzy Link tells us that it is false that all $N$ particles are in the box, even as it tells us that, for each of the $N$ particles, it is true that \emph{that} particle is in the box.

So how did that happen? We can see what is going on in the following way. The proposition `all $N$ particles are in the box' is by nature compositional: it is in some sense definitionally equivalent to `particle 1 is in the box and particle 2 is in the box and \ldots and particle $N$ is in the box.' But there are two ways to understand this compositionality:
\begin{enumerate} 
\item The actual, true-or-false, proposition `all $N$ particles are in the box' is equivalent to the conjunction of the $N$ propositions `particle 1 is in the box' `particle 2 is in the box', etc. So it is true iff each of those propositions is true. In turn, via Fuzzy Link semantics each one of those propositions is true iff $|\op{P}_i\ket{\psi}-\ket{\psi}|<p.$
\item The proposition `all $N$ particles are in the box' is associated, via the one-to-one correspondence between propositions and projectors, with some projector; since that one-to-one correspondence respects the compositional structure of propositions, the proposition is associated with that projector \op{P} which is the logical product of the $N$ projectors corresponding to `particle 1 is in the box', `particle 2 is in the box', \etc. Once we have \op{P}, we can use the Fuzzy Link to determine that the proposition is true iff $|\op{P}\ket{\Psi}-\ket{\Psi}|<p$ --- that is, iff $\Pi_i|\op{P}_i\ket{\psi}-\ket{\psi}|<p$.
\end{enumerate}
In both cases, we construct the proposition `all the marbles are in the box' via conjunction of the $N$ component propositions. But in case (1) we apply this conjunction at the level of the actual propositions \emph{after} having applied the Fuzzy Link to extract a propositional truth condition from a projector; in case (2) we do it the other way around. And the two procedures \emph{do not commute}.

That might suggest an obvious remedy to the Counting Anomaly: fix one way round --- fairly obviously (1), given that it seems forced on us by the semantics of ordinary language --- and declare it correct. Which amounts to the following: restrict the Fuzzy Link to our basic propositions (those describing the properties of individual particles), and then allow ordinary truth-functional semantics to dictate truth conditions for compound propositions. On this strategy (call it the single-particle fuzzy link) (2) becomes a derived and approximate truth --- something which in fact holds in all conceivable circumstances in the actual world, but is not logically true.

\citeN{cliftonmonton1}, in their discussion of the Counting Anomaly, consider and reject this view, for instructive reasons:
\begin{quote}
[T]his strategy would require that the wavefunction collapse theorist not simply weaken the eigenstate-eigenvalue link between proof and probability 1, but sever this link entirely. And if one is willing to entertain the thought that events in a quantum world can happen without being mandated or made overwhelmingly likely by the wavefunction, then it is no longer clear why one should need to solve the measurement problem by collapsing wavefunctions! Another reason not to [accept the single-particle fuzzy link strategy] is that it seems arbitrary to apply a semantic rule for quantum states to a single-particle-system, but not to a multi-particle-system. Indeed, to the extent that one supposes there to be a plausible intuitive connection between an event's having high probability according to a theory, and the event actually occurring, one is hard pressed to resist the intuition in the multi-particle case.
\end{quote}
However, this fails to recognise that for collapse theorists, the wavefunction is a physical entity. It is not some sort of probability distribution which makes events `overwhelmingly likely', it is the microscopic stuff out of which macroscopic objects  --- including the constituents of events ---are made. Furthermore, Clifton and Monton are too quick to accept (in their discussion of a `semantic rule') that there must be a link between the macroscopic properties of a system and the projectors onto that system's Hilbert space. As we have seen, this is a consequence of the eigenstate-eigenvalue link that we would be wise to reject. If we are serious about taking a realist attitude to the wavefunction, then macroscopic properties may turn out to supervene on any microscopic properties we like --- including directly on single-particle fuzzy-link-defined positions --- and need  not have any particularly useful direct relation to large-dimensional projectors.

In fact, the single-particle fuzzy link has found little favour, partly for technical reasons (to be fair, this too was anticipated by Clifton and Monton). However, a conceptually rather closely related  response to the Counting Anomaly has been widely accepted: the Mass Density Link \cite{ghirardigrassibenati}. The Mass Density Link makes a clean break with eigenvector/eigenvalue link semantics: commendably from the viewpoint of this chapter, it grants no particular semantic status at all to the `observables' in general. Instead, it defines the following \emph{mass density observer} for an $n$-particle system:
\be
\op{M}(\vctr{x})=\sum_i m_i \op{N}_i(\vctr{x})
\ee
where the sum is over all particles and $m_i$ and $\op{N}_i(\vctr{x})$ are the number density operators for the $i$th particle. For instance, if there is just one particle of mass $m$ under consideration then the mass density operator for that particle is
\be
\op{M}(\vctr{x})=m \proj{\vctr{x}}{\vctr{x}}.
\ee
(It is apparent from this that the mass density is a distributional operator, rigorously defined only when smeared over some finite volume.) The \emph{mass density} of state \ket{\psi}  is then defined just as
\be
\rho_\psi(\vctr{x})=\matel{\psi}{\op{M}(\vctr{x})}{\psi}.
\ee

In more intuitively understandable terms, the mass density is the sum of the mass-weighted `probability' densities for finding each particle at $x$; for a one-particle wavefunction $\psi(x)$, $\rho(x)$ is just $m \times |\psi(x)|^2$.

According to the Mass Density Link, a particle is in the box if some sufficiently high fraction $(1-p)$ of its mass is in the box (again we postpone questions as to what fixes the value of $p$). The meaning of `all $N$ particles are in the box' is, uncomplicatedly, `particle 1 is in the box and particle 2 is in the box and \ldots', and the truth conditions of that proposition are just that it is true iff all of the component propositions are true. The interpretation provides no alternative, possibly-incompatible way to access its truth value, and the Counting Anomaly is avoided.\footnote{Ghirardi \emph{et al} have a further requirement on the mass density: that it be \emph{accessible}; see the references above, and especially \citeN[pp.\,86--92]{bassighirardireview}, for the meaning of this claim, and \citeN{montonmassdensity} for a criticism.}

Lewis \citeyear{lewis2003,lewisfuzzy,lewis2005} is unconvinced. He argues (to some extent following \citeNP{cliftonmonton2}) that the Mass Density Link avoids the Counting Anomaly at the cost of a comparably unintuitive result, which he calls the \emph{location anomaly}. This anomaly arises when we consider the process of \emph{looking} at the box and physically counting the number of particles in it. The ordinary quantum measurement theory --- which the GRW theory is supposed to reproduce --- then predicts that the expected number of particles found in the box will be somewhat less than $N$. Lewis claims that this clash between the predictions of how many particles are \emph{found} in the box and how many are \emph{actually} in the box ``violates the entailments of everyday language'' \cite[p.\,174]{lewis2005}.

Ghirardi and Bassi \citeyear{bassighirardi99,bassighirardireview} are bemused by this criticism, for reasons that I share: we have a theory which (a) gives a perfectly well-defined description of how many particles are in the box; (b) allows a precise description, in terms acceptable to the realist, of the measurement process by which we determine how many particles are in the box; (c) predicts that if the number of particles is sufficiently (\iec, ridiculously) large there will be tiny deviations between the actual number of particles and the recorded number of particles. They, and I, fail to see what the problem is here; I leave readers to reach their own conclusions.

\subsection{The status of the link principles}

Perhaps, however, we are beginning to lose sight of the wood for the trees. The GRW theory is normally presented simply as a mathematical modification of the dynamics of the wavefunction, in which case the theory's mathematical consistency is not in question. So how did we even \emph{begin} to worry that the theory was internally contradictory? 
Answering this question requires us to consider: what actually is the status of these link principles (whether the ``fuzzy link'' or the ``mass density link'')? Often, discussions of them are most naturally read as regarding the principles as a piece of physical law: that is, to specify a dynamical-collapse theory we must not only give the modifications to the Schr\"{o}dinger equation but also state the link principle. On this reading ``fuzzy-link'' GRW and ``mass-density'' GRW are different physical theories; so, too, are two versions of fuzzy-link GRW which disagree about the value of $p$ in (\ref{DMWWfuzzylink}). \citeN{montonmassdensity} argues explicitly for this reading of the link principles (mostly on the grounds that he wishes to reject wavefunction realism altogether), and Allori \emph{et al}~\citeyear{goldsteinontology} explore its consequences \emph{in extenso}, but other authors seem to write as though they adopted this reading. \citeN{bassighirardireview}, for instance,  describe the mass density as the `beable' of the GRW theory (pp.\,94--95); \citeN{lewisfuzzy} proposes empirical tests to measure the value of $p$ in the Fuzzy Link.

How should we understand the ontology of a theory which treats the link principle as physical law? As far as I can see, such theories must be understood \emph{dualistically}: in addition to the nonlinearly evolving wavefunction (to be understood, perhaps, as a field on 3N-dimensional space; cf section \ref{DMWWbaretheory}) there is a 3-dimensional world of mass densities, or possibly of fuzzy-link-defined classical properties. The 3-dimensional world has no dynamical effect on the wavefunction, and conversely it is entirely determined by the wavefunction.\footnote{By contrast in hidden variable theories the hidden variables are fixed at best probabilistically by the wavefunction.} In philosophy-of-mind terms this is a \emph{property dualism}: the wavefunction has certain properties which are picked out by non-dynamical principles (the link principles in this case) as in some sense special (in philosophy of mind, for subserving conscious experience; in dynamical-collapse theories, for subserving macroscopic objects). On this ontology, the counting anomaly (though not, I think, the location anomaly) must  be taken seriously, for it entails that our property-ascription rule ascribes contradictory properties to a given system.

However, this property-dualist ontology is unattractive. For one thing, it is explicitly anti-functionalist (cf section \ref{DMWWeverettdecoherence}), since it requires that higher-level ontology supervene on only a primitively-selected subset of structural and dynamical properties of the world; for another, it effectively introduces new physical constants, such as $p$ in the case of the Fuzzy Link. Hence, in explicit \emph{discussions} of the status of the link principles an alternative view is more common: that the ontology of the theory consists of the wavefunction alone and the link principles are just perspicacious ways of picking out certain relevant properties of that wavefunction. \citeN{albertloewer1996} cash this out with reference to language use:
\begin{quote}
Our everyday language will supervene only vaguely (as it always has) on the micro-language of particle positions, and \ldots that language will itself supervene only vaguely \ldots on the fundamental language of physics.

And note (and this is important) that swallowing this additional vagueness will leave physics not one whit less of an empirical science than it has ever been. The fundamental language, the language of wavefunctions, the language of the stuff of which (on these theories) the world actually consists, is absolutely precise. \cite[p.\,90]{albertloewer1996}
\end{quote}
\citeN[p.\,716]{cliftonmonton1} and \citeN[p.\,168]{lewis2003} give similar accounts. Such accounts can be regarded as functionalist in spirit: the fundamental ontology is given by the wavefunction alone, and our higher-level talk supervenes on properties of that wavefunction picked out not \emph{a priori} (as would be the case if the link principles were fundamental) but by considerations of how our language describes these properties --- which means, ultimately, by considerations of the structural and dynamical function played by these 
properties. On this reading, the counting anomaly is  of no real import: it represents at most a failure of our linguistic conventions to operate as we might wish in a truly bizarre, and certainly never-reached-in-practice, physical situation.

However, if the link principles are to be understood in this way we will have come full circle, back to the \emph{original} problem of tails, which I called the problem of `structured tails' above. If regarded as \emph{fundamental} principles, both the Fuzzy Link and the Mass Density Link deal perfectly satisfactorily with states like
\be  \ket{\psi}=\sqrt{1-\epsilon^2}\ket{\mathrm{\mbox{live cat}}}+\epsilon\ket{\mathrm{\mbox{dead cat}}}:\ee
the Fuzzy Link says directly that the cat is alive because \ket{\psi} is very close to being an eigenstate of the `cat is alive' projector with eigenvalue 1; the Mass Density Link entails that the cat's cells are localised in the spatial regions corresponding to a living cat.

However, for all that its amplitude is tiny, the dead-cat term in the superposition is just as `real' as the live-cat term. (Recall: we are treating the wavefunction as \emph{physical}: the amplitude of a term in a superposition has nothing to do with the probability of that term, except indirectly via its role in the stochastic dynamics.)  As such, if the link principles are just a matter of descriptive convenience then what prevents us regarding observers as being just as present in the dead-cat term as in the live-cat term? After all, if we do accept functionalism then the dead-cat term is as rich in complex structure as the live-cat term.

Taken to its logical conclusion, this seems to suggest that the GRW theory with a functionalist reading of the link principles is just as much a `many-worlds' theory as is the Everett interpretation (a point  made by \cite{cordero}). But the matter has received rather little critical discussion, and it may well be that the problem is solvable either via identifying a conceptual error in the argument, or by modification of the GRW dynamics so as to suppress the structure in the low-weight branches.\footnote{The most straightforward way to make such a modification would be to replace the Gaussian used in the collapse process with a wave-packet of compact support (this does nothing to address the problem of \emph{bare} tails but does annihilate the structured tails). }

\subsection{Further Reading}

\citeN{bassighirardireview} provides a detailed review of the GRW and CSL theories; \citeN{ghirardiencyclopedia} is a briefer introduction. \citeN{lewis2005} is a good starting point for the Counting Anomaly.

There is a very different approach to the ontology of the GRW theory which has received some attention from physicists but gone almost unnoticed amongst philosophers: take the ontology to be just the collapse centres themselves, and treat the wavefunction as secondary. See \citeN{dowkerherbauts} for a technical development, and Allori \emph{et al}~\citeyear{goldsteinontology} for philosophical discussion.

Still another, and very different, approach, is the ``transactional interpretation'' developed primarily by J.\,G.\,Cramer (see, \egc,\citeNP{cramer86,cramer88}), in which (roughly speaking) the collapse propagates \emph{backwards in time} along the past light-cone of the particle. \citeN{pricebook} discusses and defends the conceptual consequences of such a theory.

One of the most exciting features of dynamical collapse is that it is in principle testable. \citeN{leggett} and \citeN{schlosshauer} consider (from very different perspectives) the prospect of empirical tests for the failure of unitarity.

\section{Hidden Variable Theories}\label{DMWWhidden}

Hidden variable theories take seriously the second half of Bell's dilemma: if QM is right, maybe it is not everything. The most famous such theory ---  the de Broglie-Bohm theory --- is now over fifty years old; other hidden variable theories are of more recent vintage (and in particular, the so-called ``modal interpretation'' is now generally recognised as a form of hidden-variable theory). Here I shall first explore some of the general requirements on hidden-variable theories, then discuss the de Broglie-Bohm theory and the modal interpretation, and then consider some of the open conceptual questions facing hidden-variable theories.

\subsection{Hidden variables for classical physics: a parable}\label{DMWWhvintro}

Suppose, fancifully, that for some reason we only had \emph{classical} mechanics in its statistical-mechanical form: as a theory concerning the evolution of some measure $\rho(\vctr{q},\vctr{p})$ over phase space, in accordance with the usual dynamical equation
\be\label{DMWWpoisson}
\dot{\rho}=\pb{\rho}{H}=\pbp{\rho}{q^i}\pbp{H}{p_i}-\pbp{\rho}{p_i}\pbp{H}{q^i};
\ee
suppose also that we have a Classical Algorithm for extracting empirical predictions from the theory, which tells us that the probability of getting a result in the vicinity of $(\vctr{q},\vctr{p})$ on making a phase-space measurement of the system is proportional to $\rho(\vctr{q},\vctr{p})$. 

If we were asked to provide an interpretation of this theory, and we were having an off day, we might well begin by taking $\rho$ as a physical entity, evolving in a $2N$-dimensional space; we might further worry about how such a highly delocalised entity can correspond to our experiences of systems having definite positions and momenta; we might even toy with modifying (\ref{DMWWpoisson}) in some nonlinear and stochastic way so as to concentrate $\rho$ periodically on much smaller regions of phase space.

Of course, we would be missing the point. There is a \emph{hidden variable} theory for classical statistical mechanics. These ``hidden variables'' are the positions and momenta of a swarm of pointlike particles, obeying the dynamical equation
\be\label{DMWWclassmech}
\ddt(\vctr{q},\vctr{p})=\left(\pbp{H}{\vctr{p}},-\pbp{H}{\vctr{q}}\right).
\ee
$\rho$ is not a physical entity at all: it is a probability distribution, summarising our ignorance of the actual values of the hidden variables, and its ``dynamical equation'' (\ref{DMWWpoisson}) is just the result of pushing that distribution forwards through time via the real dynamical equation (\ref{DMWWclassmech}). If we actually know the values of the hidden variables we can dispense with $\rho$ altogether; it is only because in practice we do \emph{not} know them (hence ``hidden'') that in practice we often fall back on using $\rho$.

The \emph{original} hope for hidden-variable theories was that they would work in just this way. The quantum state, after all, does serve to generate probability distributions (or equivalently, expectation values) over possible results of measurements, 
evolving in time via
\be\label{DMWWquantumexpect}
\ddt\langle\op{X}\rangle=\langle\comm{\op{X}}{\op{H}}\rangle
\ee
and (notwithstanding the criticisms of section \ref{DMWWagainst}) such measurements are traditionally associated with possessed quantities. So the hope was that actual quantum systems have ``hidden'' determinate values of each quantity, that the quantum state is just a shorthand way of expressing a probability distribution over the various values of each of those quantities, and that some underlying (stochastic or deterministic) law for the hidden variables generates (\ref{DMWWquantumexpect}) just as (\ref{DMWWclassmech}) generates (\ref{DMWWpoisson}). Let us call this an \emph{eliminativist} hidden-variable strategy: `eliminativist' because it seeks to eliminate the wavefunction entirely from the formalism, and recover it only as a probabilistic average over hidden variables.

Half a century of work has made it clear that any eliminativist hidden-variable theory must possess some highly undesirable features. Nonlocality is the least of these: Bell's work \cite{bellsocks} shows that hidden variable theories must be nonlocal, but it is fairly generally accepted (\citeNP[98--106]{redheadbook},\citeNP{maudlinbook}) that these conclusions apply equally to dynamical-collapse theories --- so if we want to be realists and to avoid the Everett interpretation, nonlocality is probably unavoidable. More seriously, the Bell-Kochen-Specker (BKS) theorem (\cite{bell1966,kochenspecker}; see \citeN[pp.\,118--138]{redheadbook} or \citeN[pp.\,187--212]{peres} for a discussion)  tells us that any hidden-variable theory which assigns values to all properties represented by projectors must be \emph{contextual}. That is: whether or not a system is found, on measurement, to possess a given property must depend on what \emph{other} properties are measured simultaneously. Contextuality seems well-nigh inconsistent with the idea that systems determinately do or do not possess given properties and that measurements simply determine whether or not they do.

In the light of the BKS theorem, there seem to be four strategies for hidden-variable theorists to follow. 
\begin{enumerate}
\item Construct contextual hidden-variable theories, and try to come to terms with the contextuality. This strategy does not seem to have been widely explored, presumably because failure of contextuality is so pathological (\citeN{spekkens}  is an interesting exception.)
\item Maintain the idea that the quantum state is just shorthand for a set of probability distributions, but abandon the idea that any underlying microdynamics can be found. This allows us to say, for instance, that $40\%$ of particles had position $x$ and that $35\%$ had momentum $y$, but forbids us to say anything about the correlations between the two or to refer to the position or momentum of any particular particle. Such interpretations (which are \emph{pure} interpretations in the sense of section \ref{DMWWinterpretation}) are often called \emph{ensemble} interpretations, and have been defended by, \egc, Ballentine (\citeyearNP{ballentinereview,ballentine}) and \citeN{taylorghost}. It seems fairly clear that these interpretations are essentially variants of the `operationalist interpretation' of section \ref{DMWWoperationalist}, using the ensemble just to make conceptual sense of the probabilities. I am less clear, however, whether the proponents of the ensemble interpretation would accept this.
\item Abandon classical logic, on which the BKS theorem relies. Certain versions of quantum-logic interpretations can perhaps best be understood as eliminativist hidden-variable theories built on a non-classical logic: in particular, this description seems a good fit to the quantum-logic interpretation discussed in \citeN{dicksonlogic}. The requirement for a dynamics then translates into the requirement for a logic of multi-time propositions (such as ``the particle is currently at location $x$ $\wedge$ in $t$ seconds it will be at location $y$''). In general, quantum-logic approaches lie beyond the scope of this review so I will say nothing further here about them.
\item Abandon the idea that the hidden variables must include determinate values (or values at all) for \emph{all} properties. If they only have values of, say, position, then the BKS theorem does not apply, since it relies on the existence of sets of properties not all of whose associated projectors commute. 
\end{enumerate}

It is the fourth strategy which is adopted by the vast majority of hidden-variable theories currently discussed in philosophy of QM. In fact, the restriction in practice has to be severe: if one fixes a single non-degenerate observable \op{X} and requires that the hidden variables have definite values of \op{X}, the only other values which they can possess are values of functions of \op{X}.\footnote{This statement is only true for generic choices of state vector (those which overlap with all eigenspaces of \op{X}). A very elegant theorem of Bub and Clifton (\citeNP{bubclifton}, \citeNP{bubcliftongoldstein}; see also \citeNP[chapter 4]{bubbook}) places more precise restrictions on exactly which properties can be included in a hidden-variable theory before non-contextuality fails..}

What is perhaps less obvious  is that the fourth strategy sharply constrains \emph{eliminativist} hidden-variable theories of all but the most trivial kind, for it is nearly impossible to establish empirically adequate dynamics for such theories. This follows from the fact that probability distribution over the hidden variables now badly underdetermines the quantum state. 

For instance, suppose that the hidden variables are associated with some operator $\op{X}$ (with eigenstates $\{\ket{x}\}$ satisfying $\op{X}\ket{x}=x\ket{x}$). If the probability of the hidden variables having value $x$ is R(x) then we know the state vector must have form
\be
\ket{\psi}=\sum_x R(x)\exp(i \theta(x))\ket{x},
\ee
but we have no information about the phase factors $\theta(x)$. Since these phases affect the  dynamics of \ket{\psi}, they affect the future probability distribution of the hidden variables --- hence, their current distribution underdetermines their future distribution.

In the light of this, it is unsurprising to observe that essentially all hidden-variable theories currently studied are what might be called \emph{dualist hidden-variable theories}: their formalism contains not only the hidden variables, but the quantum state, with the latter playing a dynamical role in determining the evolution of the former.\footnote{Nelson's theory \cite{nelsonarticle,nelsonbook} is a partial counterexample: it includes hidden variables with definite positions and an additional field which encodes the phase information about the wavefunction, but does not include the wavefunction itself. Still, it remains very far from eliminativist.} The remainder of this section will be concerned with such theories.

\subsection{General constraints on hidden-variable theories}\label{DMWWhvconstraints}

To be precise, formally such theories are specified by giving, for whatever system is under study,
\begin{enumerate}
\item The quantum state \ket{\psi(t)} (evolving unitarily via the Schr\"{o}dinger equation).
\item Some preferred set of observables $\op{X}_i$ (chosen sufficiently small that no Bell-Kochen-Specker paradox arises; in practice this means that the $\op{X}_i$ must all commute with one another).\footnote{\citeN{buschmodal} considers the possibility of using POVMs rather than ``old-fashioned'' observables.}
\item The actual possessed values $x_i$ of these observables. 
\item A dynamical equation for the $x_i$, which we might write schematically as
\be \label{DMWWhvdyn}\ddt x_i(t)= \mc{F}_i(x_1(t),\ldots,x_N(t);\ket{\psi(t)})
\ee
and which may be deterministic or stochastic.
\end{enumerate} 
It is then possible to speak of the `determinate properties' for such a theory: namely, all the properties whose defining projectors  are eigenprojectors of the $\op{X}_i$ (and therefore whose value is fixed by the values of the $x_i$).

The idea of such theories is that the observable world is in some sense represented by the hidden variables, rather than (or at least: as well as) the state vector.  (As such, in dualistic hidden-variable theories it is actually rather odd to call the variables ``hidden'':  if anything it is the state vector that is hidden. \citeN{bellqmforcosmologists} suggests that we might do better to refer to them as the exposed variables!)

As we shall see, there are a variety of ways to cash this out. However, it is possible to place some empirical constraints on these theories, for to resolve the measurement problem they must allow us to reproduce the Quantum Algorithm. Here it will be useful again to adopt the decoherent histories formalism: quasi-classical histories can be identified with certain sequences of projectors $\op{P}_{i_k}(t_k)$; instantaneous quasi-classical \emph{states} can be identified with single such projectors. From this we derive
\begin{quote}\textbf{First constraint on a hidden-variable theory:} There must exist a quasi-classical decoherent history space, fine-grained enough that any two empirically distinguishable histories of the world correspond to distinct histories in the space, and such that any projector $\op{P}_{i_k}(t_k)$ in that space is determinate at time $t_k$.
\end{quote}
This first constraint guarantees that if we know the values of the hidden variables, we know which macroscopic state is represented by the theory. We need to go further, however: the Quantum Algorithm is probabilistic in nature, and those probabilities must be represented somewhere in the hidden variable theory. This leads to the
\begin{quote}
\textbf{Second constraint on a hidden-variable theory (weak version):} If $\op{P}$ is a time-$t$ projector from the decoherent history space of the first constraint, then the probability of the hidden variables being such that  $\op{P}$ is determinately possessed by the system at time $t$, conditional on the universal state at time $t$ being \ket{\psi}, must be $\matel{\psi}{\op{P}}{\psi}$.
\end{quote}
This guarantees the empirical accuracy of the Born rule for macroscopic states at a given time. In view of the usual interpretation of the `hidden variables' giving the actual value of the preferred quantities, it is normal to require a stronger condition:
\begin{quote}
\textbf{Second constraint on a hidden-variable theory (strong version):} If $\op{P}$ is \emph{any} projector whose value is determinate at time $t$ then the probability of the hidden variables being such that  $\op{P}$ is determinately possessed by the system at time $t$, conditional on the universal state at time $t$ being \ket{\psi}, must be $\matel{\psi}{\op{P}}{\psi}$.
\end{quote}

Now, the second constraint (in either form) requires us to place a certain probability distribution over hidden variables at time $t$. The dynamical equation for the variables will then determine a probability distribution over them at all other times; if that probability distribution is not the one required by the second constraint, we will have a contradiction. This gives  our
\begin{quote}
\textbf{Third constraint on a hidden-variable theory:} if the second constraint is satisfied at one time, the dynamics are such that it is satisfied at all other times. (This constraint on the dynamics is sometimes called \emph{equivariance}.)
\end{quote}
It might seem that this list of constraints is sufficient to recover the Quantum Algorithm; not so. For that Algorithm involves  ``collapse of the wavefunction'': it permits us to interpret macroscopic superpositions as probabilistic in nature, and so to discard all terms in the superposition except that corresponding to the actually-observed world. In hidden-variable theories, though, the state continues to evolve unitarily at all times and no such collapse occurs --- in principle, all terms in the superposition, not just the one corresponding to the quasi-classical world picked out by the hidden variables, can affect the evolution of the hidden variables via the dynamical equation. To prevent a clash with the Quantum Algorithm, we need to rule this out explicitly:
\begin{quote}
\textbf{Fourth constraint on a hidden-variable theory:} if \op{P} is a time-$t$ projector in the decoherent history space of the first constraint, and if the hidden variables are such that the property corresponding to \op{P} is possessed by the system at time $t$, then the dynamics of the hidden variables after $t$ are the same whether we take as universal state \ket{\psi} or $\op{P}\ket{\psi}/\|\op{P}\ket{\psi}\|$.; in terms of the notation of equation (\ref{DMWWhvdyn}), we are requiring that
\be
\mc{F}(x_1, \ldots x_N; \ket{\psi})=
\mc{F}(x_1, \ldots x_N; \op{P}\ket{\psi}/\|\op{P}\ket{\psi}\|).
\ee 
\end{quote}
This assumption might be called \emph{locality}: the hidden variables are affected only by ``their'' branch of the state vector. 

If a hidden-variable theory satisfies these four constraints, then it will solve the measurement problem ---\emph{provided} that we can actually understand the mathematical formalism of the theory in such a way as to justify the claim that it represents the single approximately-classical world picked out by  the hidden variables. As such, the problem is two-fold: to construct such theories in the first place, and then to interpret them. 

\subsection{Specific Theories I: Modal interpretations}

In trying to construct actual hidden-variable theories which conform to these four constraints, we once again run up against the inherent approximateness of decoherence. If there existed a clean, precise statement of what the decoherent histories actually were, we could just \emph{stipulate} that the preferred quantities were precisely the projectors in the decoherent histories. (Effectively, of course, this would be to return us to something like the ``solution that isn't'' of section \ref{DMWWsolutionthatisnt}) But since decoherence is imprecisely defined and  approximately satisfied, this strategy is not available. 

One way around this problem is to define some cleanly-stateable, state-dependent rule which approximately picks out the decoherence-preferred quantities. A concrete way to do this was  developed by Kochen, Healey, Dieks and others, following a proposal of \citeN{vanfraassenquantum}; theories of this form are normally called \emph{modal interpretations}, although the term is used in a variety of conflicting ways by different authors. 

Modal interpretations assign definite properties to \emph{subsystems} of a given isolated system. They do so via the so-called ``Schmidt decomposition'' of the quantum state: given a Hilbert space $\mc{H}=\mc{H}_A\otimes \mc{H}_B$ and a state \ket{\psi} in that space, it is always possible to find orthonormal bases $\{\ket{A_i}\}$, $\{\ket{B_j}\}$ for $\mc{H}_A$ and $\mc{H}_B$ such that 
\be 
\ket{\psi}=\sum_i \lambda_i \tpk{A_i}{B_i};
\ee
furthermore, generically (\iec, except when two of the $\lambda_i$ are equal)  this decomposition is unique. Modal interpretations take the projectors $\proj{A_i}{A_i}$ to define the preferred properties of the subsystem represented by $\mc{H}_A$. 

Notice: the reduced state $\denop_A$ for $\mc{H}_A$ is 
\be \denop_A=\sum_i |\lambda_i|^2 \proj{A_i}{A_i}.;\ee
hence, the basis $\{\ket{A_i}\}$ diagonalises the reduced state. If the system described by $\mc{H}_A$ is macroscopic, and decohered by its environment, then the decoherence basis will approximately diagonalise $\denop_A$; hence, it is at least plausible that the modal interpretation's preferred quantity is close to being decoherence-preferred. If we add an equivariant dynamics, it appears that we have a hidden-variable interpretation which satisfies the four constraints, and thus is a candidate for solving the measurement problem. And in fact it has proved possible to construct such dynamics: see \citeN{bacciagaluppidynamics} for a discussion.

I should note that my presentation here differs significantly from the actual historical development of modal interpretations. Originally,  the Schmidt decomposition rule was proposed in order that quantum \emph{measurements} should actually succeed in measuring the right quantity: if \op{X} is some observable for $\mc{H}_A$ (with eigenstates $\ket{x_i}$)and $\mc{H}_B$ is the Hilbert space of a measurement device intended to measure \op{X}, then we might model the measurement process by
\be \tpk{x_i}{\mathrm{\mbox{ready}}}\longrightarrow \tpk{x_i}{\mathrm{\mbox{measure }}x_i},
\ee
in which case measurement on a superposition gives
\be
\sum_i \lambda_i \ket{x_i} \otimes \ket{\mathrm{\mbox{ready}}}
\longrightarrow
\sum_i \lambda_i \tpk{x_i}{\mathrm{\mbox{measure }}x_i}.
\ee
Plainly, the Schmidt decomposition applied to the post-measurement state gives $\op{X}$ as a determinate observable for $\mc{H}_A$ and 
\be
\sum_i x_i \proj{\mathrm{\mbox{measure }}x_i}{\mathrm{\mbox{measure }}x_i}
\ee
(the ``\op{X}-measurement observable'') as a determinate observable for $\mc{H}_B$; if the distribution of hidden variable values satisfies the Second Constraint then the system being measured will have value $x_i$ for \op{X} if and only if the measurement device determinately possesses the property of having measured $x_i$. 

However, most measurements are not `ideal' in this sense; many, for instance  totally destroy the system being measured, which we might represent as
\be
\tpk{x_i}{\mathrm{\mbox{ready}}}\longrightarrow \tpk{x_0}{\mathrm{\mbox{measure }}x_i},
\ee
(in optics, for instance, \ket{x_0} might be a no-photon state). It follows that the definite property for $\mc{H}_B$ in this case is the property of being in the superposition 
\be\label{DMWWmodal}
\sum_i \lambda_i\ket{\mathrm{\mbox{measure }}x_i},
\ee
which is decidedly non-classical. It was recognised by \citeN{bacciagaluppihemmo} that this problem is resolved when decoherence is taken into account, since the state (\ref{DMWWmodal}), once the environment is considered, rapidly decoheres into an entangled state like
\be
\sum_i \lambda_i\tpk{\mathrm{\mbox{measure }}x_i}{\epsilon_i},
\ee
with $\bk{\epsilon_i}{\epsilon_j}\simeq \delta_{ij}$, and so it is plausible that the definite-measurement-outcome observable is again  determinate.

Modal interpretations, however, have run into some serious difficulties, both conceptual and technical.  For one thing, it is very unclear what the ontology of the theory is intended to be. Even in the presence of decoherence, the properties picked out by the modal-interpretation rule will not \emph{exactly} be properties like position and momentum; it will determinately be true not that a system is localised in region $R$, but only that it is approximately localised in region $R$. And it has been argued (initially by \citeN{albertloewer1990}; see also \citeN{reutsche}) that this is insufficient to solve the measurement problem: superpositions with very uneven amplitudes are still superpositions.

My own view is that this should not be taken very seriously as an objection, however. It stems once again from the unmotivated assumption that classical properties have to be associated with projectors, but there is an alternative strategy available: if $\proj{A_i}{A_i}$ corresponds to the maximally fine-grained property that a system is supposed to have, just take the ontology of the theory to be the state vector (the wavefunction, if you like) \ket{A_i}, understood as a physical entity. If that physical entity is not an eigenstate of any particularly easily-described operator, so be it. (The parallel to my objections to the Fuzzy Link in the GRW theory and to the Bare Theory should be plain; see also \citeN{arntzeniusmodal} for a somewhat related response to this objection.)

More serious conceptual threats to modal interpretations arise when we consider composite systems. For one thing, the modal interpretation appears to require a preferred decomposition of the universal Hilbert space into subsystems: as has been shown by \citeN{bacciagaluppi1995}, if we apply the modal rule to all decompositions, it leads to contradiction. But even if we suppose that we are given some such preferred decomposition, trouble looms. Suppose we have a system with \emph{three} subsystems, with Hilbert space $\mc{H}=\mc{H}_A\otimes\mc{H}_B\otimes\mc{H}_C$ and state \ket{\psi}. We can determine the determinate properties of $\mc{H}_A$ in two ways: by applying the Schmidt decomposition rule to $\mc{H}_A$ directly as a subsystem of \mc{H}, or by applying it first to $\mc{H}_A \otimes \mc{H}_B$ as a subsystem of \mc{H} and then to $\mc{H}_A$ as a subsystem of $\mc{H}_A \otimes \mc{H}_B$.

However, the Schmidt decomposition theorem does not generalise: there is no guarantee that \ket{\psi} can be decomposed like
\be\label{DMWWtridecomposition}
\ket{\psi}=\sum_i \lambda_i \ket{A_i}\otimes \ket{B_i}\otimes \ket{C_i}
\ee
with $\bk{A_i}{A_j}=\bk{B_i}{B_j}=\bk{C_i}{C_j}=\delta_{ij}$.
This means, in turn, that there is no guarantee that the two methods for finding the determinate properties of $\mc{H}_A$ will give the same --- or even approximately the same --- answer. This \emph{perspectivalism} of properties was noted by \citeN{arntzenius1990} (see also \citeN{arntzeniusmodal}, \citeN{cliftonmodal96}).

Two technical results alleviate the problem. Firstly, it can easily be proved that if a property determinately \emph{is} possessed by a system from one perspective, it will not be found determinately \emph{not} to be possessed from any perspective. Secondly, in the presence of decoherence decompositions like (\ref{DMWWtridecomposition}) turn out to be \emph{approximately} true. But it is far from clear that this is sufficient: non-perspectivalism has the feel of a conceptual necessity, not just an approximately-true empirical fact. (What does it even \emph{mean} to say that some system has the property that its subsystem has property $p$, if not that the subsystem just has that property?)

There are basically two strategies to deal with perspectivalism: either accept it as a counter-intuitive but not formally contradictory property of QM (which will lead to some sort of  non-classical logic), or adopt an `atomic' theory according to which the modal rule applies only to some preferred decomposition of the system into atomic subsystems, and where the properties of compound systems are by definition given by the properties of their atomic components (as suggested by \citeN{bacciagaluppidickson}). See \citeN{vermaas} for a comparison of the two strategies; note also the close similarity between these issues and the counting anomaly discussed in section \ref{DMWWcounting}.

So much for the conceptual objections. A potentially fatal \emph{technical} objection exists: in realistic models, it seems that decoherence does not after all guarantee that the determinate properties pick out quasi-classical histories. For just because a given basis \emph{approximately} diagonalises the density operator of a system, there is no guarantee that \emph{approximately} that same basis \emph{exactly} diagonalises it. In fact, where the density operator is approximately degenerate, the exactly-diagonalising basis (\iec, the one picked out by the modal rule) can be wildly different from the approximately-diagonalising basis (\iec, the quasiclassical one picked out by decoherence). See \citeN{bacciagaluppidonaldvermaas}, \citeN{donaldmodal}, and \citeN{bacciagaluppi2000} for further details. It is unclear whether the modal interpretation remains viable at all in the face of this problem.

\subsection{Specific Theories II: The de Broglie-Bohm theory}

In view of the difficulties faced by the modal interpretation, it has fallen somewhat from favour. An older strategy for constructing hidden-variable theories remains popular, though (at least amongst philosophers of physics): rather than trying to approximate the decoherence rule itself, just pick an observable which \emph{in fact} is approximately decoherence-preferred. The obvious choice is position: we take as hidden variables, for an $N$-particle system, the positions $\vctr{q}_1, \cdots \vctr{q}_N$ of each of the $N$ particles, and represent our ignorance of their precise values via the probability measure
\be
\Pr(\langle \vctr{q}_1, \cdots \vctr{q}_N\rangle\in \mc{S})=\int_\mc{S}\dr{\vctr{q}}_1 \cdots \dr{\vctr{q}}_N |\psi(\vctr{q}_1, \cdots \vctr{q}_N)|^2.
\ee
 We then specify an equivariant, configuration-space-local differential equation for the evolution of the $\vctr{q}_1, \cdots \vctr{q}_N$. There are a number of available choices, but the standard choice is the \emph{guidance equation}:
\be
\dbd{\vctr{q}_i}{t}=\frac{1}{m}\mathrm{Im} \frac{\vctr{\nabla}_i \psi (\vctr{q}_1, \cdots \vctr{q}_N)}{\psi(\vctr{q}_1, \cdots \vctr{q}_N)}.
\ee
The resulting theory is known as the de Broglie-Bohm theory (DBB), or sometimes as Bohmian mechanics or the ``Pilot-wave theory''; its first clear and widely accessible statement was \citeN{bohm}.

DBB has a number of conceptual advantages over modal interpretations. The ontology of its hidden variables seems fairly clear: they are simply point particles, interacting with one another via a spatially nonlocal but configuration-space-local dynamical equation, and following well-defined trajectories in space (following \citeN{brownlockwood}, and to distinguish the hidden variables from other senses of `particle', I will refer to these point particles as \emph{corpuscles}). Manifestly, there is no perspectivalism in DBB: the hidden variables assigned to any set of $M$ 1-particle systems are just the $M$ corpuscles associated with those systems.\footnote{Note here that the theory has no problem with identical particles: the corpuscles of a system of identical particles can be identified with a point in the reduced configuration space, so that the corpuscles too are identical. In fact, this framework gives some insight into the origin of fermionic and bosonic statistics \cite{brownstatistics}.}

On the technical side, for DBB to solve the measurement problem we require that the corpuscles track the decohered macroscopic degrees of freedom of large systems. It is at least plausible that they do: coarse-grainings of the position components of the macroscopic degrees of freedom generate an approximately decoherent space of histories, and since the guidance equation is configuration-space-local, it seems that Constraints 1--4 are all satisfied. However, the matter is somewhat more complicated: as I noted in section \ref{DMWWhistoryframework}, the decoherent-history framework is not a perfect fit to the actual phenomenology of decoherence in macroscopic systems, where the ``pointer basis'' is overcomplete and where decoherence between distinct states in the pointer basis is not an all-or-nothing matter. Further plausibility arguments have been constructed (\eg \citeN[section 4]{bellqmforcosmologists}, \citeN[pp.\,336--50]{holland}, \citeN[pp.\,39--41]{durrincushing}, and some simple models have been studied; at present, it seems likely that the corpuscles do track the quasiclassical trajectories sufficiently well for DBB to solve the measurement problem, but there exists no full proof of this.

(As with the modal interpretation, my route to deriving and justifying DBB --- choose position as the hidden variable because it is approximately decohered and write down local dynamical equations for it --- is not exactly that seen in the literature. The more common justification of why it suffices to take  position as preferred is that ``the only observations we must consider are position observations'' \cite{bellimpossible}, and hence if we wish to reproduce the predictions of QM for all measurements it is sufficient to reproduce them for position measurements. The equivalent slogan for my approach would be ``the only macroscopic superpositions we must consider are superpositions of states with different positions''.)

DBB is substantially the most studied, and the most popular, extant hidden-variable theory. While the remainder of this section will consider general conceptual problems with hidden-variable theories, DBB will normally be used when an example is required, and in fact much of the literature study on these issues has dealt specifically with DBB.

\subsection{Underdetermination of the theory}

At least at the level of the mathematics, hidden-variable theories reproduce the predictions of QM by picking out a particular decoherent history as actual. It follows that the details of such theories are underdetermined by the empirical data, in two ways:
\begin{enumerate}
\item It is impossible to determine the values of the hidden variables more precisely than by determining which decoherent history they pick out. (For instance, it is widely recognised that the location of the de Broglie-Bohm corpuscles cannot be fixed more precisely than by determining which wave-packet they lie in.) This means that the details of what the hidden-variable observables \emph{are} cannot be determined empirically, even in principle; many different choices would lead to exactly the same macroscopic phenomenology. (In this sense, if no other, the hidden variables are indeed `hidden'!)
\item Similarly, any two hidden-variable dynamical equations that reproduce the macroscopic dynamics --- that is, that are equivariant in the sense of Constraint 3 and local in the sense of Constraint 4 --- will be empirically indistinguishable even in principle.  
\end{enumerate}

We have seen that both forms of underdetermination loom large for the modal interpretations, in the controversy over perspectivalism and in the  wide variety of different dynamical schemes which have been suggested. In DBB there is a stronger consensus that the guidance equation gives the true dynamics\footnote{There is a certain amount of disagreement as to the correct \emph{form} of the guidance equation. Bohm originally proposed a second-order form of the equation, with the wavefunction's action on the corpuscles represented by the so-called `quantum potential'; \citeN{bohmhiley} continue to prefer this formalism. It is currently more common  (mostly following Bell's expositions of the theory) to use the first-order form. However, both versions of the equation ultimately generate the same dynamics for the corpuscles.} but many variants are possible, both deterministic (\citeNP{deottoghirardi}) and stochastic  (\citeNP{bacciagaluppinelson}). There is also 
some disagreement over the hidden variables once we try to generalise the theory from spinless non-relativistic particles. In the case of particles with spin, some  wish to include spin as a further hidden variable; others  prefer a more minimalist version of the theory where corpuscles have only position. In the relativistic case, for various reasons it has been proposed that only fermions \cite{bellbeablesqft} or only bosons \cite{westman} have associated corpuscles.

\citeN[fn.\,16]{bassighirardireview} argue that underdetermination is more serious for hidden-variable theories than for dynamical-collapse theories. There are, to be sure, a great many dynamical-collapse theories compatible with the observed success of QM (consider, for instance, all the variants on the GRW theory produced by tweaking the collapse rate and the width of the post-collapse wavefunction). But these theories are empirically distinguishable, at least in principle. By contrast, no empirical test at all can distinguish different hidden-variable theories: eventually, the results of any such test would have to be recorded macroscopically, and equivariance guarantees that the probability of any given macroscopic record being obtained from a given experiment is determinable from the wavefunction alone, independent of the details of the hidden-variable theory.\footnote{At least, this is so insofar as the hidden-variable theory reproduces quantum probabilities; cf. section \ref{DMWWHVTprob}.}

As such, supporters of hidden-variable theories (and in particular of DBB) have generally tried to advance \emph{non}-empirical reasons to prefer one formulation or another --- for instance, D\"{u}rr \emph{et al}~(\citeyearNP{durr1992}, pp.\,852--854) argue that the guidance equation is the simplest equivariant, configuration-space-local and Gallilean-covariant dynamical equation. In discussing which particles actually have corpuscles, Goldstein \emph{et al}~\citeyear{goldstein05} appeal to general considerations of scepticism to argue that we should associate corpuscles to all species of particle. It is interesting at any rate to notice that if DBB really is a viable solution to the measurement problem, then it appears to offer abundant examples of underdetermination of theory by data (\citeNP[pp.\,40--3]{NSunderdetermination}; \citeNP[pp.\,162--182]{psillosbook}).

\subsection{The origin of the probability rule}\label{DMWWHVTprob}

Probability enters hidden-variable theories in much the same way is it enters classical statistical mechanics: the actual hidden variables have determinate but unknown values; the probability distribution represents our ignorance of that value. As in statistical mechanics, so in hidden-variable theories, we can ask: what justifies the \emph{particular} probability distribution which we assume in order to make empirical predictions? 

In both cases, we postulate a probability measure over possible values of the dynamical variables; in both cases, there are both conceptual and technical questions about that measure. And in fact, some of the same strategies are seen in both cases. (Most, but not all, of the literature discussion has been in the context of DBB.)

In particular, \citeN[pp.\,36--41]{durrincushing} propose an essentially cosmological approach to the problem: they suggest that we adopt as a postulate that the probability of the  de Broglie-Bohm corpuscles having initial locations in the vicinity $(q^1, \ldots q^N)$ is just (proportional to) $|\Psi(q^1,\ldots q^N)|^2$, where $|\Psi|^2$ is the initial wavefunction of the Universe. They argue that this is a sort of ``typicality'' assumption, tantamount to assuming that the actual corpuscle distribution is typical (in the $|\Psi|^2$ measure)  amongst the set of all \emph{possible} corpuscle distributions. (There is an obvious generalisation to other hidden variable theories.) They acknowledge that there is an apparent \emph{non sequitur} here, since distributions typical with respect to one measure may be highly atypical with respect to another, but they argue that the equivariance of the $|\Psi|^2$ measure makes it a natural choice (just as, in classical statistical mechanics, it is common to argue that the equivariance of the Liouville measure makes \emph{it} natural). For a defence of this `typicality' framework in the general statistical-mechanical context, see \citeN{goldsteinstatmech}; for a critique, see \citeN[p.\,122--3]{dicksonbook}.

A non-cosmological alternative exists: rather than postulate that the initial-state probability distribution was the $|\Psi|^2$ distribution, we could postulate that it was some \emph{other} distribution, and try to show that it evolves into the $|\Psi|^2$ distribution reasonably quickly. This proposal has been developed primarily by Antony Valentini (\citeNP{valentinicushing}, \citeNP{valentini01}, \citeNP{valentiniwestman}); again, he draws analogies with statistical mechanics, calling the $|\Psi|^2$ distribution ``quantum equilibrium'' and aiming to prove that arbitrary probability distributions ``converge to equilibrium''.

Whether this convergence occurs in practice obviously depends on the dynamics of the hidden-variable theory in question: it is not \emph{prima facie}  obvious that an arbitrary hidden-variable dynamics (or even one which satisfies the Third and Fourth constraints) would have this property. However, there is fairly good evidence that in fact the convergence does occur for at least a wide class of hidden-variable theories. \citeN[p.208]{bacciagaluppidynamics} claims that convergence occurs for certain natural dynamics definable within the modal interpretation, while Valentini (ibid.) proves what he calls a ``quantum H-theorem''  which establishes convergence in DBB. (As with Boltzmann's own `H-theorem', though, Valentini's result must be taken as at most indicative of convergence to equilibrium.)

One fascinating corollary of these dynamical strategies is that the Universe --- or at least some subsystems of it --- might not be in ``quantum equilibrium'' at all. This would create \emph{observable} violations of the predictions of QM, and might provide a context in which hidden-variable theories could actually be tested \cite{valentini01,valentiniblackhole}.

\subsection{The ontology of the state vector}\label{DMWWhvontology}

Perhaps the most serious conceptual problem with hidden-variable theories is their dualistic nature. Recall that in eliminativist hidden-variable theories it is unproblematically true that the hidden variables alone represent physical reality --- the state vector is a mere book-keeping device, a shorthand way of expressing our ignorance about the values of the hidden variables.  But as we saw in section \ref{DMWWhvintro}, in actual hidden-variable theories the state vector plays an essential part in the formalism. Until we can say what it is, then, we do not have a truly satisfactory understanding of hidden-variable theories, nor a solution of the measurement problem. To adapt the terminology of section \ref{DMWWinterpretation}, a technically satisfactory hidden-variable theory gives us a `bare hidden-variable formalism', and a `Hidden-variable Algorithm' whose predictions agree with those of the Quantum Algorithm, but what we need is a `pure interpretation' of the bare hidden-variable formalism from which we can derive the Hidden-variable algorithm. And we cannot achieve this unless we are in a position to comment on the ontological status of the state vector.

This question has been addressed in the literature in two ways, which we might call ``bottom-up''  and ``top-down''.  Bottom-up approaches look at particular properties of quantum systems. For instance, in DBB we might want to ask whether spin, or charge, or mass, should be considered as a property of the corpuscle, of the wave-packet, or of both. 

Much of the bottom-up literature has drawn conclusions from particular (thought- or actual) experiments: in particular, the so-called ``fooled detector'' experiments \cite{englertetal,dewdneyetal,brownfooleddetectors,aharonovvaidman,hileyfooleddetectors} suggest that not all ``position'' measurements actually measure corpuscle position. However, I shall focus here on the ``top-down'' approaches, which consider and criticise general theses about the ontological status of the wavefunction. The most obvious such thesis is \emph{wavefunction realism}: the state vector is a physical entity, which interacts with the hidden variables. As was noted in section \ref{DMWWbaretheory}, this was Bell's position; other advocates include \citeN{albertmetaphysics} and Valentini.

One problem with wavefunction realism is that it violates the so-called ``action-reaction principle'': the wavefunction acts on the hidden variables without being acted on in turn. Opinions and intuitions differ on the significance of this: \citeN{anandanbrown} regard it as an extremely serious flaw in the theory; at the other extreme, \citeN[section 11]{durr95} suggest that the  principle is just a generalisation of Newton's 3rd Law, and so is inapplicable to a theory whose dynamical equations are first order.

A potentially more serious flaw arises from the so-called ``Everett-in-denial''\footnote{``[P]ilot-wave theories are parallel-universe theories in a state of chronic denial'' \cite[p.\,225]{deutschlockwood}.} objection to realism (\citeNP{deutschlockwood,zehbohm,brownwallace}). This objection begins with the observation that if the wavefunction is physical, and evolves unitarily, then the only difference between the Everett interpretation and the hidden-variable interpretation under consideration is the additional presence of the hidden-variables. (Note that we are concerned with the `pure-interpretation' form of the Everett interpretation considered in sections \ref{DMWWbaretheory}--\ref{DMWWprobabilityquantitative}, not with the Many-Exact-Worlds or Many-Minds theories discussed in section \ref{DMWWmodification}.) Advocates of the Everett interpretation claim that, (given functionalism) the decoherence-defined quasiclassical histories in the unitarily evolving physically real wavefunction describe --- \emph{are} --- a multiplicity of almost-identical quasiclassical worlds; if that same unitarily-evolving physically real wavefunction is present in DBB (or any other hidden-variable theory) then so is that multiplicity of physically real worlds, and all the hidden variables do is point superfluously at one of them. 

So far as I can see, hidden-variable theorists have two possible responses to the Everett-in-denial objection (other than denying wavefunction realism). Firstly, they can reject functionalism (\citeN{brownlockwood} implicitly makes this recommendation to Bohmians, when he argues that the ultimate role of the de Broglie-Bohm corpuscles is to act as a supervenience base for consciousness). Secondly, they can accept functionalism as a general metaphysical principle but argue  (\emph{contra} the arguments presented in section \ref{DMWWeverettdecoherence}) that  it does not entail that a unitarily-evolving wavefunction subserves a multiplicity of quasi-classical worlds. 

Either response, interestingly, is metaphysically \emph{a priori}: functionalism, if true at all, is an \emph{a priori} truth, and it follows that it is an \emph{a priori} question whether a unitarily evolving wavefunction does or does not subserve multiple worlds. At the least, the Everett-in-denial objection seems to support the claim that it is not a \emph{contingent} fact which of the Everett and hidden-variables strategies is correct.

Wavefunction realism is not the only --- perhaps even not the most popular --- interpretation of the state vector within hidden-variable theories. Space does not permit a detailed discussion, but I list the main proposals briefly.
\begin{description}
\item[The state vector is an expression of a law.] This  can be understood best by analogy with classical physics: the gravitational potential, just like the quantum wavefunction, is a function on configuration space which determines the dynamics of the particles, but we do not reify the gravitational potential, so why reify the wavefunction? Some advocates of this proposal (such as \citeN{durr95}) believe that it entails further technical work (for instance, to remove the apparent contingency); others, such as \citeN{montonconfiguration}, claim no such work is needed. (See \citeN{brownwallace} for my own --- rather negative --- view on this strategy.)
\item[The state vector is an expression of possibilities] (so that the state vector determines what is (naturally, or physically) \emph{possible} whereas the hidden variables determine what is \emph{actual}) This is a common interpretation of the state vector within modal interpretations (hence ``modal'', in fact); the obvious worry is that the merely possible is not normally allowed to have dynamical influences on the actual, nor to be physically contingent itself (as the wavefunction is normally taken to be). 
\item[The state vector is a property of the hidden variables] \cite{montonconfiguration}. Monton bases this proposal on the eigenvalue-eigenvector link: if the properties of a quantum system are given by the projectors on its Hilbert space, then \ket{\psi} will always instantiate the property represented by \proj{\psi}{\psi}. However, it is unclear to me whether treating the state vector as a ``real property'' is essentially different from treating it as a physical thing: by analogy, insofar as it is coherent at all to regard the electromagnetic field as a ``property'' of the charged particles, it does not seem particularly relevant to its reality.
\end{description}

\subsection{Further reading}

\citeN{holland}  and \citeN{cushingbohmbook} both provide detailed accounts of the de Broglie-Bohm theory (the former is a single-authored monograph, the latter a collection of articles). \citeN{dieksvermaas} is an excellent collection of articles on modal interpretations. 

\section{Relativistic quantum physics}\label{DMWWRQM}

A great deal is now known about the constraints which observed phenomena put on \emph{any} attempt to construct a relativistic quantum theory. In outline:
\begin{itemize}
\item Bell's Theorem shows us that any realist theory which is in a certain sense also ``local'' has a certain maximum level of correlation which it permits between measurement results obtained at distinct locations. 
\item QM predicts that this maximum level will be exceeded; therefore any realist theory which reproduces the results of QM is in a certain sense "non-local".
\item In any case, actual experiments have produced correlations between spatially separated detectors which violate this maximum level; therefore --- irrespective of quantum theory --- the true theory of the world is in a certain sense ``non-local''.
\end{itemize}
For some recent reviews of the issue, see \citeN{butterfieldnonlocality}, \citeN{maudlinbook}, \citeN[pp.\,148--186]{peres}, \citeN[part 2]{dicksonbook} and \citeN[chapter 2]{bubbook}. 

In this section I shall pursue a rather different line. In practice we actually \emph{have} a well-developed relativistic quantum theory: quantum field theory (QFT). So as well as asking what constraints are placed on interpretations of QM by relativity \emph{in general}, we should ask whether a given interpretative strategy can recover the \emph{actual} empirical predictions of the concrete relativistic quantum theory we have. Perhaps such a strategy will produce a theory which is not Lorentz-covariant at the \emph{fundamental} level, but given the enormous empirical successes of QFT, our strategy had better reproduce those successes at the observable level.

Firstly, though, I shall discuss the conceptual status of QFT itself. As we shall see, it is not without its own foundational problems.

\subsection{What is quantum field theory?}

The traditional way to describe a quantum field theory is through so-called \emph{operator-valued fields}: formally speaking, these are functions from points in spacetime to operators on a Hilbert space, normally written as (for instance) $\op{\psi}(x^\mu)$. Just as quantum \emph{particle} mechanics is specified by giving certain preferred operators (\op{Q} and \op{P}, say) on a Hilbert space, and a Hamiltonian defined in terms of them, so a quantum field theory is specified by giving certain operator-valued fields and a Hamiltonian defined in terms of \emph{them}. The simplest field theory, (real) \emph{Klein-Gordon field theory}, for instance, is given by two operator-valued fields $\op{\phi}(x^\mu)$ and $\op{\pi}(x^\mu)$; their mathematical structure is given (at least formally) by the \emph{equal-time commutation relations}
\be
\comm{\op{\phi}(\vctr{x},t)}{\op{\pi}(\vctr{y,t})}= i \hbar \delta(\vctr{x}-\vctr{y})
\ee
by analogy to the particle-mechanics commutation relations $[\op{Q},\op{P}]=i\hbar$, and the dynamics is generated by the Hamiltonian 
\be\label{DMWWKGhamil}
\op{H}=\frac{1}{2}\int \dr{x^3}\left(\op{\pi}^2(\vctr{x})+\nabla \op{\phi}^2(\vctr{x})+m^2 \op{\phi}^2(\vctr{x})\right).
\ee
(Field theories defined via equal-time commutation relations are called \emph{bosonic}; there is, however, an equally important class of field theories, the \emph{fermionic} theories, which are specified by equal-time \emph{anti-}commutation relations.)

Note that, as is customary in discussing field theory, we adopt the Heisenberg picture, in which it is observables rather than states which evolve under the Schr\"{o}dinger equation:
\be
\op{Q}(t)=\exp(-i   \op{H}t/\hbar)\op{Q}(0)\exp(+i/\op{H}t\hbar)
\ee
rather than
\be
\ket{\psi(t)}=\exp(-i  \op{H}t/\hbar )\ket{\psi(0)}.
\ee
This makes the covariance of the theory rather more manifest; it leads, however, to an unfortunate temptation, to regard the operator-valued fields as analogous to the classical fields (so that quantum field theory is a field theory because it deals with operator-valued fields, just as classical field theory is a field theory because it deals with real-valued fields). This is a serious error: treating the operator-valued fields as part of the \emph{ontology} of the theory is no more justified than treating the operators \op{Q} and \op{P} as part of the ontology of quantum particle mechanics.\footnote{There is a concrete ontological proposal for QM which \emph{does} treat the operators as part of the ontology (in QM and QFT both): the reading of Heisenberg offered by \citeN{deutschhayden}. For critical discussion of their proposals see \citeN{wallacetimpsonshort}.}

This being noted, what \emph{is} the ontology of quantum field theory? This is of course a heavily interpretation-dependent question, but for the moment let us consider it in the context of state-vector realism (as would perhaps be appropriate for an Everett interpretation of QFT, or for some hypothetical dynamical-collapse or hidden-variable variant on QFT). In section \ref{DMWWbaretheory} we saw that the most common approach to state-vector realism is wavefunction realism, according to which the $N$-particle wavefunction is a complex-valued function on $3N$-dimensional space. The analogous strategy in bosonic QFT interprets the state vector as a wavefunction over \emph{field configuration space}: it assigns a complex number to every field configuration at a given time. Mathematically at least this requires a preferred foliation of spacetime and seems to break the covariance of QFT; it is also unclear how to understand it when dealing with \emph{fermionic} quantum fields. An alternative strategy (briefly mentioned in section \ref{DMWWbaretheory}) assigns a density operator to each spacetime region (defining the expectation values of all observables definable in terms of the field operators within that spacetime region). The issue has so far received very little foundational attention: what little attention there has so far been to state-vector realism has been largely restricted to the nonrelativistic domain.

There is another complication in QFT, however: perhaps it should not be understood ontologically as a \emph{field} theory at all. At least in the case of a free field theory, there is a natural interpretation of the theory in terms of \emph{particles}: the Hilbert space possesses a so-called ``Fock-space decomposition''
\be
\mc{H}=\bigoplus_{i=0}^\infty \mc{H}_n
\ee
where $\mc{H}_n$ has a natural isomorphism to a space of $n$ identical free particles. (That is, $\mc{H}_n$ is isomorphic to either the symmetrised or the antisymmetrised $n$-fold tensor product of some 1-free-particle system, and the isomorphism preserves energy and momentum, and so in particular the Hamiltonian.)

Given this, it might be tempting to interpret QFTs ontologically as particle theories, and regard the `field' aspect as merely a heuristic --- useful in constructing the theory, but ultimately to be discarded (a proposal developed in detail by \citeN{weinberg}, and with links to the so-called `Segal quantisation' approach to QFT pioneered by \citeN{segal64},  \citeN{segal67} and explored philosophically by Saunders (\citeyearNP{saundersnegative,saunderscomplexnumbers}). However, it is at best highly unclear whether it can be sustained. Part of the reason for this is purely conceptual: as we have seen, the spatiotemporally localised field observables at least provide some kind of basis for understanding the ontology of a quantum field theory. But a particle ontology seems to require a different set of observables: in particular, it seems to require \emph{position observables} for each particle. And unfortunately it appears that no such observables actually exist (see Saunders (\citeyearNP{saunderscomplexnumbers}, \citeyearNP{saunderslocality}) and references therein).

In any case, there is a more serious  reason to be very skeptical about a particle ontology: namely that it does not seem to account for the status of particles in  \emph{interacting} quantum field theories, to which we now turn.

\subsection{Particles and quasiparticles}\label{DMWWparticles}

In order to understand the status of particles in interacting QFT, it is helpful to digress via solid-state physics, where the conceptual and mathematical issues are less controversial. At the \emph{fundamental} level, the ontology of a solid-state system --- a crystal, say --- is uncontroversial: it consists of a large collection of atoms in a regular lattice, and if we describe the (no doubt highly entangled) joint state of all of those atoms then we have said all there is to say about the crystal's quantum state. Nevertheless, we can describe the crystal in quantum-field-theoretic terms --- the ``field operators'' associated to a point $\vctr{x}$ are the  $x-$, $y-$ and $z-$ components of the displacement from equilibrium of the atom whose equilibrium location is centred on point $x$ (or the closest atom if no atom is exactly at $x$), together with the associated conjugate momenta. 

If the crystal is ``harmonic'' --- that is, if its Hamiltonian is quadratic in positions and momenta --- the ``quantum field'' produced in this fashion is \emph{free}, and has an exact (formal) interpretation in terms of particles, which can be understood as quantised, localisable, propagating disturbances in the crystal. These `particles' are known as \emph{phonons}.

What if it is \emph{not} harmonic? The Hamiltonian of the crystal can often be separated into two terms ---
\be
\op{H}=\op{H}_{free}+\op{H}_{int}
\ee
such that $\op{H}_{free}$ is quadratic and $\op{H}_{int}$ is small enough to be treated as a perturbation to $\op{H}_{free}$. It is then possible to understand the crystal in terms of \emph{interacting} phonons --- their free propagation determined by the first term in the Hamiltonian, their interactions, spontaneous creations, and spontaneous decays determined by the second term. The exact division into ``free'' and ``interacting'' terms is \emph{not} unique, but is chosen to minimise the scale of the interaction term \emph{in the actual system under study} (so that the choice of division is state-dependent). 

If there exists such a division of this sort, then phonons will provide a very useful analytic tool to study the crystal --- so that various of its properties, such as the heat capacity, can be calculated by treating the crystal as a gas of fairly-weakly-interacting phonons. 
The usefulness of the particle concept in the practical analysis of the crystal decreases as the interaction term becomes larger; in circumstances where it becomes so large as to render the particle concept useless, we must either seek a different division of the Hamiltonian into free and interacting terms, or give up on particle methods altogether.

This method is completely ubiquitous in solid-state physics. Vibrations are described in terms of ``phonons''; magnetic fields in terms of ``magnons''; propagating charges in plasmas in terms of ``plasmons, and so forth. The general term for the ``particles'' used in such analyses is \emph{quasi-particles}.
The particular quasi-particles to be used will vary from situation to situation: in one circumstance the system is best understood in terms of plasmons of a particular mass and charge; in another the temperature or density or whatever is such that a different division of the Hamiltonian into free and interacting terms is more useful and so we assign a different mass and charge to the plasmons (hence we talk about `temperature-dependence' of quasi-particle mass); in still another the plasmons are not useful at all and we look for a different quasi-particle analysis. 
Quasi-particles are emergent entities in solid-state physics: not ``fundamental'', not precisely defined, but no less real for all that. 

At least formally, ``particles'' in interacting quantum field theories turn out to be closely analogous to quasi-particles. They are described by a division of the Hamiltonian of an interacting QFT into ``free'' and ``interaction'' terms: their `intrinsic' properties --- notably mass and charge --- are set by the parameters in the free part of the Hamiltonian, and the interacting part determines the parameters that govern scattering of particle off particle. As with solid-state physics, this leads to a situation-dependence of the properties, which can be seen in a variety of contexts in particle physics, such as:
\begin{itemize}
\item The masses and charges of fundamental particles are often described as `scale-dependent'. What this really means is that the most useful division of the Hamiltonian into free and interacting parts depends on the characteristic scale at which the particles under consideration are actually interacting. Actually (as can be seen from the Feynman-diagram system used to analyse interacting QFTs), essentially \emph{any} choice of mass or charge will be admissible, provided one is prepared to pay the price in increasing complexity of interactions.
\item A sufficiently substantial shift of situation does not just change the parameters of particles, it changes the particles themselves. In nucleon physics at very short ranges, the approximately-free particles are \emph{quarks} (this is referred to as \emph{asymptotic freedom}); at longer ranges, the interactions between quarks become far stronger and it becomes more useful to treat \emph{nucleons} --- neutrons and protons --- as the approximately-free particles. (There is of course a sense in which a neutron is `made from' three quarks, but the matter is a good deal more subtle than popular-science treatments might suggest!) 
\end{itemize}
This strongly suggests that, in particle physics as in solid-state physics, the particle ontology is an emergent, higher-level phenomena, derivative on a lower-level field-theoretic ontology.

\subsection{QFT and the measurement problem}

Whether the Lagrangian or algebraic approach to QFT is ultimately found correct, the result will be the same: a Lorentz-covariant\footnote{Or possibly \emph{effectively} Lorentz covariant; cf.\,\citeN[section 3.4]{wallaceconceptualqft}.} unitary quantum theory, in which the primary dynamical variables are spacetime-local operators like field strengths and in which particles are approximate and emergent. This provides a Bare Quantum Formalism in the sense of section \ref{DMWWinterpretation}, as well as the resources to define macroscopically definite states in the sense of the Quantum Algorithm: they will be states which approximate states in non-relativistic QM (that is, states of definite particle number for `ordinary' particles like electrons and atomic nuclei, with energies low compared to the masses of those particles) and are in addition macroscopically definite according to the definitions used by non-relativistic QM. And  although decoherence has been relatively little studied in the relativistic domain (see \citeN{zurekqft} for an interesting exception) it seems reasonably clear that again decoherence picks out these macroscopically definite states as pointer-basis states (so that in particular pointer-basis states are definite-particle-number states at least as regards the number of nuclei and electrons).

This means that applying the Quantum Algorithm to QFT seems to be fairly straightforward, and should suffice to reproduce both the results of non-relativistic QM in the appropriate limit and to recover the actual methods used by field theorists to calculate  particle-production rates in particle-accelerator experiments.  This means that so long as we are concerned with \emph{pure interpretations} of QM (that is: the Everett interpretation; operationalism; the consistent-histories framework; the New Pragmatism; the original Copenhagen interpretation; and at least some variants of quantum logic) there are no essentially new issues introduced by QFT. If we can find a satisfactory pure interpretation of nonrelativistic QM, it should go through to QFT \emph{mutatis mutandis}.

Things are otherwise when we try to solve the measurement problem by modifying the formalism. The plain truth is that there are currently \emph{no} hidden-variable or dynamical-collapse theories which are generally accepted to reproduce the empirical predictions of any interacting quantum field theory. This is a separate matter to the \emph{conceptual} problems with such strategies, discussed in sections \ref{DMWWdynamicalcollapse} and \ref{DMWWhidden}. We do not even have QFT versions of these theories to have conceptual problems with.

Suppose that we tried to construct one; how should we go about it? Observe that in dynamical-collapse and hidden-variable theories alike, some ``preferred observables'' must be selected: either to determine the hidden variables, or to determine which sorts of superposition are to be dynamically suppressed. And in nonrelativistic physics there is a natural choice in both cases: position. GRW collapses suppress superpositions of positionally delocalised states; the de Broglie-Bohm hidden variables have definite positions. 

It is less clear what the `natural' choice would be in QFT. One possibility is field configuration --- so that, for instance, the QFT analogues of the de Broglie-Bohm corpuscles would be classical field configurations (see, \egc, \citeN{valentinicushing}, \citeN{kaloyerou}) . There are some technical difficulties with these proposals: in particular, it is unclear what ``classical field configurations'' are in the case of fermionic fields. But more seriously, it is debatable whether field-based modificatory strategies will actually succeed in reproducing the predictions of QM. For recall: as I argued in sections \ref{DMWWGRW} and \ref{DMWWhvconstraints}, it is crucial for these strategies that they are compatible with decoherence: that is, that the preferred observable is also decoherence-preferred. A dynamical-collapse theory which regards pointer-basis states as ``macroscopic superpositions'' will fail to suppress the \emph{right} superpositions; a hidden-variable theory whose hidden-variables are not decoherence-preferred will fail the Second and Fourth Constraints on hidden-variable strategies, and so will fail to recover effective quasiclassical dynamics. And in QFT (at least where fermions are concerned) the pointer-basis states are states of definite particle number, which in general are not diagonal in the field observables. (See \citeN{saundersbeables} for further reasons to doubt that field configurations are effective hidden variables for DBB.) 

This suggests an alternative choice: the preferred observables should include particle number. In a QFT version of DBB, for instance, there could be a certain number of corpuscles (with that number possibly time-dependent), one for each particle present in a given term in the wavefunction; in a dynamical-collapse theory, the collapse mechanism could suppress superpositions of different-particle-number states (see, \egc, \citeN{bellbeablesqft}, \citeN{bohmhiley}, D\"{u}rr \emph{et al}~(\citeyearNP{goldsteinqft03,goldsteinqft04})). This strategy faces a different problem, however: as was demonstrated in section \ref{DMWWparticles}, particles in QFT appear to be effective, emergent, and approximately defined concepts --- making them poor candidates for direct representation in the microphysics.

These remarks are not meant to imply that no modificatory strategy \emph{can} successfully reproduce the experimental predictions of QFT --- they are meant only to show that no such strategy has yet succeeded in reproducing them, and that there are some general reasons to expect that it will be extremely difficult. QFT, therefore, is significantly more hostile to solutions to the measurement problem that are \emph{not} pure interpretations. 

Michael Dickson, comparing pure interpretations with modificatory strategies, observes that
\begin{quote}
[I]t is not clear that `no new physics' is a virtue at all. After all, we trust QM as it happens to be formulated primarily because it is empirically very successful. Suppose, however, that some \emph{other} theory were equally successful, and were equally explanatory. To reject it because it is not the same as QM is, it seems, to be too much attached to the particular historical circumstances that gave rise to the formulation of QM. (\citeNP[p.\,60]{dicksonbook})
\end{quote}
QFT shows the true virtue of `no new physics'. It is not that we should prefer our existing physics to some equally-successful rival theory; it is rather that, in the relativistic domain at any rate, no such theory has been found.

\subsection{Further reading}

Of the great number of textbook discussions of QFT, \citeN{peskinschroeder} is particularly lucid; \citeN{cao} and \citeN{tellerbook} discuss QFT from a philosophical perspective. 

An issue glossed over in this section is the `renormalisation problem' or `problem of infinitites': mathematically, QFT seems ill-defined and calculations seem to have to be fixed up in an \emph{ad hoc} way to avoid getting pathological results. One response to the problem is to try to reformulate QFT on mathematically rigorous foundations (the so-called ``algebraic QFT program''); \citeN{haag} is a good introduction. This strategy seems to be popular with philosophers despite its failure thus far to reproduce any of the concrete predictions of `ordinary' QFT. Another response is to try to make sense of the apparent pathology; this is the mainstream position among physicists today (see \citeN{wilson}, Binney \emph{et al}~(\citeNP{binney}) and \citeN{wallaceconceptualqft}).

See \citeN[part IV]{bassighirardireview} for a review of progress in constructing relativistic dynamical-collapse theories; see also \citeN{myrvoldcollapse} for arguments that such theories could be Lorentz-covariant even if nonlocal.

\section{Conclusion}\label{DMWWconclusion}

The predictions of QM may be deduced from an rather well-defined mathematical formalism via an rather badly defined algorithm. Solving the measurement problem may be done in one of two ways: either we must provide an interpretation of the mathematical formalism which makes it a satisfactory physical theory and which entails the correctness of the algorithm; or we must invent a different formalism and a different algorithm which gives the same results, and then give a satisfactory interpretation of \emph{that} algorithm.

Interpreting formalisms is a distinctively philosophical project. Perhaps the most important theme of this review is that questions of interpretation depend on very broad philosophical positions, and so there are far fewer interpretations of a given formalism than meets the eye. In particular, if we are prepared to be both \emph{realist} and \emph{functionalist} about a given physical theory, and if we are prepared to accept classical logic as correct, there is exactly one interpretation of any given formalism, although we may be wrong about what it is!\footnote{Depending on one's general attitude to metaphysics, there may be \emph{some} questions not settled by this prescription: questions of the fundamental nature of the wavefunction, for instance, or whether particulars are bundles of properties. I side with \citeN{vanfraassenempirical} and \citeN{ladymanbook} in regarding these as largely non-questions, but in any case they do not seem to affect the validity or otherwise of a given solution of the measurement problem.} (And if we are realists but not functionalists, in effect we have further technical work to do, in picking out the \emph{non}-functional properties of our theory which are supposed to act as a supervenience base for consciousness or other higher-level properties.)

One sees this most clearly with the pure interpretations: your  general philosophical predilections lead you to one interpretation or another. Unapologetic instrumentalists and positivists are led naturally to Operationalism. Those who are more apologetic, but who wish to hold on to the insight that ``no phenomenon is a phenomenon until it is an observed phenomenon'' will adopt a position in the vicinity of Bohr's. Those willing to reject classical logic will (depending on the details of their proposed alternative) adopt some form of quantum-logic or consistent-histories interpretation. Functionalist realists will become Everettians, or else abandon pure interpretation altogether. (As noted above, non-functionalist realists are already in effect committed to a modificatory strategy.) 

In criticising a given pure interpretation, one can object to specific features (one may argue against Everett on grounds of probability, or against quantum logic on grounds of intertemporal probability specifications) but one is as likely to reject it because of disagreements with the general philosophical position (so those who regard positivism as dead are unlikely to be impressed by the purely \emph{technical} merits of Operationalism).

In the case of modificatory strategies, there ought to be rather less purely philosophical dispute, since these strategies are generally pursued in the name of realism. But we have seen that general philosophical problems are still highly relevant here: the nature of higher-level ontology, for instance, and the validity and implications of a general functionalism. 

If the \emph{interpretation} of a formalism is a distinctively philosophical task, designing new formalisms is physics, and at its hardest. The empirical success of the Quantum Algorithm --- never mind its foundational problems  --- is tremendous, underpinning tens of thousands of breakthroughs in twentieth- and twenty-first-century physics. No wonder, then that the new formalisms that \emph{have} been constructed are very closely based on the bare quantum formalism, supplementing it only by dynamical modifications which serve to eliminate all but one quasiclassical history, or by hidden variables which pick out one such history. We have seen that in practice this is achieved because one of the dynamically fundamental variables in NRQM --- position --- also suffices to distinguish different quasiclassical histories. We have also seen that in QFT this strategy fails, making it an as-yet-unsolved problem to construct alternatives to the bare formalism of QFT. 

Sometimes it is easy to forget how grave a problem the `measurement problem' actually is. One can too-easily slip into a mindset where there is one theory --- quantum mechanics --- and a myriad empirically-equivalent interpretations of that theory. Sometimes, indeed, it can seem that the discussion is carried out on largely \emph{aesthetic} grounds: do I find this theory's stochasticity more distressing than that interpretation's ontological excesses  or the other theory's violation of action-reaction?

The truth is very different. Most philosophers of physics are realists, or at least sympathetic to realism. At present we know of \emph{at most} one realist (and classical-logic) solution to the measurement problem: the Everett interpretation. If the Everett interpretation is incoherent for one reason or another (as is probably the mainstream view among philosophers of physics, if not among physicists) then currently we have \emph{no} realist solutions to the measurement problem. There are interesting \emph{research programs}, which (disregarding their potential conceptual problems)  have successfully reproduced the predictions of non-relativistic physics, but a research program is not a theory.

\citeN{penroseroadtoreality} regards ``measurement problem'' as too anodyne a term for this conceptual crisis in physics. He proposes ``measurement paradox''; perhaps philosophers would do well to follow his lead.


\begin{thebibliography}{}

\bibitem[\protect\citeauthoryear{Aharonov and Vaidman}{Aharonov and
  Vaidman}{1996}]{aharonovvaidman}
Aharonov, Y. and L.~Vaidman (1996).
\newblock About position measurements which do not show the {B}ohmian particle
  position.
\newblock In \citeN{cushingbohmbook}, pp.\  141--154.

\bibitem[\protect\citeauthoryear{Albert}{Albert}{1992}]{albertqmbook}
Albert, D.~Z. (1992).
\newblock {\em Quantum mechanics and experience}.
\newblock Cambridge, Massachussets: Harvard University Press.

\bibitem[\protect\citeauthoryear{Albert}{Albert}{1996}]{albertmetaphysics}
Albert, D.~Z. (1996).
\newblock Elementary quantum metaphysics.
\newblock In \citeN{cushingbohmbook}, pp.\  277--284.

\bibitem[\protect\citeauthoryear{Albert and Loewer}{Albert and
  Loewer}{1988}]{albertloewermm}
Albert, D.~Z. and B.~Loewer (1988).
\newblock {I}nterpreting the {M}any {W}orlds {I}nterpretation.
\newblock {\em Synthese\/}~{\em 77}, 195--213.

\bibitem[\protect\citeauthoryear{Albert and Loewer}{Albert and
  Loewer}{1990}]{albertloewer1990}
Albert, D.~Z. and B.~Loewer (1990).
\newblock {W}anted {D}ead or {A}live: {T}wo {A}ttempts to {S}olve
  {S}chr{\"o}dinger's {P}aradox.
\newblock In A.~Fine, M.~Forbes, and L.~Wessels (Eds.), {\em Proceedings of the
  1990 Biennial Meeting of the Philosophy of Science Association}, Volume~1,
  pp.\  277--285. East Lansing, Michigan: Philosophy of Science Association.

\bibitem[\protect\citeauthoryear{Albert and Loewer}{Albert and
  Loewer}{1996}]{albertloewer1996}
Albert, D.~Z. and B.~Loewer (1996).
\newblock {T}ails of {S}chr{\"o}dinger's {C}at.
\newblock In R.~Clifton (Ed.), {\em Perspectives on Quantum Reality}, pp.\
  81--92. Dordrecht: Kluwer Academic Publishers.

\bibitem[\protect\citeauthoryear{Allori, Goldstein, Tumulka, and Zanghi}{Allori
  et~al.}{2007}]{goldsteinontology}
Allori, V., S.~Goldstein, R.~Tumulka, and N.~Zanghi (2007).
\newblock On the common structure of {B}ohmian mechanics and the
  {G}hirardi-{R}imini-{W}eber theory.
\newblock Forthcoming in \emph{British Journal for the Philosophy of Science};
  available online at http://arxiv.org/abs/quant-ph/0603027.

\bibitem[\protect\citeauthoryear{Anandan and Brown}{Anandan and
  Brown}{1995}]{anandanbrown}
Anandan, J. and H.~R. Brown (1995).
\newblock On the reality of space-time geometry and the wavefunction.
\newblock {\em Foundations of Physics\/}~{\em 25}, 349--360.

\bibitem[\protect\citeauthoryear{Anglin and Zurek}{Anglin and
  Zurek}{1996}]{zurekqft}
Anglin, J.~R. and W.~H. Zurek (1996).
\newblock Decoherence of quantum fields: Pointer states and predictability.
\newblock {\em Physical Review D\/}~{\em 53}, 7327--7335.

\bibitem[\protect\citeauthoryear{Armstrong}{Armstrong}{1968}]{armstrongmind}
Armstrong, D. (1968).
\newblock {\em A Materialist Theory of the Mind}.
\newblock London: Routledge and Kegan Paul.

\bibitem[\protect\citeauthoryear{Arntzenius}{Arntzenius}{1990}]{arntzenius1990}
Arntzenius, F. (1990).
\newblock {K}ochen's interpretation of quantum mechanics.
\newblock {\em Proceedings of the Philosophy of Science Association\/}~{\em 1},
  241--249.

\bibitem[\protect\citeauthoryear{Arntzenius}{Arntzenius}{1998}]{arntzeniusmoda%
l}
Arntzenius, F. (1998).
\newblock Curioser and curioser: A personal evaluation of modal
  interpretations.
\newblock In \citeN{dieksvermaas}, pp.\  337--377.

\bibitem[\protect\citeauthoryear{Bacciagalupp}{Bacciagalupp}{2005}]{bacciagalu%
ppiencyclopedia}
Bacciagalupp, G. (2005).
\newblock The role of decoherence in quantum mechanics.
\newblock In the Stanford Encyclopedia of Philosophy (Summer 2005 edition),
  Edward N. Zalta (ed.), available online at
  http://plato.stanford.edu/archives/sum2005/entries/qm-decoherence.

\bibitem[\protect\citeauthoryear{Bacciagaluppi}{Bacciagaluppi}{1995}]{bacciaga%
luppi1995}
Bacciagaluppi, G. (1995).
\newblock {K}ochen-{S}pecker theorem in the modal interpretation of quantum
  mechanics.
\newblock {\em International Journal of Theoretical Physics\/}~{\em 34},
  1206--1215.

\bibitem[\protect\citeauthoryear{Bacciagaluppi}{Bacciagaluppi}{1998}]{bacciaga%
luppidynamics}
Bacciagaluppi, G. (1998).
\newblock {B}ohm-{B}ell dynamics in the modal interpretation.
\newblock In \citeN{dieksvermaas}, pp.\  177--212.

\bibitem[\protect\citeauthoryear{Bacciagaluppi}{Bacciagaluppi}{1999}]{bacciaga%
luppinelson}
Bacciagaluppi, G. (1999).
\newblock {N}elsonian mechanics revisited.
\newblock {\em Foundations of Physics Letters\/}~{\em 12}, 1--16.

\bibitem[\protect\citeauthoryear{Bacciagaluppi}{Bacciagaluppi}{2000}]{bacciaga%
luppi2000}
Bacciagaluppi, G. (2000).
\newblock Delocalized properties in the modal interpretation of quantum
  mechanics.
\newblock {\em Foundations of Physics\/}~{\em 30}, 1431--1444.

\bibitem[\protect\citeauthoryear{Bacciagaluppi and Dickson}{Bacciagaluppi and
  Dickson}{1999}]{bacciagaluppidickson}
Bacciagaluppi, G. and M.~Dickson (1999).
\newblock Dynamics for modal interpretations.
\newblock {\em Foundations of Physics\/}~{\em 29}, 1165--1201.

\bibitem[\protect\citeauthoryear{Bacciagaluppi, Donald, and
  Vermaas}{Bacciagaluppi et~al.}{1995}]{bacciagaluppidonaldvermaas}
Bacciagaluppi, G., M.~J. Donald, and P.~E. Vermaas (1995).
\newblock Continuity and discontinuity of definite properties in the modal
  interpretation.
\newblock {\em Helvetica Physica Acta\/}~{\em 68}, 679--704.

\bibitem[\protect\citeauthoryear{Bacciagaluppi and Hemmo}{Bacciagaluppi and
  Hemmo}{1996}]{bacciagaluppihemmo}
Bacciagaluppi, G. and M.~Hemmo (1996).
\newblock Modal interpretations, decoherence and measurements.
\newblock {\em Studies in the History and Philosophy of Modern Physics\/}~{\em
  27}, 239--277.

\bibitem[\protect\citeauthoryear{Baker}{Baker}{2007}]{baker}
Baker, D. (2007).
\newblock Measurement outcomes and probability in everettian quantum mechanics.
\newblock {\em Studies in the History and Philosophy of Modern Physics\/}~{\em
  38\/}(38), 153--169.

\bibitem[\protect\citeauthoryear{Ballentine}{Ballentine}{1970}]{ballentinerevi%
ew}
Ballentine, L.~E. (1970).
\newblock The statistical interpretation of quantum mechanics.
\newblock {\em Reviews of Modern Physics\/}~{\em 42}, 358--381.

\bibitem[\protect\citeauthoryear{Ballentine}{Ballentine}{1990}]{ballentine}
Ballentine, L.~E. (1990).
\newblock {\em Quantum Mechanics}.
\newblock Englewood Cliffs: Prentice Hall.

\bibitem[\protect\citeauthoryear{Barbour}{Barbour}{1994}]{barbour2}
Barbour, J.~B. (1994).
\newblock {T}he {T}imelessness of {Q}uantum {G}ravity: {II}. {T}he {A}ppearence
  of {D}ynamics in {S}tatic {C}onfigurations.
\newblock {\em Classical and Quantum Gravity\/}~{\em 11}, 2875--2897.

\bibitem[\protect\citeauthoryear{Barbour}{Barbour}{1999}]{barbour99}
Barbour, J.~B. (1999).
\newblock {\em The End of Time}.
\newblock London: Weidenfeld and Nicholson.

\bibitem[\protect\citeauthoryear{Barnum, Caves, Finkelstein, Fuchs, and
  Schack}{Barnum et~al.}{2000}]{barnumetal}
Barnum, H., C.~M. Caves, J.~Finkelstein, C.~A. Fuchs, and R.~Schack (2000).
\newblock {Q}uantum {P}robability from {D}ecision {T}heory?
\newblock {\em Proceedings of the Royal Society of London\/}~{\em A456},
  1175--1182.
\newblock Available online at http://arXiv.org/abs/quant-ph/9907024.

\bibitem[\protect\citeauthoryear{Barrett}{Barrett}{1998}]{barrettbare}
Barrett, J.~A. (1998).
\newblock The bare theory and how to fix it.
\newblock In \citeN{dieksvermaas}, pp.\  319--326.

\bibitem[\protect\citeauthoryear{Barrett}{Barrett}{1999}]{barrettbook}
Barrett, J.~A. (1999).
\newblock {\em The quantum mechanics of minds and worlds}.
\newblock Oxford: Oxford University Press.

\bibitem[\protect\citeauthoryear{Bassi and Ghirardi}{Bassi and
  Ghirardi}{1999}]{bassighirardi99}
Bassi, A. and G.~Ghirardi (1999).
\newblock More about dynamical reduction and the enumeration principle.
\newblock {\em British Journal for the Philosophy of Science\/}~{\em 50}, 719.

\bibitem[\protect\citeauthoryear{Bassi and Ghirardi}{Bassi and
  Ghirardi}{2003}]{bassighirardireview}
Bassi, A. and G.~Ghirardi (2003).
\newblock Dynamical reduction models.
\newblock {\em Physics Reports\/}~{\em 379}, 257.

\bibitem[\protect\citeauthoryear{Bassi and Ghirardi}{Bassi and
  Ghirardi}{2000}]{ghirardiconsistent}
Bassi, A. and G.~C. Ghirardi (2000).
\newblock Decoherent histories and realism.
\newblock {\em Journal of Statistical Physics\/}~{\em 98}, 457--494.
\newblock Available online at http://arxiv.org/abs/quant-ph/9912031.

\bibitem[\protect\citeauthoryear{Bell}{Bell}{1981a}]{bellsocks}
Bell, J. (1981a).
\newblock Bertlmann's socks and the nature of reality.
\newblock {\em Journal de Physique\/}~{\em 42}, C2 41--61.
\newblock Reprinted in \citeN{bellbook}, pp.\,139--158.

\bibitem[\protect\citeauthoryear{Bell}{Bell}{1966}]{bell1966}
Bell, J.~S. (1966).
\newblock On the problem of hidden variables in quantum mechanics.
\newblock {\em Reviews of Modern Physics\/}~{\em 38}, 447--452.
\newblock Reprinted in \citeN{bellbook}, pp.\,1--13.

\bibitem[\protect\citeauthoryear{Bell}{Bell}{1981b}]{bellqmforcosmologists}
Bell, J.~S. (1981b).
\newblock {Q}uantum {M}echanics for {C}osmologists.
\newblock In C.~J. Isham, R.~Penrose, and D.~Sciama (Eds.), {\em Quantum
  Gravity 2: a second {O}xford Symposium}, Oxford. Clarendon Press.
\newblock Reprinted in \citeN{bellbook}, pp.\,117--138.

\bibitem[\protect\citeauthoryear{Bell}{Bell}{1982}]{bellimpossible}
Bell, J.~S. (1982).
\newblock On the impossible pilot wave.
\newblock {\em Foundations of Physics\/}~{\em 12}, 989--999.
\newblock Reprinted in \citeN{bellbook}, pp.\,159--168.

\bibitem[\protect\citeauthoryear{Bell}{Bell}{1984}]{bellbeablesqft}
Bell, J.~S. (1984).
\newblock Beables for quantum field theory.
\newblock CERN preprint CERN-TH 4035/84. Reprinted in \citeN{bellbook},
  pp.\,173--180.

\bibitem[\protect\citeauthoryear{Bell}{Bell}{1987}]{bellbook}
Bell, J.~S. (1987).
\newblock {\em Speakable and Unspeakable in Quantum Mechanics}.
\newblock Cambridge: Cambridge University Press.

\bibitem[\protect\citeauthoryear{Binney, Dowrick, Fisher, and Newman}{Binney
  et~al.}{1992}]{binney}
Binney, J.~J., N.~J. Dowrick, A.~J. Fisher, and M.~E.~J. Newman (1992).
\newblock {\em The Theory of Critical Phenomena : an introduction to the
  renormalisation group}.
\newblock Oxford: Oxford University Press.

\bibitem[\protect\citeauthoryear{Bohm}{Bohm}{1952}]{bohm}
Bohm, D. (1952).
\newblock {A} {S}uggested {I}nterpretation of {Q}uantum {T}heory in {T}erms of
  ``{H}idden'' {V}ariables.
\newblock {\em Physical Review\/}~{\em 85}, 166--193.

\bibitem[\protect\citeauthoryear{Bohm and Hiley}{Bohm and
  Hiley}{1993}]{bohmhiley}
Bohm, D. and B.~J. Hiley (1993).
\newblock {\em The Undivided Universe: An Ontological Interpretation of Quantum
  Theory}.
\newblock London: Routledge and Kegan Paul.

\bibitem[\protect\citeauthoryear{Bricmont}{Bricmont}{2001}]{bricmont}
Bricmont, J.~e. (Ed.) (2001).
\newblock {\em Chance in Physics: Foundations and Perspectives}, London.
  Springer.

\bibitem[\protect\citeauthoryear{Brown}{Brown}{1996}]{brownlockwood}
Brown, H.~R. (1996).
\newblock Comment on {L}ockwood.
\newblock {\em British Journal for the Philosophy of Science\/}~{\em 47},
  189--248.

\bibitem[\protect\citeauthoryear{Brown, Dewdney, and Horton}{Brown
  et~al.}{1995}]{brownfooleddetectors}
Brown, H.~R., C.~Dewdney, and G.~Horton (1995).
\newblock {B}ohm particles and their detection in the light of neutron
  interferometry.
\newblock {\em Foundations of Physics\/}~{\em 25}, 329--347.

\bibitem[\protect\citeauthoryear{Brown, Sj{\"o}qvist, and Bacciagaluppi}{Brown
  et~al.}{1999}]{brownstatistics}
Brown, H.~R., E.~Sj{\"o}qvist, and G.~Bacciagaluppi (1999).
\newblock Remarks on identical particles in de {B}roglie-{B}ohm theory.
\newblock {\em Physics Letters A\/}~{\em 251}, 229--235.

\bibitem[\protect\citeauthoryear{Brown and Wallace}{Brown and
  Wallace}{2005}]{brownwallace}
Brown, H.~R. and D.~Wallace (2005).
\newblock Solving the measurement problem: de {B}roglie-{B}ohm loses out to
  {E}verett.
\newblock {\em Studies in the History and Philosophy of Modern Physics\/}~{\em
  35}, 517--540.

\bibitem[\protect\citeauthoryear{Bub}{Bub}{1997}]{bubbook}
Bub, J. (1997).
\newblock {\em Interpreting the Quantum World}.
\newblock Cambridge: Cambridge University Press.

\bibitem[\protect\citeauthoryear{Bub and Clifton}{Bub and
  Clifton}{1996}]{bubclifton}
Bub, J. and R.~Clifton (1996).
\newblock A uniqueness theorem for ``no collapse'' interpretations of quantum
  mechanics.
\newblock {\em Studies in the History and Philosophy of Modern Physics\/}~{\em
  27}, 181--219.

\bibitem[\protect\citeauthoryear{Bub, Clifton, and Goldstein}{Bub
  et~al.}{2000}]{bubcliftongoldstein}
Bub, J., R.~Clifton, and S.~Goldstein (2000).
\newblock Revised proof of the uniqueness theorem for `no collapse'
  interpretations of quantum mechanics.
\newblock {\em Studies in the History and Philosophy of Modern Physics\/}~{\em
  31}, 95.

\bibitem[\protect\citeauthoryear{Bub, Clifton, and Monton}{Bub
  et~al.}{1998}]{bubcliftonmonton}
Bub, J., R.~Clifton, and B.~Monton (1998).
\newblock The bare theory has no clothes.
\newblock In G.~Hellman and R.~Healey (Eds.), {\em Quantum Measurement: Beyond
  Paradox}, pp.\  32--51. Minneapolis: University of Minnesota Press.

\bibitem[\protect\citeauthoryear{Busch}{Busch}{1998}]{buschmodal}
Busch, P. (1998).
\newblock Remarks on unsharp observables, objectification, and modal
  interpretations.
\newblock In \citeN{dieksvermaas}, pp.\  279--288.

\bibitem[\protect\citeauthoryear{Busch, Lahti, and Mittelstaedt}{Busch
  et~al.}{1996}]{buschmeasurement}
Busch, P., P.~J. Lahti, and P.~Mittelstaedt (1996).
\newblock {\em The Quantum Theory of Measurement\/} (2nd revised ed.).
\newblock Berlin: Springer-Verlag.

\bibitem[\protect\citeauthoryear{Butterfield}{Butterfield}{1992}]{butterfieldn%
onlocality}
Butterfield, J.~N. (1992).
\newblock {B}ell's theorem: What it takes.
\newblock {\em British Journal for the Philosophy of Science\/}~{\em 43},
  41--83.

\bibitem[\protect\citeauthoryear{Butterfield}{Butterfield}{1996}]{butterfielde%
verett}
Butterfield, J.~N. (1996).
\newblock {W}hither the {M}inds?
\newblock {\em British Journal for the Philosophy of Science\/}~{\em 47},
  200--221.

\bibitem[\protect\citeauthoryear{Cao}{Cao}{1997}]{cao}
Cao, T.~Y. (1997).
\newblock {\em Conceptual Developments of 20th Century Field Theories}.
\newblock Cambridge: Cambridge University Press.

\bibitem[\protect\citeauthoryear{Caves, Fuchs, Mann, and Renes}{Caves
  et~al.}{2004}]{cavesetalgleason}
Caves, C.~M., C.~A. Fuchs, K.~Mann, and J.~M. Renes (2004).
\newblock {G}leason-type derivations of the quantum probability rule for
  generalized measurements.
\newblock {\em Foundations of Physics\/}~{\em 34}, 193.

\bibitem[\protect\citeauthoryear{Caves, Fuchs, and Schack}{Caves
  et~al.}{2002}]{cavesetalprobability}
Caves, C.~M., C.~A. Fuchs, and R.~Schack (2002).
\newblock Quantum probabilities as {B}ayesian probabilities.
\newblock {\em Physical Review A\/}~{\em 65}, 022305.

\bibitem[\protect\citeauthoryear{Caves and Schack}{Caves and
  Schack}{2005}]{cavesschackfrequentism}
Caves, C.~M. and R.~Schack (2005).
\newblock Properties of the frequency operator do no imply the quantum
  probability postulate.
\newblock {\em Annals of Physics\/}~{\em 315}, 123--146.

\bibitem[\protect\citeauthoryear{Clifton}{Clifton}{1996}]{cliftonmodal96}
Clifton, R. (1996).
\newblock The properties of modal interpretations of quantum mechanics.
\newblock {\em British Journal for the Philosophy of Science\/}~{\em 47},
  371--398.

\bibitem[\protect\citeauthoryear{Clifton and Monton}{Clifton and
  Monton}{1999}]{cliftonmonton1}
Clifton, R. and B.~Monton (1999).
\newblock {L}osing your {M}arbles in {W}avefunction {C}ollapse {T}heories.
\newblock {\em British Journal for the Philosophy of Science\/}~{\em 50},
  697--717.

\bibitem[\protect\citeauthoryear{Clifton and Monton}{Clifton and
  Monton}{2000}]{cliftonmonton2}
Clifton, R. and B.~Monton (2000).
\newblock {C}ounting {M}arbles with `{A}ccessible' {M}ass {D}ensity: a reply to
  {B}assi and {G}hirardi.
\newblock {\em British Journal for the Philosophy of Science\/}~{\em 51},
  155--164.

\bibitem[\protect\citeauthoryear{Cordero}{Cordero}{1999}]{cordero}
Cordero, A. (1999).
\newblock Are {GRW} tails as bad as they say?
\newblock {\em Philosophy of Science\/}~{\em 66}, S59--S71.

\bibitem[\protect\citeauthoryear{Cramer}{Cramer}{1986}]{cramer86}
Cramer, J.~G. (1986).
\newblock The transactional interpretation of quantum mechanics.
\newblock {\em Reviews of Modern Physics\/}~{\em 58\/}(58), 647--687.

\bibitem[\protect\citeauthoryear{Cramer}{Cramer}{1988}]{cramer88}
Cramer, J.~G. (1988).
\newblock An overview of the transactional interpretation of quantum mechanics.
\newblock {\em International Journal of Theoretical Physics\/}~{\em 27\/}(27),
  227--236.

\bibitem[\protect\citeauthoryear{Cushing}{Cushing}{1994}]{cushingcopenhagen}
Cushing, J.~T. (1994).
\newblock {\em Quantum Mechanics: Historical Contingency and the {C}openhagen
  Hegemony}.
\newblock Chicago: University of Chicago Press.

\bibitem[\protect\citeauthoryear{Cushing, Fine, and Goldstein}{Cushing
  et~al.}{1996}]{cushingbohmbook}
Cushing, J.~T., A.~Fine, and S.~Goldstein (Eds.) (1996).
\newblock {\em Bohmian Mechanics and Quantum Theory: An Appraisal}, Dordrecht.
  Kluwer Academic Publishers.

\bibitem[\protect\citeauthoryear{Davies and Brown}{Davies and
  Brown}{1986}]{ghostatom}
Davies, P. and J.~Brown (Eds.) (1986).
\newblock {\em The Ghost in the Atom}, Cambridge. Cambridge University Press.

\bibitem[\protect\citeauthoryear{Dennett}{Dennett}{1991}]{dennettrealpatterns}
Dennett, D.~C. (1991).
\newblock Real patterns.
\newblock {\em Journal of Philosophy\/}~{\em 87}, 27--51.
\newblock Reprinted in \textit{Brainchildren}, D.\, Dennett, (London: Penguin
  1998) pp.\, 95--120.

\bibitem[\protect\citeauthoryear{Dennett}{Dennett}{2005}]{dennettsweetdreams}
Dennett, D.~C. (2005).
\newblock {\em Sweet Dreams: Philosophical Objections to a Science of
  Consciousness}.
\newblock Cambridge, MA: MIT Press.

\bibitem[\protect\citeauthoryear{Deotto and Ghirardi}{Deotto and
  Ghirardi}{1998}]{deottoghirardi}
Deotto, E. and G.~Ghirardi (1998).
\newblock {B}ohmian mechanics revisited.
\newblock {\em Foundations of Physics\/}~{\em 28}, 1--30.
\newblock Available online at http://arxiv.org/abs/quant-ph/9704021.

\bibitem[\protect\citeauthoryear{Deutsch}{Deutsch}{1985}]{deutsch85}
Deutsch, D. (1985).
\newblock {Q}uantum {T}heory as a {U}niversal {P}hysical {T}heory.
\newblock {\em International Journal of Theoretical Physics\/}~{\em 24\/}(1),
  1--41.

\bibitem[\protect\citeauthoryear{Deutsch}{Deutsch}{1986}]{deutschghost}
Deutsch, D. (1986).
\newblock Interview.
\newblock In \citeN{ghostatom}, pp.\  83--105.

\bibitem[\protect\citeauthoryear{Deutsch}{Deutsch}{1996}]{deutschlockwood}
Deutsch, D. (1996).
\newblock {C}omment on {L}ockwood.
\newblock {\em British Journal for the Philosophy of Science\/}~{\em 47},
  222--228.

\bibitem[\protect\citeauthoryear{Deutsch}{Deutsch}{1999}]{deutschprob}
Deutsch, D. (1999).
\newblock Quantum theory of probability and decisions.
\newblock {\em Proceedings of the Royal Society of London\/}~{\em A455},
  3129--3137.

\bibitem[\protect\citeauthoryear{Deutsch and Hayden}{Deutsch and
  Hayden}{2000}]{deutschhayden}
Deutsch, D. and P.~Hayden (2000).
\newblock Information flow in entangled quantum systems.
\newblock {\em Proceedings of the Royal Society of London\/}~{\em A456},
  1759--1774.

\bibitem[\protect\citeauthoryear{Dewdney, Hardy, and Squires}{Dewdney
  et~al.}{1993}]{dewdneyetal}
Dewdney, C., L.~Hardy, and E.~J. Squires (1993).
\newblock How late measurements of quantum trajectories can fool a detector.
\newblock {\em Physics Letters\/}~{\em 184A}, 6--11.

\bibitem[\protect\citeauthoryear{DeWitt and Graham}{DeWitt and
  Graham}{1973}]{dewittgraham}
DeWitt, B. and N.~Graham (Eds.) (1973).
\newblock {\em The many-worlds interpretation of quantum mechanics}.
\newblock Princeton: Princeton University Press.

\bibitem[\protect\citeauthoryear{Dickson}{Dickson}{1998}]{dicksonbook}
Dickson, M. (1998).
\newblock {\em Quantum Chance and Non-Locality: Probability and Non-locality in
  the Interpretations of Quantum Mechanics}.
\newblock Cambridge: Cambridge University Press.

\bibitem[\protect\citeauthoryear{Dickson}{Dickson}{2001}]{dicksonlogic}
Dickson, M. (2001).
\newblock Quantum logic is alive $\wedge$ (it is true $\vee$ it is false).
\newblock {\em Philosophy of Science\/}~{\em 68}, S274--S287.

\bibitem[\protect\citeauthoryear{Dieks and Vermaas}{Dieks and
  Vermaas}{1998}]{dieksvermaas}
Dieks, D. and P.~E. Vermaas (Eds.) (1998).
\newblock {\em The Modal Interpretation of Quantum Mechanics}, Dordrecht.
  Kluwer Academic Publishers.

\bibitem[\protect\citeauthoryear{Donald}{Donald}{1990}]{donald90}
Donald, M.~J. (1990).
\newblock Quantum theory and the brain.
\newblock {\em Proceedings of the Royal Society of London A\/}~{\em 427},
  43--93.

\bibitem[\protect\citeauthoryear{Donald}{Donald}{1992}]{donald92}
Donald, M.~J. (1992).
\newblock A priori probability and localized observers.
\newblock {\em Foundations of Physics\/}~{\em 22}, 1111--1172.

\bibitem[\protect\citeauthoryear{Donald}{Donald}{1998}]{donaldmodal}
Donald, M.~J. (1998).
\newblock Discontinuity and continuity of definite properties in the modal
  interpretation.
\newblock In \citeN{dieksvermaas}, pp.\  213--222.

\bibitem[\protect\citeauthoryear{Donald}{Donald}{2002}]{donald02}
Donald, M.~J. (2002).
\newblock Neural unpredictability, the interpretation of quantum theory, and
  the mind-body problem.
\newblock Available online at http://arxiv.org/abs/quant-ph/0208033.

\bibitem[\protect\citeauthoryear{Dowker and Herbauts}{Dowker and
  Herbauts}{2005}]{dowkerherbauts}
Dowker, F. and I.~Herbauts (2005).
\newblock The status of the wave function in dynamical collapse models.
\newblock {\em Foundations of Physics Letters\/}~{\em 18}, 499--518.

\bibitem[\protect\citeauthoryear{Dowker and Kent}{Dowker and
  Kent}{1996}]{dowkerkent}
Dowker, F. and A.~Kent (1996).
\newblock On the consistent histories approach to quantum mechanics.
\newblock {\em Journal of Statistical Physics\/}~{\em 82}, 1575--1646.

\bibitem[\protect\citeauthoryear{D{\"u}rr, Goldstein, Tumulka, and
  Zanghi}{D{\"u}rr et~al.}{2004}]{goldsteinqft03}
D{\"u}rr, D., S.~Goldstein, R.~Tumulka, and N.~Zanghi (2004).
\newblock {B}ohmian mechanics and quantum field theory.
\newblock {\em Physical Review Letters\/}~{\em 93}, 090402.

\bibitem[\protect\citeauthoryear{D{\"u}rr, Goldstein, Tumulka, and
  Zanghi}{D{\"u}rr et~al.}{2005}]{goldsteinqft04}
D{\"u}rr, D., S.~Goldstein, R.~Tumulka, and N.~Zanghi (2005).
\newblock {B}ell-type quantum field theories.
\newblock {\em Journal of Physics\/}~{\em A38}, R1.

\bibitem[\protect\citeauthoryear{D{\"u}rr, Goldstein, and Zanghi}{D{\"u}rr
  et~al.}{1996}]{durrincushing}
D{\"u}rr, D., S.~Goldstein, and N.~Zanghi (1996).
\newblock {B}ohmian mechanics as the foundation of quantum mechanics.
\newblock In \citeN{cushingbohmbook}, pp.\  21--44.

\bibitem[\protect\citeauthoryear{D{\"u}rr, Goldstein, and Zanghi}{D{\"u}rr
  et~al.}{1997}]{durr95}
D{\"u}rr, D., S.~Goldstein, and N.~Zanghi (1997).
\newblock {B}ohmian mechanics and the meaning of the wave function.
\newblock In R.~S. Cohen, M.~Horne, and J.~Stachel (Eds.), {\em Potentiality,
  Entanglement and Passion-at-a-Distance --- Quantum Mechanical Studies in
  Honor of {A}bner {S}himony}. Dordrecht: Kluwer.
\newblock Available online at http://arxiv.org/abs/quant-ph/9512031.

\bibitem[\protect\citeauthoryear{D{\"u}rr, Goldstein, and Zhangi}{D{\"u}rr
  et~al.}{1992}]{durr1992}
D{\"u}rr, D., S.~Goldstein, and N.~Zhangi (1992).
\newblock Quantum equilibrium and the origin of absolute uncertainty.
\newblock {\em Journal of Statistical Physics\/}~{\em 67}, 843--907.

\bibitem[\protect\citeauthoryear{Englert, Scully, Sussmann, and
  Walther}{Englert et~al.}{1992}]{englertetal}
Englert, B.~G., M.~O. Scully, G.~Sussmann, and H.~Walther (1992).
\newblock Surrealistic {B}ohm trajectories.
\newblock {\em Zeitschrift f{\"u}r Naturforschung\/}~{\em 47A}, 1175--1186.

\bibitem[\protect\citeauthoryear{Everett}{Everett}{1957}]{everett}
Everett, H. (1957).
\newblock {R}elative {S}tate {F}ormulation of {Q}uantum {M}echanics.
\newblock {\em Review of Modern Physics\/}~{\em 29}, 454--462.
\newblock Reprinted in \citeN{dewittgraham}.

\bibitem[\protect\citeauthoryear{Farhi, Goldstone, and Gutmann}{Farhi
  et~al.}{1989}]{fgg}
Farhi, E., J.~Goldstone, and S.~Gutmann (1989).
\newblock How probability arises in quantum-mechanics.
\newblock {\em Annals of Physics\/}~{\em 192}, 368--382.

\bibitem[\protect\citeauthoryear{Foster and Brown}{Foster and
  Brown}{1988}]{brownondeutsch}
Foster, S. and H.~R. Brown (1988).
\newblock On a recent attempt to define the interpretation basis in the many
  worlds interpretation of quantum mechanics.
\newblock {\em International Journal of Theoretical Physics\/}~{\em 27},
  1507--1531.

\bibitem[\protect\citeauthoryear{Fuchs and Peres}{Fuchs and
  Peres}{2000a}]{fuchsperes}
Fuchs, C. and A.~Peres (2000a).
\newblock Quantum theory needs no ``interpretation''.
\newblock {\em Physics Today\/}~{\em 53\/}(3), 70--71.

\bibitem[\protect\citeauthoryear{Fuchs and Peres}{Fuchs and
  Peres}{2000b}]{fuchsperesreply}
Fuchs, C.~A. and A.~Peres (2000b).
\newblock Fuchs and peres reply.
\newblock {\em Physics Today\/}~{\em 53}, 14.

\bibitem[\protect\citeauthoryear{Gell-Mann and Hartle}{Gell-Mann and
  Hartle}{1990}]{gellmannhartle}
Gell-Mann, M. and J.~B. Hartle (1990).
\newblock {Q}uantum {M}echanics in the {L}ight of {Q}uantum {C}osmology.
\newblock In W.~H. Zurek (Ed.), {\em Complexity, Entropy and the Physics of
  Information}, pp.\  425--459. Redwood City, California: Addison-Wesley.

\bibitem[\protect\citeauthoryear{Gell-Mann and Hartle}{Gell-Mann and
  Hartle}{1993}]{gellmannhartle93}
Gell-Mann, M. and J.~B. Hartle (1993).
\newblock Classical equations for quantum systems.
\newblock {\em Physical Review D\/}~{\em 47}, 3345--3382.

\bibitem[\protect\citeauthoryear{Ghiardi}{Ghiardi}{2002}]{ghirardiencyclopedia}
Ghiardi, G.~C. (2002).
\newblock Collapse theories.
\newblock In the Stanford Encyclopedia of Philosophy (Summer 2002 edition),
  Edward N. Zalta (ed.), available online at
  http://plato.stanford.edu/archives/spr2002/entries/qm-collapse .

\bibitem[\protect\citeauthoryear{Ghirardi, Grassi, and Benatti}{Ghirardi
  et~al.}{1995}]{ghirardigrassibenati}
Ghirardi, G., R.~Grassi, and F.~Benatti (1995).
\newblock Describing the macroscopic world: Closing the circle within the
  dynamical reduction program.
\newblock {\em Foundations of Physics\/}~{\em 25}, 5--38.

\bibitem[\protect\citeauthoryear{Ghirardi, Rimini, and Weber}{Ghirardi
  et~al.}{1986}]{grw}
Ghirardi, G., A.~Rimini, and T.~Weber (1986).
\newblock {U}nified {D}ynamics for {M}icro and {M}acro {S}ystems.
\newblock {\em Physical Review D\/}~{\em 34}, 470--491.

\bibitem[\protect\citeauthoryear{Goldstein}{Goldstein}{2002}]{goldsteinstatmec%
h}
Goldstein, S. (2002).
\newblock {B}oltzmann's approach to statistical mechanics.
\newblock In \citeN{bricmont}.
\newblock Available online at http://arxiv.org/abs/cond-mat/0105242.

\bibitem[\protect\citeauthoryear{Goldstein, Taylor, Tumulka, and
  Zaghi}{Goldstein et~al.}{2005}]{goldstein05}
Goldstein, S., J.~Taylor, R.~Tumulka, and N.~Zaghi (2005).
\newblock Are all particles real?
\newblock {\em Studies in the History and Philosophy of Modern Physics\/}~{\em
  36}, 103--112.

\bibitem[\protect\citeauthoryear{Greaves}{Greaves}{2004}]{greaves}
Greaves, H. (2004).
\newblock Understanding {D}eutsch's probability in a deterministic multiverse.
\newblock {\em Studies in the History and Philosophy of Modern Physics\/}~{\em
  35}, 423--456.

\bibitem[\protect\citeauthoryear{Greaves}{Greaves}{2007}]{greavesepistemic}
Greaves, H. (2007).
\newblock On the {E}verettian epistemic problem.
\newblock {\em Studies in the History and Philosophy of Modern Physics\/}~{\em
  38}, 120--152.

\bibitem[\protect\citeauthoryear{Griffiths}{Griffiths}{1984}]{griffiths}
Griffiths, R.~B. (1984).
\newblock Consistent histories and the interpretation of quantum mechanics.
\newblock {\em Journal of Statistical Physics\/}~{\em 36}, 219--272.

\bibitem[\protect\citeauthoryear{Griffiths}{Griffiths}{2002}]{griffithsbook}
Griffiths, R.~B. (2002).
\newblock {\em Consistent Quantum Theory}.
\newblock Cambridge: Cambridge University Press.

\bibitem[\protect\citeauthoryear{Haag}{Haag}{1996}]{haag}
Haag, R. (1996).
\newblock {\em Local Quantum Theory: Fields, Particles, Algebras}.
\newblock Berlin: Springer-Verlag.

\bibitem[\protect\citeauthoryear{Hemmo and Pitowsky}{Hemmo and
  Pitowsky}{2007}]{hemmopitowsky07}
Hemmo, M. and I.~Pitowsky (2007).
\newblock Quantum probability and many worlds.
\newblock {\em Studies in the History and Philosophy of Modern Physics\/}~{\em
  38}, 333--350.

\bibitem[\protect\citeauthoryear{Hiley, Callaghan, and Maroney}{Hiley
  et~al.}{2000}]{hileyfooleddetectors}
Hiley, B.~J., R.~E. Callaghan, and O.~J. Maroney (2000).
\newblock Quantum trajectories, real, surreal, or an approximation to a deeper
  process?
\newblock Available online at http://arxiv.org/abs/quant-ph/0010020.

\bibitem[\protect\citeauthoryear{Hofstadter and Dennett}{Hofstadter and
  Dennett}{1981}]{hofstadterdennett}
Hofstadter, D.~R. and D.~C. Dennett (Eds.) (1981).
\newblock {\em The Mind's {I}: Fantasies and Reflections on Self and Soul}.
\newblock London: Penguin.

\bibitem[\protect\citeauthoryear{Holland}{Holland}{1993}]{holland}
Holland, P. (1993).
\newblock {\em The Quantum Theory of Motion}.
\newblock Cambridge: Cambridge University Press.

\bibitem[\protect\citeauthoryear{Joos, Zeh, Kiefer, Giulini, Kubsch, and
  Stametescu}{Joos et~al.}{2003}]{joosetal}
Joos, E., H.~D. Zeh, C.~Kiefer, D.~Giulini, J.~Kubsch, and I.~O. Stametescu
  (2003).
\newblock {\em Decoherence and the Appearence of a Classical World in Quantum
  Theory\/} (2nd ed.).
\newblock Berlin: Springer.

\bibitem[\protect\citeauthoryear{Kaloyerou}{Kaloyerou}{1996}]{kaloyerou}
Kaloyerou, P.~N. (1996).
\newblock An ontological interpretation of boson fields.
\newblock In \citeN{cushingbohmbook}, pp.\  155--168.

\bibitem[\protect\citeauthoryear{Kent}{Kent}{1990}]{kent}
Kent, A. (1990).
\newblock {A}gainst {M}any-{W}orlds {I}nterpretations.
\newblock {\em International Journal of Theoretical Physics\/}~{\em A5}, 1764.
\newblock Revised version available at http://www.arxiv.org/abs/gr-qc/9703089.

\bibitem[\protect\citeauthoryear{Kent}{Kent}{1996a}]{kenthistory}
Kent, A. (1996a).
\newblock Quasiclassical dynamics in a closed quantum system.
\newblock {\em Physical Review A\/}~{\em 54}, 4670--5675.

\bibitem[\protect\citeauthoryear{Kent}{Kent}{1996b}]{kentbohmhistory}
Kent, A. (1996b).
\newblock Remarks on consistent histories and {B}ohmian mechanics.
\newblock In \citeN{cushingbohmbook}, pp.\  343--352.

\bibitem[\protect\citeauthoryear{Kim}{Kim}{1998}]{kim98}
Kim, J. (1998).
\newblock {\em Mind in a physical world}.
\newblock Cambridge, Massachusets: MIT Press/Bradford.

\bibitem[\protect\citeauthoryear{Kochen and Specker}{Kochen and
  Specker}{1967}]{kochenspecker}
Kochen, S. and E.~Specker (1967).
\newblock The problem of hidden variables in quantum mechanics.
\newblock {\em Journal of Mathematics and Mechanics\/}~{\em 17}, 59--87.

\bibitem[\protect\citeauthoryear{Leggett}{Leggett}{2002}]{leggett}
Leggett, A.~J. (2002).
\newblock Testing the limits of quantum mechanics: Motivation, state of play,
  prospects.
\newblock {\em Journal of Physics: Condensed Matter\/}~{\em 14}, R415--R451.

\bibitem[\protect\citeauthoryear{Levin}{Levin}{2004}]{levinencyclopedia}
Levin, J. (2004).
\newblock Functionalism.
\newblock In the Stanford Encyclopedia of Philosophy (Fall 2004 edition),
  Edward N. Zalta (ed.), available online at
  http://plato.stanford.edu/archives/fall2004/entries/functionalism.

\bibitem[\protect\citeauthoryear{Lewis}{Lewis}{1974}]{lewisradical}
Lewis, D. (1974).
\newblock Radical interpretation.
\newblock {\em Synthese\/}~{\em 23}, 331--44.
\newblock Reprinted in David Lewis, \textit{Philosophical Papers}, Volume I
  (Oxford University Press, Oxford, 1983).

\bibitem[\protect\citeauthoryear{Lewis}{Lewis}{1980}]{lewischance}
Lewis, D. (1980).
\newblock A subjectivist's guide to objective chance.
\newblock In R.~C. Jeffrey (Ed.), {\em Studies in Inductive Logic and
  Probability}, Volume~II. Berkeley: University of California Press.
\newblock Reprinted in David Lewis, \textit{Philosophical Papers}, Volume II
  (Oxford University Press, Oxford, 1986).

\bibitem[\protect\citeauthoryear{Lewis}{Lewis}{1986}]{lewispapers2}
Lewis, D. (1986).
\newblock {\em Philosophical Papers, Vol.\,II}.
\newblock Oxford: Oxford University Press.

\bibitem[\protect\citeauthoryear{Lewis}{Lewis}{1997}]{lewis94}
Lewis, P.~J. (1997).
\newblock {Q}uantum {M}echanics, {O}rthogonality, and {C}ounting.
\newblock {\em British Journal for the Philosophy of Science\/}~{\em 48},
  313--328.

\bibitem[\protect\citeauthoryear{Lewis}{Lewis}{2003}]{lewis2003}
Lewis, P.~J. (2003).
\newblock Counting marbles: Reply to critics.
\newblock {\em British Journal for the Philosophy of Science\/}~{\em 54},
  165--170.

\bibitem[\protect\citeauthoryear{Lewis}{Lewis}{2004a}]{lewisconfiguration}
Lewis, P.~J. (2004a).
\newblock Life in configuration space.
\newblock {\em British Journal for the Philosophy of Science\/}~{\em 55},
  713--729.

\bibitem[\protect\citeauthoryear{Lewis}{Lewis}{2004b}]{lewisfuzzy}
Lewis, P.~J. (2004b).
\newblock Quantum mechanics and ordinary language: the fuzzy link.
\newblock {\em Philosophy of Science\/}~{\em 70\/}(55), 713--729.

\bibitem[\protect\citeauthoryear{Lewis}{Lewis}{2005}]{lewis2005}
Lewis, P.~J. (2005).
\newblock Interpreting spontaneous collapse theories.
\newblock {\em Studies in the History and Philosophy of Modern Physics\/}~{\em
  36}, 165--180.

\bibitem[\protect\citeauthoryear{Lewis}{Lewis}{2007}]{lewissu}
Lewis, P.~J. (2007).
\newblock Uncertainty and probability for branching selves.
\newblock {\em Studies in the History and Philosophy of Modern Physics\/}~{\em
  38}, 1--14.

\bibitem[\protect\citeauthoryear{Lockwood}{Lockwood}{1989}]{lockwoodbook}
Lockwood, M. (1989).
\newblock {\em Mind, Brain and the Quantum: the compound `I'}.
\newblock Oxford: Blackwell Publishers.

\bibitem[\protect\citeauthoryear{Lockwood}{Lockwood}{1996}]{lockwoodbjps1}
Lockwood, M. (1996).
\newblock `{M}any {M}inds' {I}nterpretations of {Q}uantum {M}echanics.
\newblock {\em British Journal for the Philosophy of Science\/}~{\em 47},
  159--188.

\bibitem[\protect\citeauthoryear{Maudlin}{Maudlin}{2002}]{maudlinbook}
Maudlin, T. (2002).
\newblock {\em Quantum Non-Locality and Relativity: Metaphysical Intimations of
  Modern Physics\/} (2nd edition ed.).
\newblock Oxford: Blackwell.

\bibitem[\protect\citeauthoryear{Monton}{Monton}{2004a}]{montonmassdensity}
Monton, B. (2004a).
\newblock The problem of ontology for spontaneous collapse theories.
\newblock {\em Studies in the History and Philosophy of Modern Physics\/}~{\em
  35}, 407--421.

\bibitem[\protect\citeauthoryear{Monton}{Monton}{2004b}]{montonconfiguration}
Monton, B. (2004b).
\newblock Quantum mechanics and 3{N}-dimensional space.
\newblock Forthcoming; available online from http://philsci-archive.pitt.edu.

\bibitem[\protect\citeauthoryear{Myrvold}{Myrvold}{2002}]{myrvoldcollapse}
Myrvold, W. (2002).
\newblock On peaceful coexistence: is the collapse postulate incompatible with
  relativity?
\newblock {\em Studies in the History and Philosophy of Modern Physics\/}~{\em
  33}, 435--466.

\bibitem[\protect\citeauthoryear{Nelson}{Nelson}{1966}]{nelsonarticle}
Nelson, E. (1966).
\newblock Derivation of the {S}chr{\"o}dinger equation from {N}ewtonian
  mechanics.
\newblock {\em Physical Review\/}~{\em 150}, 1079--1085.

\bibitem[\protect\citeauthoryear{Nelson}{Nelson}{1985}]{nelsonbook}
Nelson, E. (1985).
\newblock {\em Quantum Fluctuations}.
\newblock Princeton: Princeton University Press.

\bibitem[\protect\citeauthoryear{Newton-Smith}{Newton-Smith}{2000}]{NSunderdet%
ermination}
Newton-Smith, W.~S. (2000).
\newblock Underdetermination of theory by data.
\newblock In W.~H. Newton-Smith (Ed.), {\em A Companion to the Philosophy of
  Science}, pp.\  532--536. Oxford: Blackwell.

\bibitem[\protect\citeauthoryear{Nielsen and Chuang}{Nielsen and
  Chuang}{2000}]{nielsenchuang}
Nielsen, M.~A. and I.~L. Chuang (2000).
\newblock {\em Quantum Computation and Quantum Information}.
\newblock Cambridge: Cambridge University Press.

\bibitem[\protect\citeauthoryear{Omnes}{Omnes}{1988}]{omnes}
Omnes, R. (1988).
\newblock Logical reformulation of quantum mechanics. i. foundations.
\newblock {\em Journal of Statistical Physics\/}~{\em 53}, 893--932.

\bibitem[\protect\citeauthoryear{Omnes}{Omnes}{1994}]{omnesbook}
Omnes, R. (1994).
\newblock {\em The Interpretation of Quantum Mechanics}.
\newblock Princeton: Princeton University Press.

\bibitem[\protect\citeauthoryear{Page}{Page}{1996}]{pagesensible}
Page, D.~N. (1996).
\newblock Sensible quantum mechanics: are probabilities only in the mind?
\newblock {\em International Journal of Modern Physics\/}~{\em D5}, 583--596.

\bibitem[\protect\citeauthoryear{Papineau}{Papineau}{1996}]{papineaubjps}
Papineau, D. (1996).
\newblock {M}any {M}inds are {N}o {W}orse than {O}ne.
\newblock {\em British Journal for the Philosophy of Science\/}~{\em 47},
  233--241.

\bibitem[\protect\citeauthoryear{Parfit}{Parfit}{1984}]{parfitbook}
Parfit, D. (1984).
\newblock {\em Reasons and Persons}.
\newblock Oxford: Oxford University Press.

\bibitem[\protect\citeauthoryear{Pearle}{Pearle}{1989}]{pearle}
Pearle, P. (1989).
\newblock {C}ombining {S}tochastic {D}ynamical {S}tate-{V}ector {R}eduction
  with {S}pontaneous {L}ocalization.
\newblock {\em Physical Review A\/}~{\em 39\/}(5), 2277--2289.

\bibitem[\protect\citeauthoryear{Penrose}{Penrose}{2004}]{penroseroadtoreality}
Penrose, R. (2004).
\newblock {\em The Road to Reality: a Complete Guide to the Laws of the
  Universe}.
\newblock London: Jonathon Cape.

\bibitem[\protect\citeauthoryear{Peres}{Peres}{1993}]{peres}
Peres, A. (1993).
\newblock {\em Quantum Theory: Concepts and Methods}.
\newblock Dordrecht: Kluwer Academic Publishers.

\bibitem[\protect\citeauthoryear{Peskin and Schroeder}{Peskin and
  Schroeder}{1995}]{peskinschroeder}
Peskin, M.~E. and D.~V. Schroeder (1995).
\newblock {\em An introduction to Quantum Field Theory}.
\newblock Reading, Massachusetts: Addison-Wesley.

\bibitem[\protect\citeauthoryear{Price}{Price}{1996}]{pricebook}
Price, H. (1996).
\newblock {\em Time's Arrow and Archimedes' Point: New Directions for the
  Physics of Time}.
\newblock Oxford: Oxford University Press.

\bibitem[\protect\citeauthoryear{Psillos}{Psillos}{1999}]{psillosbook}
Psillos, S. (1999).
\newblock {\em Scientific Realism: How Science Tracks Truth}.
\newblock London: Routledge.

\bibitem[\protect\citeauthoryear{Redhead}{Redhead}{1987}]{redheadbook}
Redhead, M. (1987).
\newblock {\em Incompleteness, Nonlocality and Realism: A Prolegomenon to the
  Philosophy of Quantum Mechanics}.
\newblock Oxford: Oxford University Press.

\bibitem[\protect\citeauthoryear{Reutsche}{Reutsche}{1998}]{reutsche}
Reutsche, L. (1998).
\newblock {H}ow {C}lose is `{C}lose {E}nough'?
\newblock In \citeN{dieksvermaas}, pp.\  223--240.

\bibitem[\protect\citeauthoryear{Ross}{Ross}{2000}]{rossinformation}
Ross, D. (2000).
\newblock Rainforest realism: a {D}ennettian theory of existence.
\newblock In D.~Ross, A.~Brook, and D.~Thompson (Eds.), {\em {D}ennett's
  Philosophy: a comprehensive assessment}, pp.\  147--168. Cambridge,
  Massachusets: MIT Press/Bradford.

\bibitem[\protect\citeauthoryear{Ross and Ladyman}{Ross and
  Ladyman}{2007}]{ladymanbook}
Ross, D. and J.~Ladyman (2007).
\newblock {\em Every Thing Must Go: Metaphysics Naturalized}.
\newblock Oxford: Oxford University Press.

\bibitem[\protect\citeauthoryear{Ross and Spurrett}{Ross and
  Spurrett}{2004}]{rossspurrett}
Ross, D. and D.~Spurrett (2004).
\newblock What to say to a sceptical metaphysician: a defense manual for
  cognitive and behavioral scientists.
\newblock {\em Behavioral and Brain Sciences\/}~{\em 27}, 603--627.

\bibitem[\protect\citeauthoryear{Saunders}{Saunders}{1991}]{saundersnegative}
Saunders, S. (1991).
\newblock The negative-energy sea.
\newblock In S.~Saunders and H.~R. Brown (Eds.), {\em The Philosophy of
  Vacuum}, pp.\  65--108. Oxford: Clarendon Press.

\bibitem[\protect\citeauthoryear{Saunders}{Saunders}{1992}]{saunderscomplexnum%
bers}
Saunders, S. (1992).
\newblock {L}ocality, {C}omplex {N}umbers and {R}elativistic {Q}uantum
  {T}heory.
\newblock {\em Philosophy of Science Association 1992\/}~{\em 1}, 365--380.

\bibitem[\protect\citeauthoryear{Saunders}{Saunders}{1993}]{saundersevolution}
Saunders, S. (1993).
\newblock {D}ecoherence, {R}elative {S}tates, and {E}volutionary {A}daptation.
\newblock {\em Foundations of Physics\/}~{\em 23}, 1553--1585.

\bibitem[\protect\citeauthoryear{Saunders}{Saunders}{1995}]{saundersdecoherenc%
e}
Saunders, S. (1995).
\newblock {T}ime, {D}ecoherence and {Q}uantum {M}echanics.
\newblock {\em Synthese\/}~{\em 102}, 235--266.

\bibitem[\protect\citeauthoryear{Saunders}{Saunders}{1997}]{saundersmetaphysic%
s}
Saunders, S. (1997).
\newblock {N}aturalizing {M}etaphysics.
\newblock {\em The Monist\/}~{\em 80\/}(1), 44--69.

\bibitem[\protect\citeauthoryear{Saunders}{Saunders}{1998a}]{saunderslocality}
Saunders, S. (1998a).
\newblock A dissolution of the problem of locality.
\newblock {\em Proceedings of the Philosophy of Science Association\/}~{\em 2},
  88--98.

\bibitem[\protect\citeauthoryear{Saunders}{Saunders}{1998b}]{saundersprobabili%
ty}
Saunders, S. (1998b).
\newblock {T}ime, {Q}uantum {M}echanics, and {P}robability.
\newblock {\em Synthese\/}~{\em 114}, 373--404.

\bibitem[\protect\citeauthoryear{Saunders}{Saunders}{1999}]{saundersbeables}
Saunders, S. (1999).
\newblock The `beables' of relativistic pilot-wave theory.
\newblock In J.~Butterfield and C.~Pagonis (Eds.), {\em From Physics to
  Philosophy}, pp.\  71--89. Cambridge: Cambridge University Press.

\bibitem[\protect\citeauthoryear{Saunders}{Saunders}{2005}]{saunderscopenhagen}
Saunders, S. (2005).
\newblock Complementarity and scientific rationality.
\newblock {\em Foundations of Physics\/}~{\em 35}, 347--372.

\bibitem[\protect\citeauthoryear{Saunders and Wallace}{Saunders and
  Wallace}{2007}]{saunderswallace}
Saunders, S. and D.~Wallace (2007).
\newblock Branching and uncertainty.
\newblock Available online from http://philsci-archive.pitt.edu.

\bibitem[\protect\citeauthoryear{Schlosshauer}{Schlosshauer}{2006}]{schlosshau%
er}
Schlosshauer, M. (2006).
\newblock Experimental motivation and empirical consistency in minimal
  no-collapse quantum mechanics.
\newblock {\em Annals of Physics\/}~{\em 321}, 112--149.
\newblock Available online at http://arxiv.org/abs/quant-ph/0506199.

\bibitem[\protect\citeauthoryear{Segal}{Segal}{1964}]{segal64}
Segal, I. (1964).
\newblock Quantum field and analysis in the solution manifolds of differential
  equations.
\newblock In W.~Martin and I.~Segal (Eds.), {\em Analysis in Function Space},
  pp.\  129--153. Cambridge, MA: MIT Press.

\bibitem[\protect\citeauthoryear{Segal}{Segal}{1967}]{segal67}
Segal, I. (1967).
\newblock Representations of the canonical commutation relations.
\newblock In F.~Lurcat (Ed.), {\em {C}argese Lectures on Theoretical Physics}.
  New York: Gordon and Breach.

\bibitem[\protect\citeauthoryear{Spekkens}{Spekkens}{2007}]{spekkens}
Spekkens, R.~W. (2007).
\newblock In defense of the epistemic view of quantum states: a toy theory.
\newblock {\em Physical Review A\/}~{\em 75}, 032110.

\bibitem[\protect\citeauthoryear{Struyve and Westman}{Struyve and
  Westman}{2006}]{westman}
Struyve, W. and H.~Westman (2006).
\newblock A new pilot-wave model for quantum field theory.
\newblock {\em AIP Conference Proceedings\/}~{\em 844}, 321.

\bibitem[\protect\citeauthoryear{Styer, Sobottka, Holladay, Brun, Griffiths,
  and Harris}{Styer et~al.}{2000}]{fuchsperescomments}
Styer, D., S.~Sobottka, W.~Holladay, T.~A. Brun, R.~B. Griffiths, and P.~Harris
  (2000).
\newblock Quantum theory--interpretation, formulation, inspiration.
\newblock {\em Physics Today\/}~{\em 53}, 11.

\bibitem[\protect\citeauthoryear{Taylor}{Taylor}{1986}]{taylorghost}
Taylor, J. (1986).
\newblock Interview.
\newblock In \citeN{ghostatom}, pp.\  106--117.

\bibitem[\protect\citeauthoryear{Teller}{Teller}{1995}]{tellerbook}
Teller, P. (1995).
\newblock {\em An Interpretative Introduction to Quantum Field Theory}.
\newblock Princeton: Princeton University Press.

\bibitem[\protect\citeauthoryear{Vaidman}{Vaidman}{2002}]{vaidmanencyclopedia}
Vaidman, L. (2002).
\newblock {T}he {M}any-{W}orlds {I}nterpretation of {Q}uantum {M}echanics.
\newblock In the Stanford Encyclopedia of Philosophy (Summer 2002 edition),
  Edward N. Zalta (ed.), available online at
  http://plato.stanford.edu/archives/sum2002/entries/qm-manyworlds .

\bibitem[\protect\citeauthoryear{Valentini}{Valentini}{1996}]{valentinicushing}
Valentini, A. (1996).
\newblock Pilot-wave theory of fields, gravitation and cosmology.
\newblock In \citeN{cushingbohmbook}, pp.\  45--67.

\bibitem[\protect\citeauthoryear{Valentini}{Valentini}{2001}]{valentini01}
Valentini, A. (2001).
\newblock Hidden variables, statistical mechanics and the early universe.
\newblock In \citeN{bricmont}, pp.\  165--181.
\newblock Available online at http://arxiv.org/abs/quant-ph/0104067.

\bibitem[\protect\citeauthoryear{Valentini}{Valentini}{2004}]{valentiniblackho%
le}
Valentini, A. (2004).
\newblock Extreme test of quantum theory with black holes.
\newblock Available online at http://arxiv.org/abs/astro-ph/0412503.

\bibitem[\protect\citeauthoryear{Valentini and Westman}{Valentini and
  Westman}{2005}]{valentiniwestman}
Valentini, A. and H.~Westman (2005).
\newblock Dynamical origin of quantum probabilities.
\newblock {\em Proceedings of the Royal Society of London A\/}~{\em 461},
  187--193.

\bibitem[\protect\citeauthoryear{Van~Fraassen}{Van~Fraassen}{1980}]{vanfraasse%
nscientificimage}
Van~Fraassen, B.~C. (1980).
\newblock {\em The Scientific Image}.
\newblock Oxford: Oxford University Press.

\bibitem[\protect\citeauthoryear{van Fraassen}{van
  Fraassen}{1991}]{vanfraassenquantum}
van Fraassen, B.~C. (1991).
\newblock {\em Quantum Mechanics}.
\newblock Oxford: Oxford University Press.

\bibitem[\protect\citeauthoryear{van Fraassen}{van
  Fraassen}{2002}]{vanfraassenempirical}
van Fraassen, B.~C. (2002).
\newblock {\em The Empirical Stance}.
\newblock New Haven: Yale University Press.

\bibitem[\protect\citeauthoryear{Vermaas}{Vermaas}{1998}]{vermaas}
Vermaas, P.~E. (1998).
\newblock The pros and cons of the {K}ochen-{D}ieks and the atomic modal
  interpretation.
\newblock In \citeN{dieksvermaas}, pp.\  103--148.

\bibitem[\protect\citeauthoryear{Wallace}{Wallace}{2002}]{wallaceworlds}
Wallace, D. (2002).
\newblock Worlds in the {E}verett {I}nterpretation.
\newblock {\em Studies in the History and Philosopy of Modern Physics\/}~{\em
  33}, 637--661.

\bibitem[\protect\citeauthoryear{Wallace}{Wallace}{2003a}]{wallacestructure}
Wallace, D. (2003a).
\newblock {E}verett and {S}tructure.
\newblock {\em Studies in the History and Philosophy of Modern Physics\/}~{\em
  34}, 87--105.

\bibitem[\protect\citeauthoryear{Wallace}{Wallace}{2003b}]{decshort}
Wallace, D. (2003b).
\newblock {E}verettian rationality: defending {D}eutsch's approach to
  probability in the {E}verett interpretation.
\newblock {\em Studies in the History and Philosophy of Modern Physics\/}~{\em
  34}, 415--439.

\bibitem[\protect\citeauthoryear{Wallace}{Wallace}{2004}]{wallacebbs}
Wallace, D. (2004).
\newblock Protecting cognitive science from quantum theory.
\newblock {\em Behavioral and Brain Sciences\/}~{\em 27}, 636--637.

\bibitem[\protect\citeauthoryear{Wallace}{Wallace}{2005}]{wallacebranching}
Wallace, D. (2005).
\newblock Language use in a branching universe.
\newblock Forthcoming; Available online from http://philsci-archive.pitt.edu.

\bibitem[\protect\citeauthoryear{Wallace}{Wallace}{2006a}]{wallaceepist}
Wallace, D. (2006a).
\newblock Epistemology quantized: circumstances in which we should come to
  believe in the {E}verett interpretation.
\newblock {\em British Journal for the Philosophy of Science\/}~{\em 57},
  655--689.

\bibitem[\protect\citeauthoryear{Wallace}{Wallace}{2006b}]{wallaceconceptualqf%
t}
Wallace, D. (2006b).
\newblock In defence of naivet{\'{e}}: The conceptual status of lagrangian
  quantum field theory.
\newblock {\em Synthese\/}~{\em 151}, 33--80.

\bibitem[\protect\citeauthoryear{Wallace}{Wallace}{2006c}]{wallace3branch}
Wallace, D. (2006c).
\newblock Probability in three kinds of branching universe.
\newblock Forthcoming.

\bibitem[\protect\citeauthoryear{Wallace}{Wallace}{2007}]{wallaceprobdec}
Wallace, D. (2007).
\newblock Quantum probability from subjective likelihood: Improving on
  {D}eutsch's proof of the probability rule.
\newblock {\em Studies in the History and Philosophy of Modern Physics\/}~{\em
  38}, 311--332.

\bibitem[\protect\citeauthoryear{Wallace and Timpson}{Wallace and
  Timpson}{2007}]{wallacetimpsonshort}
Wallace, D. and C.~Timpson (2007).
\newblock Non-locality and gauge freedom in {D}eutsch and {H}ayden's
  formulation of quantum mechanics.
\newblock {\em Foundations of Physics Letters\/}~{\em 37}, 951--955.

\bibitem[\protect\citeauthoryear{Weinberg}{Weinberg}{1995}]{weinberg}
Weinberg, S. (1995).
\newblock {\em The Quantum Theory of Fields}, Volume~1.
\newblock Cambridge: Cambridge University Press.

\bibitem[\protect\citeauthoryear{Wilson and Kogut}{Wilson and
  Kogut}{1974}]{wilson}
Wilson, K.~G. and J.~Kogut (1974).
\newblock The {R}enormalization {G}roup and the $\epsilon$ {E}xpansion.
\newblock {\em Physics Reports\/}~{\em 12C}, 75--200.

\bibitem[\protect\citeauthoryear{Zeh}{Zeh}{1993}]{zeh93}
Zeh, H.~D. (1993).
\newblock There are no quantum jumps, nor are there particles!
\newblock {\em Physics Letters\/}~{\em A172}, 189.

\bibitem[\protect\citeauthoryear{Zeh}{Zeh}{1999}]{zehbohm}
Zeh, H.~D. (1999).
\newblock Why {B}ohm's quantum theory?
\newblock {\em Foundations of Physics Letters\/}~{\em 12}, 197--200.

\bibitem[\protect\citeauthoryear{Zurek}{Zurek}{1991}]{zurek91}
Zurek, W.~H. (1991).
\newblock Decoherence and the transition from quantum to classical.
\newblock {\em Physics Today\/}~{\em 43}, 36--44.
\newblock Revised version available online at
  http://arxiv.org/abs/quant-ph/0306072.

\bibitem[\protect\citeauthoryear{Zurek}{Zurek}{1998}]{zurekroughguide}
Zurek, W.~H. (1998).
\newblock Decoherence, einselection, and the quantum origins of the classical:
  the rough guide.
\newblock {\em Philosophical Transactions of the Royal Society of
  London\/}~{\em A356}, 1793--1820.

\bibitem[\protect\citeauthoryear{Zurek}{Zurek}{2003a}]{zurek01review}
Zurek, W.~H. (2003a).
\newblock Decoherence, einselection, and the quantum origins of the classical.
\newblock {\em Reviews of Modern Physics\/}~{\em 75}, 715.

\bibitem[\protect\citeauthoryear{Zurek}{Zurek}{2003b}]{zurekenvariance03}
Zurek, W.~H. (2003b).
\newblock Environment-assisted invariance, causality, and probabilities in
  quantum physics.
\newblock {\em Physical Review Letters\/}~{\em 90}, 120403.

\bibitem[\protect\citeauthoryear{Zurek}{Zurek}{2005}]{zurekenvariance05}
Zurek, W.~H. (2005).
\newblock Probabilities from entanglement, {B}orn's rule from envariance.
\newblock {\em Physical Review A\/}~{\em 71}, 052105.

\end{thebibliography}
\end{document}